**TITLE:** Halide perovskites: Is it all about the interfaces?

**AUTHORS:** Philip Schulz[1,2], David Cahen[3], Antoine Kahn[4]

**AFFILIATIONS:** [1]Institut Photovoltaïque d'Île-de-France (IPVF); [2]National Center for Photovoltaics, National Renewable Energy Laboratory; [3]Department of Materials and Interfaces, Weizmann Institute of Science; [4]Department of Electrical Engineering, Princeton University;

**ABSTRACT:** Design and modification of the interfaces, always a critical issue for semiconductor devices, has become the primary tool to harness the full potential of halide perovskite (HaP)-based ones. In particular the outstanding improvements in HaP solar cell performance and stability can be primarily ascribed to a careful choice of the interfacial layout in the layer stack. In this review we describe the unique challenges and opportunities of these approaches (section A). For this purpose, we first elucidate the basic physical and chemical properties of the exposed HaP thin film and crystal surface (section B). We then lay out the energetic alignment processes to adjacent transport and buffer layers (section C) and finally elaborate on the impact of the interface formation on how well/poor a device functions. Based on those sections we then present a road map for the next steps in interfacial design principles for HaP semiconductors (section D).

**TABLE OF CONTENTS:**

**MANUSCRIPT:**

**A. Introduction: Unique surfaces and interfaces leading to technological challenges and opportunities**

This review brings into focus the complex interfaces that lie at the core of **halide perovskite (HaP)-based** devices, which have been identified to play a crucial role for device functionality and even more so in terms of device performance limits and device stability. The intricacy of these interfaces arises from the composition of the HaP bulk material itself, which often include five to six different elemental and molecular components. As will be explained in the later part of this subsection, many HaP compounds employed in the current technological iterations are comprised of organic and inorganic building blocks; these materials hence exhibit true hybrid character, which leads to a variety of possible surface terminations and interface formation processes and related physico-chemical mechanisms.

In **section A,** we lay out the fundamental definitions of the chosen material system, including challenges and applications (A.1), our approach to measure, assess, and quantify the interface phenomena (A.2, A.3), and the role of interfaces in devices (A.4). **Section B** centers on the exposed surface of HaP thin-films and single crystals. We first introduce the chemical landscape of the HaP surface (B.1) and subsequently describe the surface's electronic properties (B.2). The section concludes with the description of single crystals as an experimental platform, which enables the most fundamental measurements of the HaP electronic structure (B.3). In the device, the HaP surface transforms into an interface with the adjacent functional layer. The main objective of this review lies in **section C** in which we describe the chemical, physical and electronic aspects of interfaces in HaP-based devices. After a brief classification of relevant types of substrate-HaP film interfaces encountered in the device, we turn to a detailed description of the impact of the substrate-HaP interaction on the HaP film (C.1). Subsequently, we introduce the interface between the HaP layer and a semiconducting charge transport layer deposited on top (C.2), as well as the direct HaP/metal contact interface (C.3). Section C closes with the analysis of interfaces between HaP films and passivation layers (C.4). In our discussion on interfaces we also touch on the effect of likely ion migration into/from the interfaces, on charge transfer as well as device functionality. We summarize the review in **section D** and sketch the road towards guidelines in interface engineering for optoelectronics that employ functional HaP layers. Eventhough there is a growing awareness that the interface in the HaP-based device is a major weak spot in terms of device performance and stability, we find that our fundamental understanding of the interface science of HaPs is limited, especially due to the large variety of potential chemical reactions at the interface. Dedicated interface studies remain scarce and often it is unclear to what extent the results can be used for the description of an operating cell. Hence, we conclude that before we can discuss targeted strategies to mitigate these interface-dictated problems in devices, we need to be able to better assess the interfaces to these semiconductor compounds, that are chemically complex and often metastable, and we suggest ways to achieve this (D.2).

### A.1. Emergence of halide perovskite (HaP) semiconductor-based optoelectronics

Following their (re-)emergence as light absorber in a novel class of thin-film photovoltaic devices, HaP semiconductors have also found use in a wide range of other optoelectronic applications.[1–8] From high efficiency solar cells to bright light emitting diodes to radiation sensors, HaP thin films offer a unique set of tunable optical and electronic properties combined with facile processing routines.[9–11]

We begin with a clear definition of the class of HaPs that recently attracted the most interest and is the focus of this review. HaPs have the perovskite crystal structure or closely related ones and the corresponding $ABX_3$ stoichiometry (see Figure 1a). For HaPs the A-site is occupied by a monovalent cation, the B-site by a divalent cation and a monovalent halide anion on the X-site. The two times smaller formal charge on the X site than for the established oxide perovskites plays a large role in the properties of these materials, due to the much smaller Coulomb interactions, the decreased ionic character of the bonds and the concomitant changes in the electronic structure. The choice of ions and their respective ionic radii critically determine the dimensionality of the resultant HaP system, which can range from isolated octahedra in 0D to corner-shared perovskite lattices in true 3D perovskites.[12] Most of the technological surge of HaPs has been with the 3D variants. For these, the possible combinations of A and B site cations is governed by the so-called Goldschmidt tolerance factor,[13] which essentially is derived from a close-packing of spherical hard ions.

The variety of ions that can occupy the X, A and B sites in perovskites, and also in those where X=halide, enables tunability of the optoelectronic properties of the material while at the same time defines a complex material system.[14] The focus of the research community and device makers has been on HaP materials with lead, and to a lesser extent tin, on the B-site. In the most common HaP compounds with this lead halide basis, which are employed in the current generation of semiconductor devices, the A-site is typically occupied by either $Cs^+$ or by a small organic molecule. Two organic molecules, methylammonium (MA, $CH_3NH_2^+$) and formamidinium (FA, $CH(NH_2)_2^+$), satisfy the Goldschmidt tolerance factor for most lead and tin halides; hence the corresponding HaPs crystallize in a 3D perovskite structure.[15] This compositional complexity in the bulk leads to an equally convoluted chemical composition for the HaP surfaces and, thus, the respective interfaces to adjacent layers,[16] with potentially drastic consequences for the system.

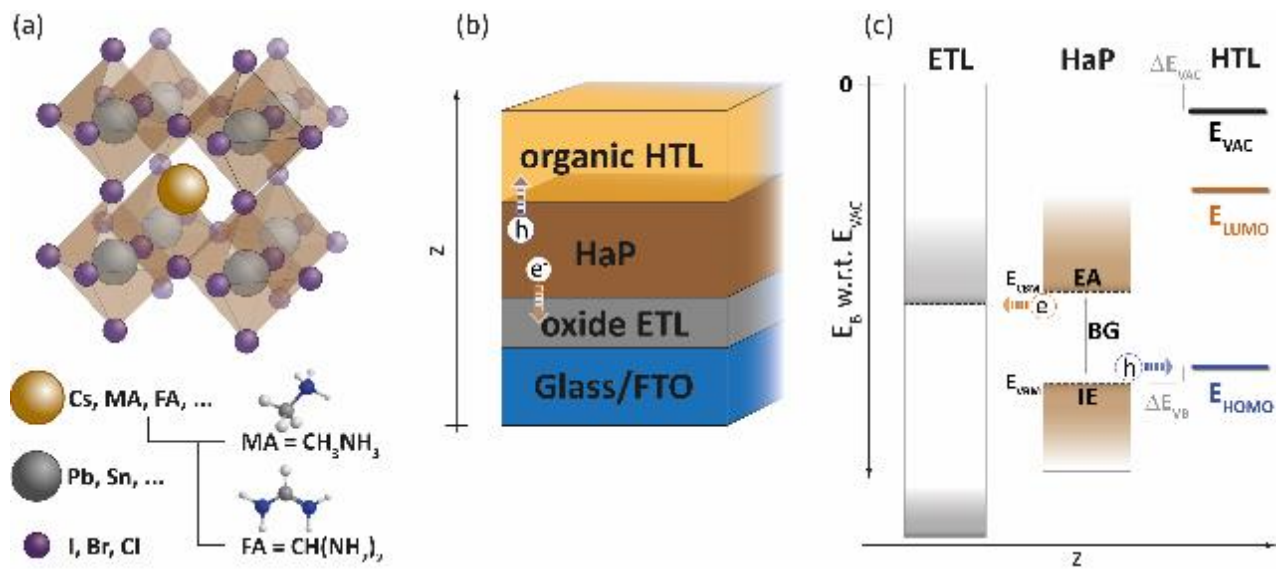

*Figure 1* – (a) $ABX_3$ crystal structure and most commonly employed elemental and molecular components of HaPs. (b) Schematic representation of layer stack as realized in a prototypical HaP-based solar cell with organic hole transport layer (HTL) and oxide electron transport layer (ETL). (c) Schematic energy level diagram for the device pictured in b. Characteristic energies, such as the ionization energy (IE) and electron affinity (EA), as well as the relative positions of the band edges ($E_{CBM}$, $E_{VBM}$) of the HaP absorber of a given band gap (BG) with respect to the energy levels of the adjacent transport levels determine charge carrier extraction at the interfaces. $\Delta E_{VB}$ indicates the band offset between the valence band maxima of two adjacent layers (here, the HaP and HOMO level in the hole transporter), and $\Delta E_{VAC}$ is the difference between the vacuum level positions of the HaP and HTL. Reproduced from Ref. 145 with permission from The Royal Society of Chemistry.

First, a HaP organic/inorganic hybrid system could have the remarkable character of exhibiting surface termination of either the organic molecular moiety or the metal halide unit, with direct implications for surface potential, electronic structure and chemical reactivity. This scenario becomes even more prevalent when the simple molecular component, MA or FA, is replaced by a larger molecular ion such as butylammonium (BA, $C_4H_9NH_2^+$) or other long-chain alkylamines ($C_nH_{(2n+1)}NH_2^+$) as in the case of layered 2D perovskites or mixed 2D/3D Ruddlesden-Popper phases.[17,18] There, the long-chain molecule acts as a spacer unit between the repeating 2D perovskite sheets and likely provides passivating surface termination. This scenario has been reported to be beneficial for both solar cell and LED applications.[19,20]

Second, the currently best performing perovskite solar cells (PSC) employ a perovskite active layer with complex stoichiometry, based on $FAPbI_3$ including cesium and methylammonium substitutions on the A-site, bromine substitution on the X-site.[21,22] Again, this complexity is mirrored at the interface to the adjacent transport layers, and likely affects device performance and stability due to interface energy alignment, trap state population and chemical reaction pathways.

With these unique surface and interface characteristics in mind, we now turn to examples of specific chemical and physical mechanisms present in HaP semiconductor devices.

## A.2. Rich chemistry and physics at the interface

The influence of passivating effects at the interface was described qualitatively in the initial optical studies of HaPs.[23,24] For instance, the long decay times of the photoluminescence signal, attributed to remarkably long charge carrier lifetimes for a semiconductor made by solution-processing near room temperature, were readily seen in MAPbI$_3$ films. This behavior was observable since the suppression of surface recombination was comparably easy to achieve using a Poly(methyl methacrylate) (PMMA) capping layer.[23] In addition to long-lived charge carriers, remarkable semiconductor properties in HaPs include unusual defect science,[25] pronounced relativistic effects on the electronic structure[26,27] along with lattice dynamics that are unusually "soft" for a high-performance semiconductor system.[28,29] A more comprehensive collection of the unique set of traits shared by most HaPs has recently been assembled in papers by Egger *et al.* and Manser *et al.*.[30,31] In summary, this combination of characteristics puts HaPs in a special class of semiconductors, within a potential application space that ranges well beyond traditional photovoltaic devices. While the HaP (bulk) material themselves seem to exhibit optoelectronic properties that enable excellent device performance (esp. photovoltaic behavior in solar cells), the interfaces between HaPs and adjacent materials likely determine the quality of the devices and distinguish good from bad ones. This formulates a clear requirement to find ways to maintain the HaP bulk behavior at their surfaces in ways that such surfaces can be transformed into interfaces to adjacent functional layers.

The case is well illustrated by the HaP's resilience to the formation of defects. First-principle calculations have been performed to unravel the nature and formation mechanisms of intrinsic point defects such as vacancies, interstitial and anti-site defects.[32] Taking MAPbI$_3$ as an example, Yin *et al.* initially calculated that formation enthalpies of defects that are energetically located in the middle of the band gap (BG) (e.g. Pb and I anti-sites and Pb interstitials) are rather high, while the more easily formed defects, such as I and MA vacancies, exhibit energies close to or immediately at the band edges.[25] This is consistent with the experimental finding that the experimentally determined band gap is relatively clean, even though the compound (i.e., it's A and X sub-lattices) is typically highly dynamically disordered (static disorder however occurs in mixed halides), in marked contrast to more traditional elemental or compound semiconductors (e.g., Si, III-Vs) for which device functionality relies on crystalline quality. In part, this apparent contradiction in properties (high optoelectronic quality and dynamic disorder, especially on the X sub-lattice), can be ascribed to the relatively benign nature of defects related to that sub-lattice. Steirer *et al.* found from X-ray photoemission spectroscopy (XPS) that some electronic properties of the HaP compounds, such as the position of the Fermi level in the electronic band gap, are resilient to marked structural and compositional changes in the film, such as loss of the volatile organic component, with the caveat that this result was obtained under vacuum conditions.[33] We note that the various defect mechanisms can be part of self-healing capabilities, as recently evidenced in $APbBr_3$ compounds.[34] Still, such resilience can be compromised at interfaces with a different material in the device if further chemical reactions occur. In that case, defective interfaces (i.e., deviating from atomically clean ones as between two organic semiconductors) could lead to the formation of recombination centers and eventually disrupt charge carrier transport across the interface. Furthermore, it was proposed that a precise control of the defect composition by adjusting the growth conditions would be the key parameter to controlling the doping type of the resulting perovskite thin-film. Initially, Yin *et al.* report that Pb vacancies would lead to p-type films while methylammonium interstitials would lead to n-type $MAPbI_3$ films,[25,35] which was further explored experimentally, e.g. by post-treatment of a HaP film.[36] However, photoemission spectroscopy measurements by Miller *et al.* and Schulz *et al.* indicated that the Fermi level position in the band gap of an $MAPbI_3$ layer could also be changed by depositing the film on various types of substrates.[37,38] A clear link between these interfacial effects, film formation and defect statistics remains yet to be uncovered.

**A.3. Relevant interface energetics**

In order to determine the key parameters that govern the behavior of HaP interfaces, we need to look back at the respective quantities in thin-film technologies from a more general point of view. In this regard, the relative positions of key electronic energy levels of two solids in contact govern electronic charge transfer mechanisms, such as electron or hole injection, and consequently impact device characteristics.[39,40] However, before we describe energy level alignment and interactions at an interface between two different compounds, we give a set of definitions for the surface, i.e. the solid-vacuum interface, which clearly is easier to assess experimentally.

With respect to the electron energetics of a surface of a solid, two energy levels constitute important reference points for the position of the electronic levels. The Fermi level ($E_F$) is defined as the energy at which the probability of occupation of an electronic state is ½. In a metal, $E_F$ cuts through the conduction band and marks the limit between occupied and unoccupied states. In an intrinsic or non-degenerately doped semiconductor, however, $E_F$ is located within the band gap. The vacuum level $E_{VAC}$ is defined as the energy above which an electron can escape from the solid into vacuum. $E_{VAC}$ does not refer to the absolute vacuum level but is defined as the local vacuum level, which deviates from the absolute one because of attractive and repulsive forces, due for example to electrostatic dipoles, that operate at the local level at the surface.[41] Thus, the composition and resulting electrostatic energy landscape at the surface strongly influences the position of $E_{VAC}$, which in turn can be changed by means of surface conditioning (e.g. through adsorption of molecules).[41] The energy difference between $E_{VAC}$ and $E_F$ defines the work function $\Phi$, i.e.

$$\Phi = E_{VAC} - E_F.$$

To a first approximation, the work functions of two isolated surfaces are useful parameters to predict the charge carrier transfer direction and estimate the energy level alignment between the two materials, when the surfaces are brought into contact to form an interface. However, two important points need to be emphasized: (1) the work function, as usually measured and assessed in experiments, is always a convolution of a bulk part and a surface part. The bulk part corresponds to the electrochemical potential relevant for the exchange of charge carriers; (2) the surface part takes into account various physical effects that are only present at the solid-vacuum interface, e.g. the evanescent electron density tailing into the vacuum, and the vacuum level shift due to surface dipoles, which is fulfilled in good approximation for solid-gas interfaces as well. Hence, the work function reported from experiments such as contact potential difference measurement with a Kelvin probe or photoemission spectroscopy measurements remains a well-defined parameter only at the exposed surface. A detailed discussion on the role and assessment of the vacuum level has been revisited recently by Kahn.[42]

In a broader picture, the classical semiconductor thin film device is comprised of multiple metal–semiconductor, insulator-semiconductor, insulator-metal, and/or semiconductor-semiconductor interfaces, all of which require distinct model descriptions and routes for characterization. Considering interfaces with HaPs, electronic transport can primarily be attributed to electrons and holes at the conduction and valence band edges, respectively, if we neglect defect-level assisted transport. Hence, knowledge of the position of $E_F$ in the semiconductor gap and the relative positions of the band edges with respect to the Fermi level and vacuum level becomes of primary interest. We, thus, focus on the role of the band offsets (e.g. $\Delta E_{VB}$ for the valence bands in figure 1) at the interface between a metal contact and a semiconductor, as well as between two semiconductors. In the latter case, the offset between the conduction band edges of the two constituent materials determines the electron transport across the interface, while the same holds true for the valence band edge offset and hole transport. In this context, we introduce two further energy quantities: the ionization energy (IE), defined as the energy difference between vacuum level and valence band maximum, and the electron affinity (EA), which is the equivalent quantity measured between vacuum level and conduction band minimum. Generally, these energies can be understood as the minimum energy required to remove a valence electron from the surface, or the minimum energy released by capturing a free electron from vacuum, respectively.

Mott and Schottky formulated an early model according to which transport across a metal-semiconductor interface is via thermionic emission through the interfacial barrier, estimated from the difference between the metal work function and IE (EA) of the p-type (n-type) semiconductor. This simple evaluation of the barrier energy corresponds to what is known as the Schottky-Mott limit. In the case of a doped semiconductor, the barrier height corresponds in principle to a mismatch between the metal and semiconductor work functions.[43,44] The equivalent model for semiconductor-semiconductor heterojunctions is the so-called Anderson rule,[45] whereby the electronic structure of the interface is defined by vacuum level alignment, and the interface energy barrier for electron (hole) transport is the difference between the electron affinities (ionization energies) of the two materials. Decades of research ensued to develop more realistic models of semiconductor heterojunctions, leading, for example, to the introduction of an interface dipole that self-consistently accounts for the difference between the semiconductor charge neutrality levels, which determine the flow of charge at a metal/semiconductor contacts.[46] Within the last 20 years in particular, the approach was expanded to include hybrid interfaces between organic and inorganic layers.[47–50] A significant mile stone is the investigation of the induced density of interface states.[51] Earlier inclusion of polaronic effects were challenged by the detailed dissection of the spatial profile of the electrostatic potential in the organic layer via the Poisson equation,[50] which yields a more universal model to distinguish the Schottky-Mott limit from cases with pronounced charge transfer upon interface formation.

It became possible to explore this description with the advent of powerful computational methods for the calculation of the electronic structure of semiconductor surfaces and interfaces, and more accurate direct and inverse photoemission spectroscopy (PES/IPES) experiments on *in situ* layer growth with atomic precision, to determine band onsets (i.e. the distance of the band extrema from $E_F$) at the surface and band offsets at the interface of two semiconductor systems.[52]

Turning now towards HaP-based devices and the typical cell stack illustrated in Figure 1b, we see a variety of interfaces between transparent oxide and HaP semiconductors as well as between HaPs and organic semiconductors, which have been the topic of several reviews and perspective pieces.[16,53–57] In the latter case, the transport levels are the lowest unoccupied molecular orbital ($E_{LUMO}$) and conduction band minimum (CBM) for electrons, and the highest occupied molecular orbital ($E_{HOMO}$) and valence band maximum (VBM) for holes. In the following, we discuss the role of the alignment of these levels in the device (Figure 1c). We describe current progress on the determination of these properties in the later chapters of this review.

### A.4. Applications: The key role of interfaces in devices

#### A.4.1 Electronic transport across interfaces in optoelectronic devices: the example of thin-film PV

In order to understand the role of interfaces in HaP-based optoelectronics, we need to consider the overwhelming importance that interfaces have in all thin-film devices. Concepts of semiconductor physics, for which many descriptions rely on the perfect periodicity of the crystal lattice, are central to discussions on thin-film semiconductor technologies, including photovoltaic devices. Clearly there is no perfect periodicity at interfaces between various functional layers, such as absorber and charge carrier transport materials, in a solar cell. This case can be made for most relevant photovoltaic technologies, where the classical homojunction between p-doped and n-doped regions of the same material has been replaced by heterojunctions between different materials. Along this disadvantage, an advantage of such heterojunctions is that they offer a higher degree of freedom for the device layout and ultimately promise higher performance than many homojunctions.

The degree of complexity within the device increases with the number of interfaces, which is particularly true for multijunction photovoltaic devices, seen as an important promise of HaP solar cells, as part of tandem devices.[58–61] This imposes stringent requirements on the design of interfaces and on analysis of their properties.[57] We note that most of the loss processes in thin-film photovoltaic devices are dominated by non-radiative electron-hole recombination via defect states that are primarily attributed to interruptions in crystal periodicity at interfaces. Within this context, we can consider the surface and interface as some of the largest 'defects' in the solid. This idea is readily demonstrated as a function of the crystalline quality of the absorber layer in thin-film silicon solar cells: for such devices, based on crystalline Si (c-Si), the lifetime of photogenerated charge carriers is on the order of milliseconds and thus vastly exceeds the carrier diffusion time to the interface with the charge transport layers, which is on the order of microseconds.[62] We can thus assume that recombination in the bulk will be negligible, compared to recombination at the interface.[63]

The main physical loss mechanisms are depicted in Figure 2. We discuss here the example of the surface (interface) recombination current (or loss current) $J_S$ of electrons at a hole contact, which is given by:

$$J_S = qSn_0 \, e^{\Delta E_F/k_B T}$$

where $q$ is the unit (electron) charge, $S$ the surface recombination velocity, $n_0$ the electron density at the hole contact in thermal equilibrium, $\Delta E_F$ the quasi Fermi level splitting and $k_B T$ the thermal energy. For any given $\Delta E_F$, $J_S$ scales proportionally with surface recombination velocities and charge carrier densities at the interface. $S$ is mainly determined by the density of surface defects acting as recombination centers, whereas $n_0$ is determined by the potential drop at the electrodes and is related to the built-in field. For a higher built-in field, $n_0$ decreases at the hole contact and so does the equilibrium hole concentration $p_0$ at the electron contact. With this expression of the recombination current, the path to minimizing the interfacial recombination current lies in designing defect-free interfaces with band alignment that retains the maximum built-in field. For c-Si solar cells, the use of passivating selective contacts, which conduct one carrier type but block the other caused by an asymmetry in carrier concentration and energy level alignment, has been a key solution.[64]

While these surface recombination considerations apply to all solar cells, they appear to be even more important for HaP perovskite solar cells (and related optoelectronic devices). As described in the previous sections, HaP light absorbers present very particular challenges with respect to interface engineering efforts. Given the relatively recent emergence of HaP-based solar cells and other HaP-based optoelectronic devices, we are still lacking comprehensive models of interface band alignment, and information on the nature, density and origin of bulk and surface defect states, which take into account the full picture of the rich chemistry and challenging physics of this material class.

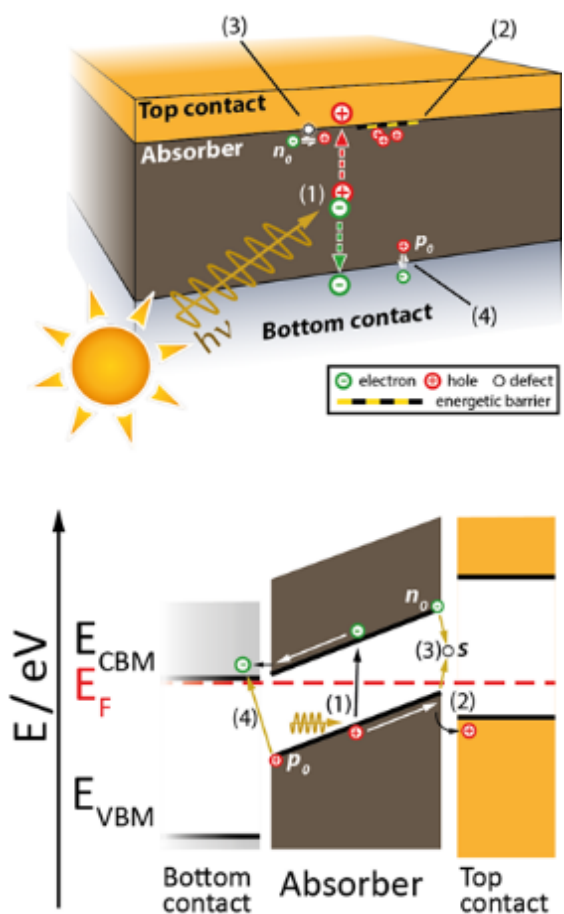

*Figure 2* – *Interface-related loss mechanisms in a simplified photovoltaic device. After photoexcitation (1) carrier transport to the contact interfaces occurs without significant losses. At the interface, carrier extraction can be impaired by (2) interfacial energy barriers due to inadequate band alignment, (3) defect-induced surface (interface) recombination velocity S and (4) back recombination of extracted carriers, which still reside in the interface region. $E_{CBM}$, $E_{VBM}$ and $E_F$ are the conduction band minimum, valence band maximum, and Fermi level, respectively. Reprinted with the permission from 57. Copyright 2018 American Chemical Society.*

In the spirit of Kroemer's dictum that the interface is the device[65], we reiterate the question posed in the title: **Halide perovskites – Is it all about the interface?** Here, the satisfactory answer would go beyond the drawing of a simple correlation between device performance and interface formation, and rather consider if the interface is the primary contributor to determine the mode of operation of devices, based on these materials.

**A.4.2 Charge transfer vs. recombination current in HaP solar cells**

Attempts have been made to better differentiate bulk from interface recombination in HaP solar cells by connecting the diode ideality factor to optical measurements, which suggest that recombination at interfaces is the main contributor to performance losses.[66] The results of impedance spectroscopy measurements to probe surface recombination and photogenerated carrier collection in PSCs suggest that interfacial recombination is the main factor for photovoltage loss (on the order of 0.3 V), on top of the thermodynamically determined losses from the radiative limit and is due in part to charge accumulation at the interfaces of these cells.[67,68]

As described in the beginning of section A.2, the role of recombination and charge transfer at interfaces can be evaluated by transient optical experiments. In this regard, decay curves from time-resolved photoluminescence measurements demonstrate the need to understand and quantify charge transfer and interfacial recombination rates in HaP semiconductors. Krogmeier *et al.* performed a quantitative analysis via numerical simulation of the transient photoluminescence of MAPbI$_3$/PC$_{61}$BM heterojunctions, from which they conclude that charge accumulation at the interface is a dominating factor.[69] Schematically, the implications of the band offsets, which are pictured in Figure 3a during a laser pulse and in Figure 3b after a delay, allow for an assessment of likely recombination pathways at the interface. The input parameters can then be used to simulate the PL decay signal (Figure 3c) and determine the differential lifetime (Figure 3d) for low and high laser fluences ($E_L$). Depending on the fluence, two different time regimes were identified, with the first interval (up to 60 ns) being dominated by charge transfer to the quencher according to the description of the charge transfer current for electrons:

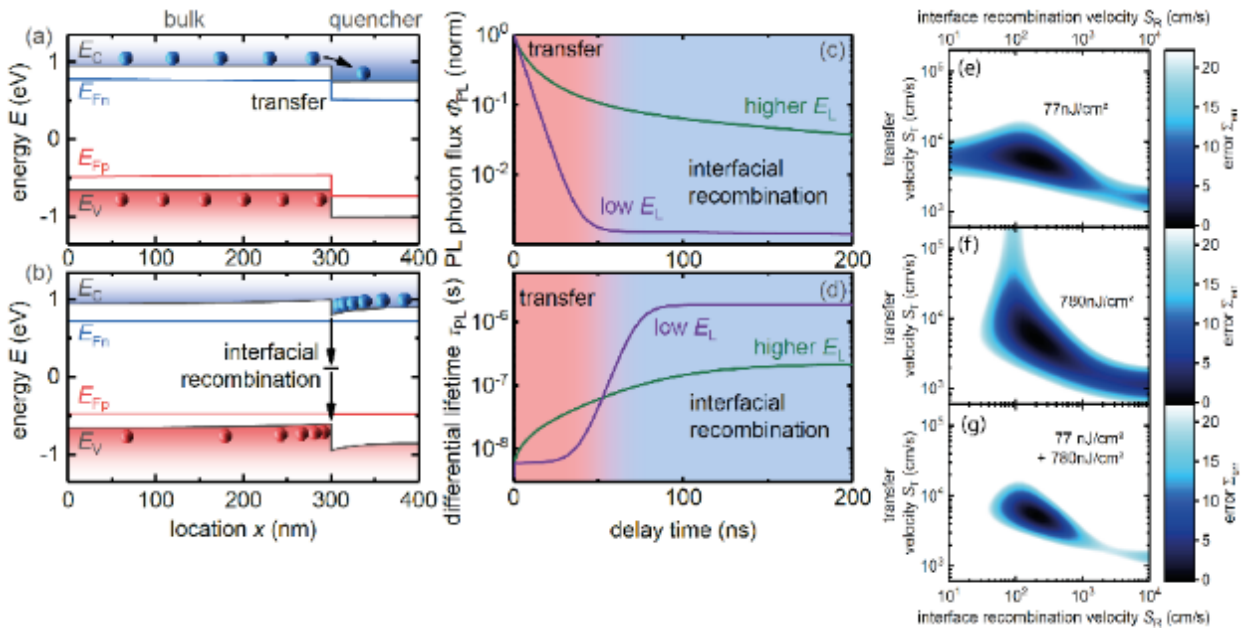

*Figure 3* – Band diagram and PL transients of simulated perovskite (bulk) and PCBM (quencher) layers. (a) Band diagram during the laser pulse. (b) Band diagram for longer delay times. (c) TRPL signal for two laser fluences $E_L$. (d) Differential lifetime showing two constant regions due to accumulation of charge carriers. (e-g) 2D error plots for the transfer and interface recombination velocities fit parameters. Reproduced from Ref. 69 with permission from The Royal Society of Chemistry.

$$J_{\text{T,n}} = qS_T \, (n_{b,int} - n_{q,int}) \, e^{\Delta W_c / k_B T}$$

with $S_T$ the charge transfer velocity, $n_{b,int}$ and $n_{q,int}$ the charge carrier densities at the interface in the HaP bulk and PCBM quencher, and $\Delta W_c$ the offset between the MAPbI$_3$ conduction band and the PCBM LUMO, i.e. $\Delta W_c = E_{\text{LUMO}} - E_{\text{CBM}}$. Interface recombination dominates the second time regime and is determined by the carrier densities at the interfaces resulting from charge accumulation at the barriers. Subsequently, the numerical model served as a set of equations to fit charge transfer and interface recombination velocities for real PL spectra acquired at various laser fluences, for which the minima in the 2D error plots yield the optimal parameter set (Figure 3e-g). In this example, a quantitative understanding of the energy levels and hence $\Delta W_c$ would serve as an important check on validity of the model for this system.

**A.4.3 Implication for interfaces beyond electronic transport**

It is important to note that further interface factors, apart from electronic transport and charge carrier extraction, outlined above, affect HaP device functionality. As a result of the HaP compounds being relatively soft, ion migration can occur in these materials,[70,71] an effect that has been proposed to be driven or suppressed, among other possible factors, also by the presence or absence of chemical passivation at the interfaces.[72,73] More generally, the interface can serve as a nucleation site for intrinsic and extrinsic defects. While such defects clearly can impact electronic transport properties, and particularly the recombination current $J_S$ described in the previous sub-section, we want to highlight further technological implications concerning the stability of the HaP layer and hence the device stability and lifetime, which has emerged as a crucially important figure of merit for the further development of HaP-based solar cells.[74]

Recent studies demonstrate that the stability of perovskite solar cells (PSC) can be improved through deliberate tailoring of interface properties. Grancini *et al.* achieved a one-year stable PSC module by employing a 2D/3D HaP heterojunction.[75] The general claim in this approach is that it yields a built-in 2D/3D HaP, that forms a phase segregated thin-film. In this graded structure, the 2D layer acts as a protective buffer against moisture at the interface and hence preserves the 3D perovskite region of the film. Alternatively, other research groups have worked on adjusting the hole or electron transport layer side by either modifying the substrate onto which the HaP layer is deposited, or the overlayer on top of the HaP film to improve the device stability.[53,76] As an example, a comprehensive approach led to the combination of a tailored $SnO_2$ bottom electrode and a newly synthesized small molecule organic hole transport layer (EH44, 9-(2-ethylhexyl)-*N*,*N*,*N*,*N*-tetrakis(4-methoxyphenyl)-9H-carbazole-2,7-diamine) capped with a $MoO_x$/Al anode. The resulting PSC could be operated under ambient conditions and without encapsulation for over a thousand hours without significant signs of degradation or decline in power conversion efficiency (Figure 4a).[77] It was suggested that this result was achieved primarily by reducing interface chemistry at the oxide/HaP interface and by inhibiting migration through HTL and electrode. We note that especially for hybrid HaPs, the interface/surface is a weak link because of potential loss of the more volatile organic component. Hence, degradation is likely to occur at the surface/interface first. Naturally, this not special for HaPs, as for most materials the bulk is more stable than the surface.

Finally, other means of tailoring and controlling HaP interfaces have been shown to be instrumental in the development of novel device classes. As an example, the development that led to the deployment of the next generation of HaP-based quantum dot solar cells was initially driven by the prospect of stabilizing $CsPbI_3$ in its cubic phase through surface passivation of quantum dots.[78] This approach was further refined by applying a surface treatment to the $CsPbI_3$ quantum dot film. Exposing the film to a formamidinium iodide salt solution resulted in better interconnections (as the smaller FA replaces the long-chained oleylamine ligands around the dots) between the quantum dots, and hence in a higher charge carrier mobility (Figure 4b).[79,80] An additional explanation could be that replacing Cs at the surface of the quantum dot with an organic compound results in a closed shell capping, and therefore reduces the density of surface states. Overall, this surface treatment approach played a central role in the realization of a device that currently holds the record value in power conversion efficiency for a quantum dot solar cell.[81]

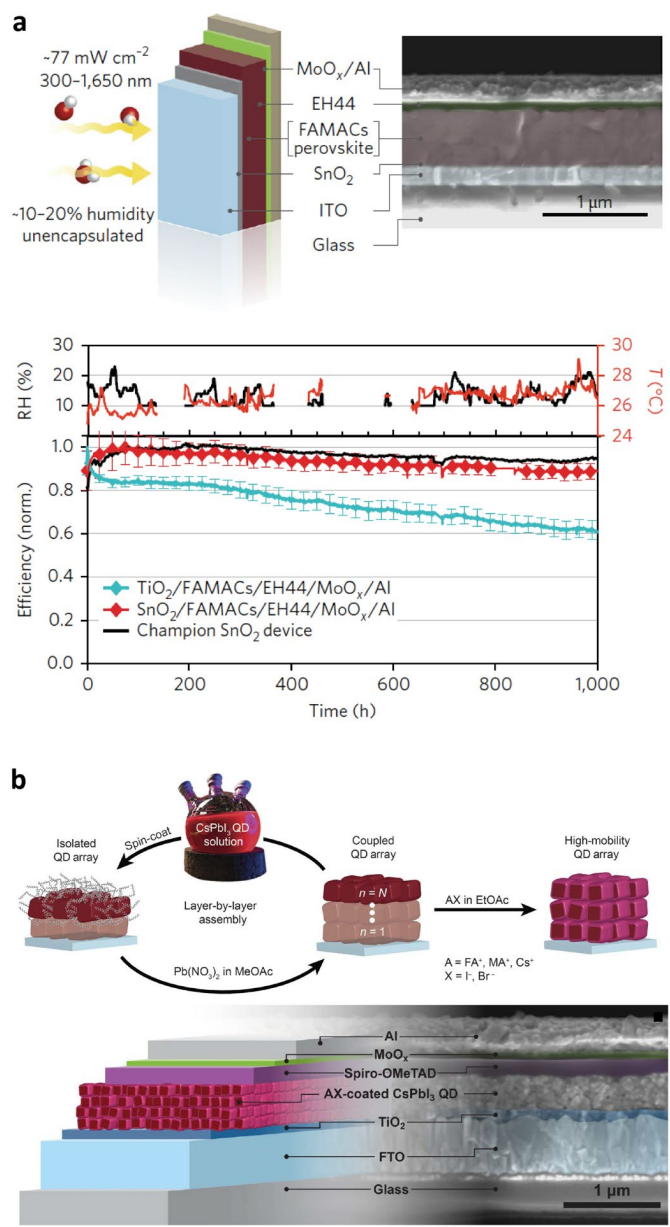

*Figure 4 –* a) Device architecture of a perovskite solar cell with specifically tailored electron transport layer (SnO$_2$), hole transport layer (EH44) as well as an MoO$_x$/Al electrode for enhanced stability. The power conversion efficiency was recorded for a non-encapsulated device in constant operation under ambient conditions for 1000 hours. Reprinted with permission from Ref. 77. Copyright YEAR by Springer Nature Publishing AG. b) Fabrication scheme and device cross section for a CsPbI$_3$ based quantum dot solar cell with FAI surface treatment leading to a better interconnected film. Reprinted figure with permission from Ref. 79. Copyright 2017 American Association for the Advancement of Science.

In summary, careful design and analysis of interfaces in HaP semiconductor devices is required to understand and improve charge transport phenomena as well as chemical processes that are critical to achieve and maintain high performance device operation. In that regard, a key question to be resolved concerns the persisting entanglement between interface and bulk effects. In the following section B, we will thus first discuss the plain exposed surface of the perovskite with a focus on the chemical surface termination and electronic structure.

**B. HaP thin-films and single crystals: The exposed surface**

The relative ease to produce HaP thin-films for photovoltaic applications has led to a wide range of HaP processing routines, with low temperature solution and vacuum deposition processes being the most prominent. Historically, spin-coating has been the method of choice, enriched with additional parameters such as solvent-annealing, hot-casting or vacuum drying, which not only affect bulk crystallization but also likely impact the final HaP film surface in terms of morphology, chemical composition and electronic properties.[82–84] We note that these methods produce HaP absorber films that exhibit different degrees of crystalline quality, culminating in record device performance. Despite these encouraging device results, little information is typically available on the surface condition of the HaP films, before deposition of the interface-forming layer onto them, with the exception of few morphological aspects, mostly gained from scanning electron and atomic force microscopy experiments. More detailed scrutiny of the surface formation process for spin-coated thin films should allow clearer assessment of the requirements in various upscaling efforts, e.g. doctor-blading or slot-die coating, which follow similar film growth processes.[85,86] In addition to these solution- and sol-gel based deposition processes, evaporation-based methods offer even more possibilities to tune the surface properties. While the vapor deposition of MAPbI$_3$ in either co-evaporation, sequential evaporation or hybrid chemical vapor deposition (CVD) techniques has been well explored early on,[87,88] it was not until recently that HaP films with multiple cations and anions were processed through this route in a reliable and reproducible fashion.[89] As the surfaces of these various HaP films define one of the two parts of the interfaces to the adjacent functional layers, we begin by exploring their properties on an atomistic scale.

### B.1. Surface termination and probable chemistry

#### B.1.1. Surface termination in non-stoichiometric thin-films

The techniques developed for HaP thin film deposition all suffer to various degrees from offering only limited control and precision over the layer formation. Even in the case of vapor phase deposition, epitaxial layer growth, which has been perfected for nearly all elemental, III-V or II-VI semiconductors, has not yet been achieved for HaP compounds. Hence, surfaces of interest are rarely stoichiometric and usually defective. Nonetheless, before classifying the surface defects, we begin the discussion with a theoretical description of the stoichiometric HaP surface of HaP in the cubic, tetragonal and orthorhombic crystal structures, based on density functional theory (DFT) slab calculations.[90–92]

Note that, to date, a large body of computational theory has been dedicated to investigations of the structural and optoelectronic properties of HaPs. The full exploration of the unique combination of materials' bulk properties is beyond the scope of this review and is covered extensively in other reviews.[32,93,94] Here, we select some examples of slab calculations that are of primary relevance for the understanding of the surface/interface formation. For our discussion of the theoretical and experimental approaches, the underlying questions always remains:

- Can one derive a comprehensive picture of the surface of HaP compounds?
- Which, if any aspects of the surface are strongly coupled to the unique combination of materials properties, observed in HaPs?
- In how far do surface properties affect those of resulting interfaces with adjacent functional layers (e.g. for charge transfer or passivation) and how significantly are their electronic properties affected?

*Theoretical modeling of stoichiometric HaP surfaces:*

We begin by looking at the intrinsic surface properties for various structures of the same HaP compound, i.e., the tetragonal, orthorhombic and cubic phases of $MAPbI_3$. Haruyama *et al.* investigated the structural stability of the (110), (001), (100) and (101) surfaces in the tetragonal phase using DFT calculations on a plane wave basis set and including van der Waals interactions. By sampling through various $PbI_x$ polyhedron coordinations, they found that for the energetically favorably (110) and (001) surfaces the formation enthalpies of the $PbI_2$-rich flat and the "vacant", i.e. MA-rich, surface terminations are similar (see Figure 5a). Hence, they suggest that there is a small window under thermodynamic equilibrium growth conditions, within which both surface terminations coexist. However, generally the vacant, i.e. MAI terminated, surface is more stable and more likely to grow in thermodynamic equilibrium.[91] Quarti *et al.* reach a similar conclusion in their DFT calculations of the (110) and (001) surfaces of $MAPbI_3$ in the tetragonal phase and describe the MAI termination to be significantly more stable than the $PbI_2$-rich one.[95]

For the stoichiometric orthorhombic phase of $MAPbI_3$, Wang and coworkers optimized surface structures of the different spatial isomers and found minimum formation enthalpies for the (100) and (001) surfaces.[92] They considered the various possible spatially and constitutionally isomeric structures by sampling through a set of orientations and connectivities of the surface Pb-I bonds. As a direct result, the stabilization mechanism was correlated to the coordination number of the Pb-centered octahedra, i.e.

the number of broken Pb-I bonds, and the number of surface iodine atoms. Wang expanded their investigation by molecular dynamics (MD) calculations to monitor structural reorganizations or isomerizations, which they eventually rule out in their study.[92]

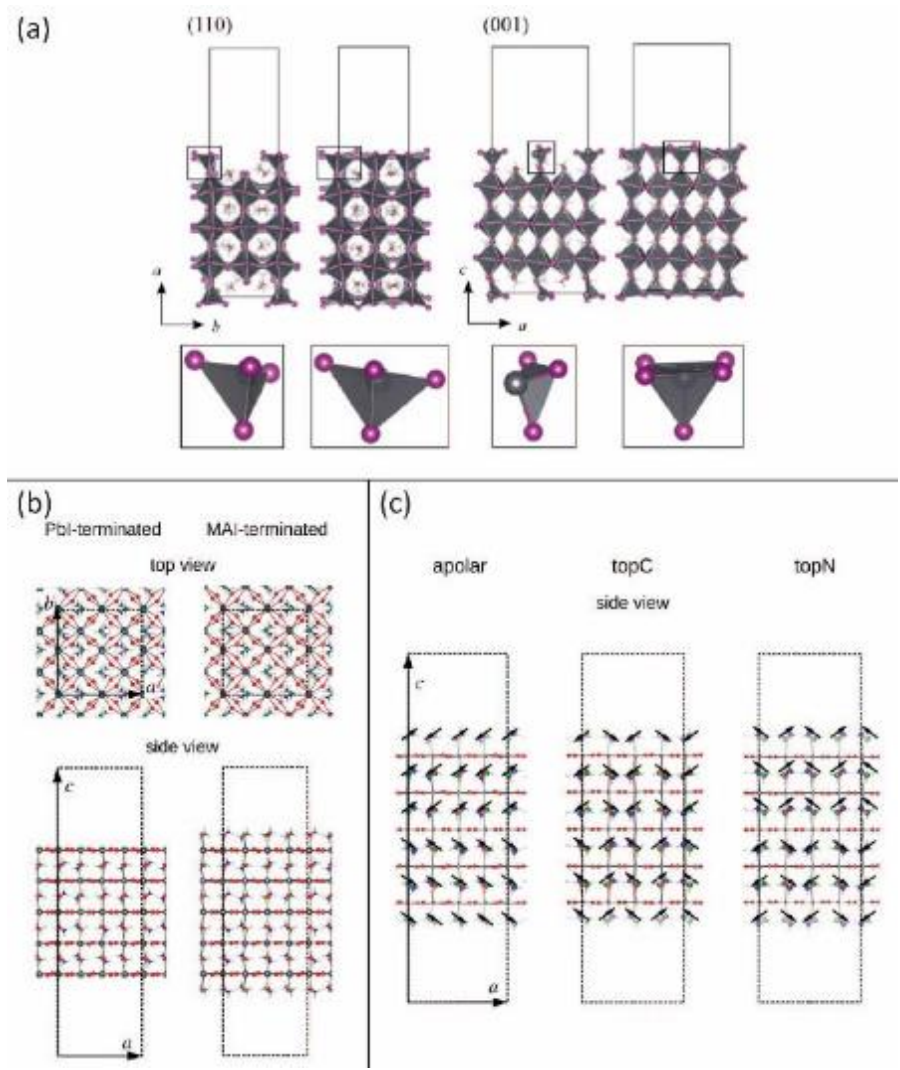

*Figure 5* – *DFT slab calculations of the tetragonal phase of MAPbI$_3$ to assess surface termination. (a) Relaxed structures for the most stable (110) and (001) surfaces with "vacant" and PbI$_2$-rich flat surface terminations. Reprinted with the permission from Ref. 91. Copyright 2014 American Chemical Society. (b) Top and side views for the classification of the PbI$_2$ and MAI surface terminations for the (001) surface (c) Different orientations (black arrows with NH$_3$ at arrow head) of the MA cations (apolar, topC, topN) with respect to the (001) surface. Reprinted with the permission from Ref 95. Copyright 2017 American Chemical Society. The color maps are: lead=gray, iodide=purple for (a) and lead=black, iodide=red, carbon=green, nitrogen=blue, hydrogen=white for (b) and (c).*

In this regard, Torres *et al.* provided calculations for the cubic phase of MAPbI$_3$,[90] the results of which are ostensibly in conflict with the finding from the preceding study. They describe the relaxation of four different surfaces in their DFT calculation of 2D slabs using pseudo-cubic unit cells and MD methods. As they characterize the initial surface morphology by the termination of the cell, i.e., either MA- or flat PbI$_2$ terminated, as well as by the orientation of the organic cation MA with respect to this PbI$_6$ cage, they observed that mainly three competing effects drive a pronounced surface relaxation: (i) on the flat PbI$_2$-rich surface, the inorganic cage can be subject to contraction; (ii) the MA group's tendency to maximize hydrogen bonds to the bridging iodine atoms; and (iii) the directional alignment of the dipole of the MA group.[90] While generally the MA$^+$ ion is believed to be free to rotate within the cage in the bulk MAPbI$_3$, its orientation on the surface is subject to stabilization mechanisms that are on the order of 1-3 k$_B$*T* per unit cell and thus to be reckoned with at room temperature. In a simple model, the orientation of the MA$^+$ moiety can be classified as - a non-polar surface, - a surface with the carbon moiety sticking out (topC), or - a surface with a protruding nitrogen moiety (topN) as depicted in figure 5b. Slab calculations yield a minimum of the surface energy for the topC configuration.[95]

*Non-stoichiometric HaP surfaces and surface reconstruction:*

In the context of searching for the lowest energy configurations, these theoretical computations are generally constrained to yield relaxed structural parameters based on a finite set of unit cells in the slab, usually limited to a few hundred atoms. Hence, important processes involving long-range structural relaxation phenomena could be missed. In this regard, surface reconstruction phenomena have been well explored for a broad set of materials in general, and for metal oxide perovskites in particular. For the latter, we find extensive phase diagrams for various surface structures, which are usually attributed to instabilities due to perpendicular macroscopic polarization of the surface layers. In those cases, a reconstruction of the polar surfaces leads to a compensation and hence depolarization, which ultimately stabilizes the top layer. These structural alterations have a pronounced effect on the surface electronic structure and energetics.[96] Generally, electron diffraction experiments (e.g. Low Energy Electron Diffraction, LEED) and scanning tunneling microscopy (STM) methods can be employed to determine changes to the surface structural parameters, and investigate the surface over sampling areas extending from a few to several 100s of nanometers. However, the typical requirements for these experiments, such as the preparation of atomically flat surfaces and measurements under ultra-high vacuum conditions, have considerably restricted their applications to HaP surfaces, with only very few studies published so far on this specific topic.[97] Unfortunately, the generation of reliable LEED data remains difficult, as the impact of the flux of electrons required for this technique, causes severe damage to the $MAPbI_3$ layer.[98] Nonetheless, successful measurements of the surface crystal structure of a $MAPbI_3$ single crystal has been reported recently and the results confirm the transition temperature from the tetragonal to cubic phase (see section B.3.1).[99]

She et al. succeeded in growing ultra-thin $MAPbI_3$ films with a thickness of nominally 10.8 monolayers on top of atomically clean Au(111) surfaces by co-deposition of $CH_3NH_3I$ and $PbI_2$.[100] Subsequent *in situ* STM experiments performed at 78 K revealed the formation of flat $MAPbI_3$ terraces on the scale of 100 nm, with steps of a monolayer of half the HaP's unit cell, i.e. one $PbX_6$ octahedron and one MA unit, could be identified (Figure 6a,b). The measurements did not reveal the presence of surface reconstruction of the $MAPbI_3$ film over multiple unit cells, but instead pointed to two distinct rearrangement patterns of the surface atoms on the unit cell scale. In the high-resolution images in figures 6c,d the "zigzag" and "dimer" patterns of the bright protrusions with a periodicity of the perovskite lattice constant can be clearly distinguished. In certain areas of the terraces, both phases co-exist in close vicinity (see figure 6e), and each phase can be switched to the other reversibly by applying a tunneling current. Based on this behavior, one can exclude that the different phases represent different surface terminations. In the chosen experimental configuration, the STM measurements probed the filled states of the surface, i.e. the valence band states, and hence the experiments were sensitive to the I 5p as well as the Pb 6s and 6p orbitals, which mark the largest contribution to the MAPbI$_3$ valence band DOS.[37,101] A flat PbI$_2$ terminated surface would exhibit a Pb-I bond distance of approximately 3.2 Å, whereas the periodicity extracted from the STM images amounts to 8.5 – 8.8 Å. Hence the bright spots are identified as iodine atoms on the top corner of the PbI$_6$ octahedra for the MAI terminated (001) surface of MAPbI$_3$; notably surface defects in terms of missing iodine appear as dark spots in the images. Complementary DFT calculations suggest that this rearrangement of surface iodine atoms originates from the orientation of the MA$^+$ cation and corresponds to a slight depolarization and stabilization of the surface, in accordance to earlier reports.[91] The implication of this effect on the surface electronic properties can be dramatic. Indeed, any ferroelectric order from the bulk material would be modulated by this surface depolarization, and the disruption of the symmetry could also lead to the proposed Dresselhaus- and Rashba-splitting in HaP compounds, which would otherwise not occur in the cubic crystal phase without inversion center (e.g. MAPbBr$_3$).[102]

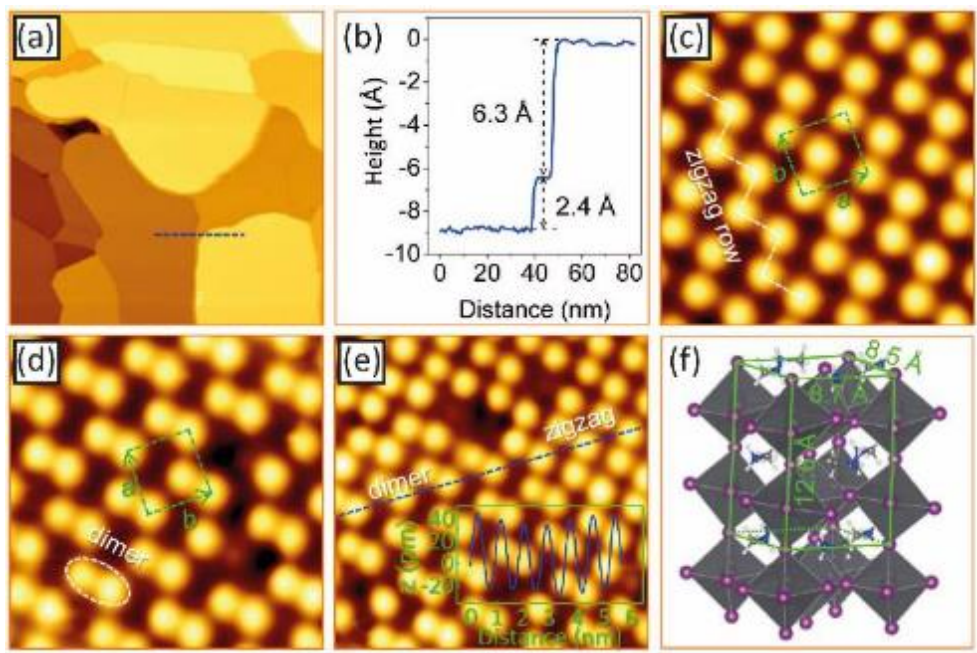

*Figure 6* – STM images of MAPbI$_3$ thin-films. (a) Large-scale image of atomically flat terraces (300 × 300 nm$^2$). (b) Height profile along the dashed line in (a), showing the step edges of a gold terrace (2.4 Å) and a MAPbI$_3$ layer (6.3 Å). (c,d) High-resolution images of the MAPbI$_3$ zigzag and dimer structures (4.3 × 4.3 nm$^2$) on a MAPbI$_3$ terrace. (e) High-resolution image of the two phases coexisting in the same region (5.6 × 5.6 nm$^2$) with the inset showing the height profile along the dashed line. (f) Orthorhombic MAPbI$_3$ unit cell. Reprinted with the permission from Ref 100. Copyright 2016 American Chemical Society.

*Surfaces of mixed halide HaPs:*

The complexity of these surfaces rises with the number of components in the ABX$_3$ compound, i.e. a more than one type of halide anion on the HaP's X-site and/or various cations such as MA, FA and Cs, on the A-site. Mixing halide compositions in HaP films is the most common approach to tune the band gap of the resulting compound but can lead to instabilities like a pronounced light-induced segregation of halide species.[103] However, only few studies, theoretical or experimental, have investigated the specifics of surfaces of mixed halide HaP systems. The large number of combinations and local alloying phases often make clear correlations between results from different studies difficult. The case is illustrated best for chlorine incorporation and surface termination in MAPbI$_{3-x}$Cl$_x$ present in the technologically relevant MAPbI$_{3-x}$Cl$_x$ compound, for which the determination of the absolute chlorine content and traces remains elusive.[104] Mosconi *et al.* calculated that the band gap for the surface of a Cl-terminated MAPbI$_3$ film should not change since the chlorine atoms would only induce additional energy levels outside the MAPbI$_3$ gap, that would only indirectly change the band gap.[105] Thus, theoretically, no major change in the band gap would be expected for traces of Cl in MAPbI$_3$ or on the MAPbI$_3$ surface. Experimentally, small amounts of chlorine could be traced by angle resolved XPS measurements confined at the TiO$_2$/HaP interface.[104] In a different example Yost *et al.* performed STM measurements at the cross-section of a cleaved MAPbI$_{3-x}$Cl$_x$ thin-film to probe the homogeneity of the electric properties and suggested a correlation to the chlorine content in the so-exposed cross-sectional surface.[106] They found domains with differing tunnel resistances, yet similar band gaps (see Figure 7a), which led them to excluded that the domains were ferroelectric features but instead correspond to variations in the surface I/Cl ratios. Indeed, aside from only changing the interface chemistry, a change in surface termination can play a significant role for charge carrier transfer rates to adjacent layers as will be discussed later in section C of this review.

*Surfaces of mixed A-site cation HaPs:*

The case for increased complexity and substantial changes in surface stabilization mechanisms becomes evident, when $MA^+$ is replaced with the smaller $Cs^+$ cation. The resultant $CsPbI_3$ compound does not satisfy the Goldschmidt tolerance factor rule, and as a consequence the bulk material does not assume a stable cubic phase at room temperature.[107] However, a phase stabilization for the cubic phase in $CsPbI_3$ is achieved by the synthesis of the material in the form of nanocrystals and quantum dot films;[78] the condensed film consists therefore of an array of quantum dots, which retain the cubic phase due to the large contribution of the surface energy linked to the increased surface-to-volume ratio.

A good example of a rich surface phase diagram, studied by first principle calculations, is afforded by the tin perovskite-based cesium and rubidium mixed cation system, $Rb_xCs_{1-x}SnI_3$, which stands at the edge of the Goldschmidt tolerance factor range ($t$ = 0.8 for $RbSnI_3$ and 0.83 for $CsSnI_3$) and hence exhibits limited phase stability in the cubic perovskite phase.[108] In particular the small rubidium cation might be too small to sustain a stable perovskite structure (e.g. in the more widely studied Pb-based HaP systems no experimental evidence for Rb on the A-site has been found), while the conjecture is that cation mixing would lead to (increased) entropic stabilization. In their approach to capture the surface thermodynamics, Jung *et al.* calculated the surface slabs of $CsSnI_3$ and $RbSnI_3$ and assumed Vegard's linear relation to extrapolate the lattice parameters between those two end points.[109] Similar to the case of the organic A-site cation, $Cs^+$ and $Rb^+$, when placed on the A-site in the calculated DOS of $Rb_xCs_{1-x}SnI_3$, do not contribute directly to the band edge DOS, as the valence band is mainly comprised of I 5p, and Sn 5s orbitals, whereas the Sn 5p dominates the conduction band DOS. However, the electronic properties, and particularly the band gap, could be changed locally by steric effects, as cations of different sizes alter the bond lengths and angles of the SnI$_6$ octahedron and hence its shape. The SnI$_2$ chemical potential ($\Delta\mu_{SnI2}$) determines the likelihood of various surface terminations and defines the phase diagram depicted in figure 7c. Jung *et al.* found that both alkali iodide terminated (100) surfaces have distinctively lower surface energies ($\Delta E > 3$ meV/Å$^2$) under CsI- and RbI-rich conditions than under SnI$_2$-rich conditions, respectively. In contrast, the surface, which is terminated by truncated octahedra (SnI$_2$-rich), has only a slightly lower surface energy ($\Delta E < 0.5$ meV/Å$^2$) under SnI$_2$-rich growth conditions than under CsI- and RbI-rich growth conditions. These findings suggest that the surface of HaPs with ABX$_3$ structure generally tends to be capped by AX surface terminations, analogous to the specific case of MAPbI$_3$. However, under excess BX$_2$ growth conditions, the AX and BX$_2$ surface terminations compete and the result depends on the immediate environment.[108] To date no further data are available for the mixed cation surface system, but the first principle calculations of the electronic structure of the individual phase pure systems, as presented for Rb$_x$Cs$_{1-x}$SnI$_3$, point to the existence of slight local fluctuations of band edge energies, which are strongly dependent on the surface termination and hence growth conditions. This also means that associated charge transfer rates across an interface with such a surface would be affected, which could complicate the selection of the ideal charge transport material.

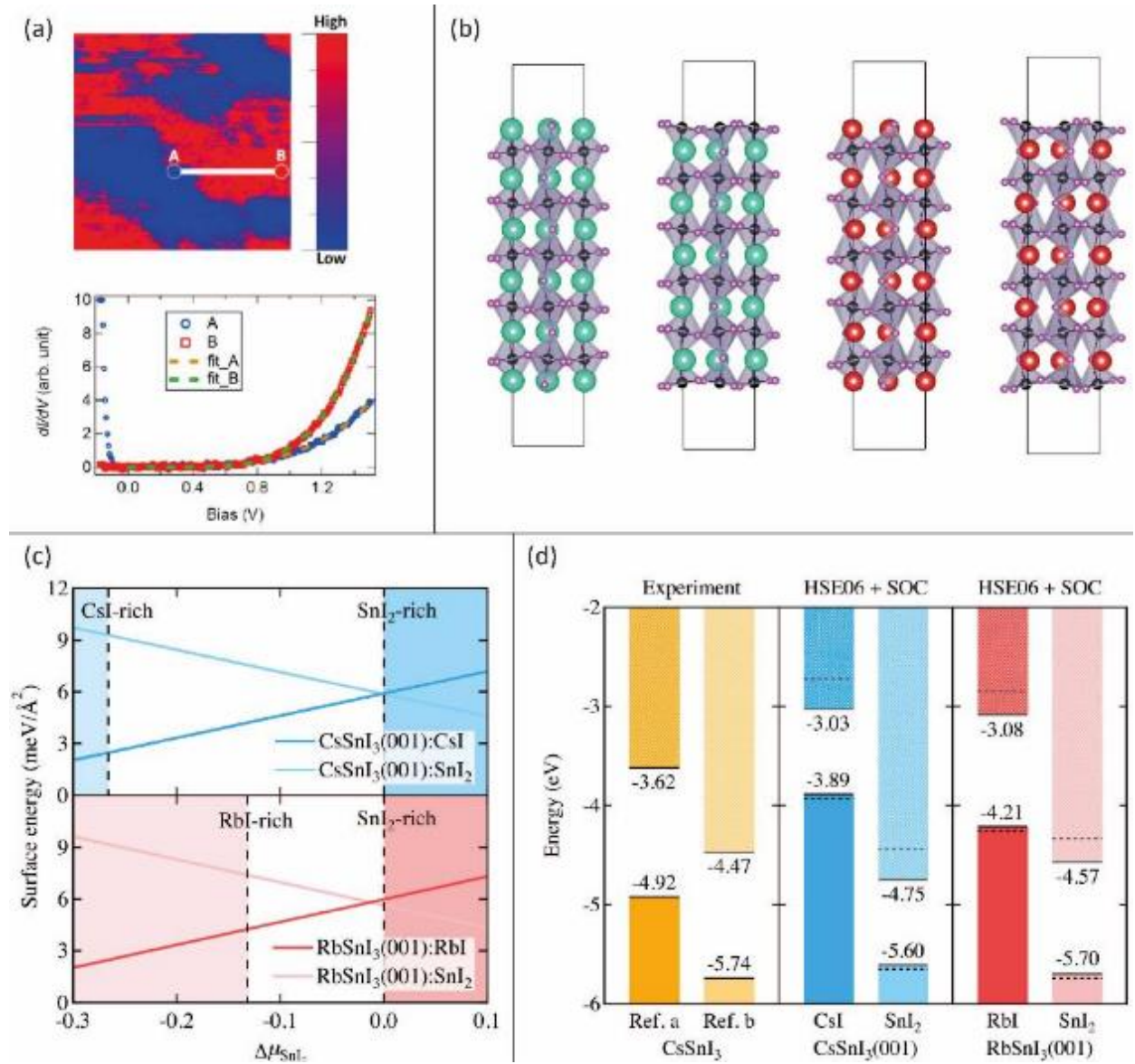

*Figure 7 – (a) Scanning tunneling microscopy dI/dV mapping at -2.0 V of a MAPbI$_{3-x}$Cl$_x$ thin-film cross-section (20 × 20 nm²) with representative dI/dV point spectra measured at low and high contrast regions in point A and point B in the image, respectively. Reprinted with the permission from Ref. 106. Copyright 2016 American Chemical Society. (b) Side view of the relaxed CsSnI$_3$(RbSnI$_3$) (001) surface slab with CsI(RbI) and flat SnI$_2$ surface terminations, respectively (Cs=teal, Rb=red). (c) Calculated surface energy as a function of the change in Δμ$_{SnI2}$ for the surface terminations from panel (b). (d) Calculated band diagram for the (001) surfaces of CsSnI$_3$(RbSnI$_3$) with respect to the vacuum level and with horizontal solid and dashed lines referring to the HSE06 calculated valence and conduction band edges with and without consideration of SOC effects. Comparison to experimental data referenced in Ref. 108. Reprinted with the permission from Ref. 108. Copyright 2017 American Chemical Society.*

Having discussed the nature of the non-stoichiometric surfaces of HaP compounds, we now turn towards the issue of interface formation, namely the adsorption and aggregation on the HaP surface of species that are not constituents of the ABX$_3$ material.

### B.1.2. Aggregation of "Extrinsic" Species

In the context of fabrication and performance of HaP-based optoelectronic devices, it is of paramount importance to understand the nature and impact of interactions between a variety of extrinsic elemental or molecular species and HaP surfaces. Various scenarios and types of interfaces should be considered:

(i) Interfaces with passivating films and buffer layers, which suppress recombination losses, form barriers for migrating ions, shield the bulk from radiation and more generally, can act as encapsulating agents for the cell.

(ii) Interaction with ions and dopants from other parts of the device (Metal, Li$^+$); these constituents of contacts or charge transport layers can induce chemical reactions or act as optoelectronic active centers once they migrate to the interface with the HaP film.

(iii) Interaction with environmental species; primarily $O_2$ and $H_2O$ play a central role due to their natural abundance, particularly for outdoor operating conditions of PSCs.

(iv) Interface formation with charge transport layers, often an organic semiconductor, a transparent conductive oxide or simply the metal electrode.

We will touch upon species that comprise buffer layers at a later point in this review, when we discuss passivation effects, and we turn now to the interaction with metals and dopant molecules. First, the migration of metal atoms from the device electrodes into the perovskite absorber layer has been identified as detrimental to the device functionality and stability. Using time of flight secondary ion mass spectrometry (ToF-SIMS) measurements, Domanski *et al.* found that the incorporation of Au atoms into $FA_{1-x-y}MA_xCs_yPbI_{3-z}Br_z$ occurs after operating and temperature cycling of a PSC layer stack. Au acts as a catalyst for irreversible decomposition of the HaP layer, leading to an irreversible loss in device performance.[110] Understanding of this chemical process, which is still at an early stage, will be subject of a more detailed discussion in section C.3, when we discuss this interface in more detail and present mitigation strategies to suppress the degradation processes.

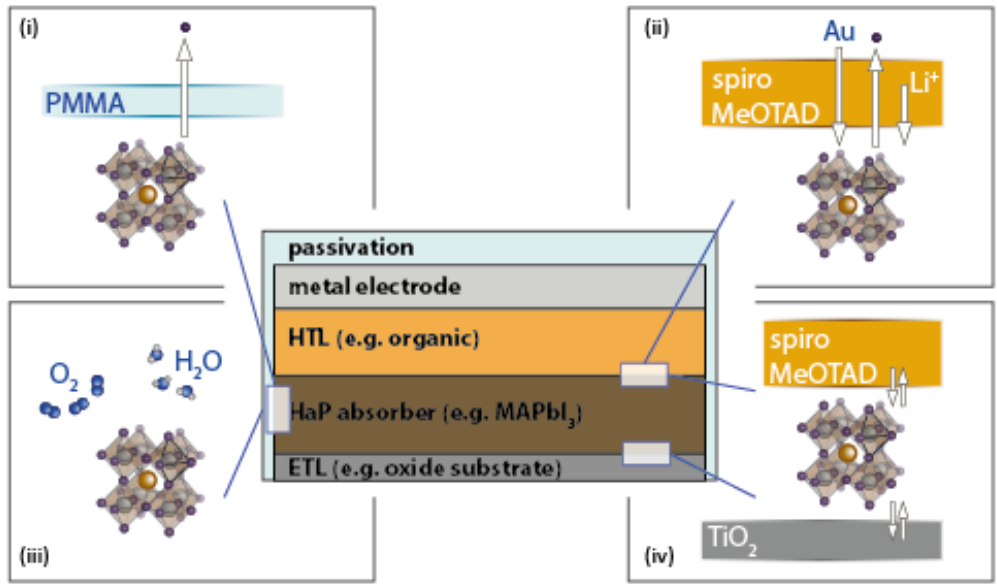

*Figure 8* – *Likely Interaction scenarios with extrinsic species at the HaP surface, including (i) passivation layers, (ii) migrated dopant molecules and metal ions, (iii) atmospheric gases and liquids and (iv) charge transport materials.*

Second, the interaction with dopant species from adjacent charge transport layers can be quite complex and only limited data are currently available to describe the implications of extrinsic dopant contributions to the HaP structural and optoelectronic properties. More in general, attempts to dope HaP semiconductors with extrinsic ions such as $Bi^{3+}$ or $Sr^{2+}$ showed limited success in increasing the density of free carriers in the film, and hence no strong electronic interactions in terms of donating or accepting free carriers from the HaP host have been proposed.[111,112] In a recent approach, Li and coworkers showed that extrinsic ions, which are usually present in the doped organic hole transport layers (e.g., $Li^+$, $H^+$, $Na^+$), can readily migrate to and through the HaP layer. In this context, it was the latter case and hence accumulation of $Li^+$ ions at the $TiO_2$ interface, that modulated carrier injection from the HaP into the $TiO_2$ layer and thus PSC performance, including the degree of hysteretic behavior in the cells current-voltage characteristics.[113]

*Organic molecular adsorbates on HaP surfaces:*

Going back to the initial hypotheses, out of the four scenarios outlined at the beginning of this subsection, the interaction with species from the ambient and the interaction with molecular species in charge carrier transport layers have been subject to additional scrutiny and explored at the atomic level.[90,95] We give here the example of $CH_3OC_6H_5$ (methoxybenzene or anisole) adsorbed on the tetragonal (001) surface of MAPbI$_3$. Historically, the organic semiconductor N',N'-octakis(4-methoxyphenyl)-9,9'-spirobi[9H-fluorene]-2,2',7,7'-tetramine (Spiro-MeOTAD or Spiro-OMeTAD), depicted in Figure 9a, has been the most commonly employed charge transport material on the hole collection side of solid state HaP solar cells.[8] Anisole is also the end group of the Spiro-MeOTAD molecule and other polytriarlyamine derivatives that are commonly employed in organic hole transport layers. In their DFT-based calculations, Torres *et al.* describe that adsorption of anisole on the (001) surface of MAPbI$_3$ occurs only for termination with exposed PbI$_6$ octahedra, which corresponds to the MAI-terminated surface. In this case, the binding energy exhibits a stable minimum for the methoxy group in the interstice site of four corner-sharing octahedra. In contrast, the flat PbI$_2$-terminated surface does not exhibit a stable anchoring site for the adsorption of the molecular compound. The calculations do not reveal any correlation between the binding energy and the orientation of the MA$^+$ cations beneath the Pb-I coordinated surfaces. Hence the adsorption mechanism is described in terms of competing repulsive electrostatic interaction between the methoxy group and the protruding iodine atoms, and the attractive electrostatic interaction between the methoxy group and the Pb$^{2+}$ ions.[90]

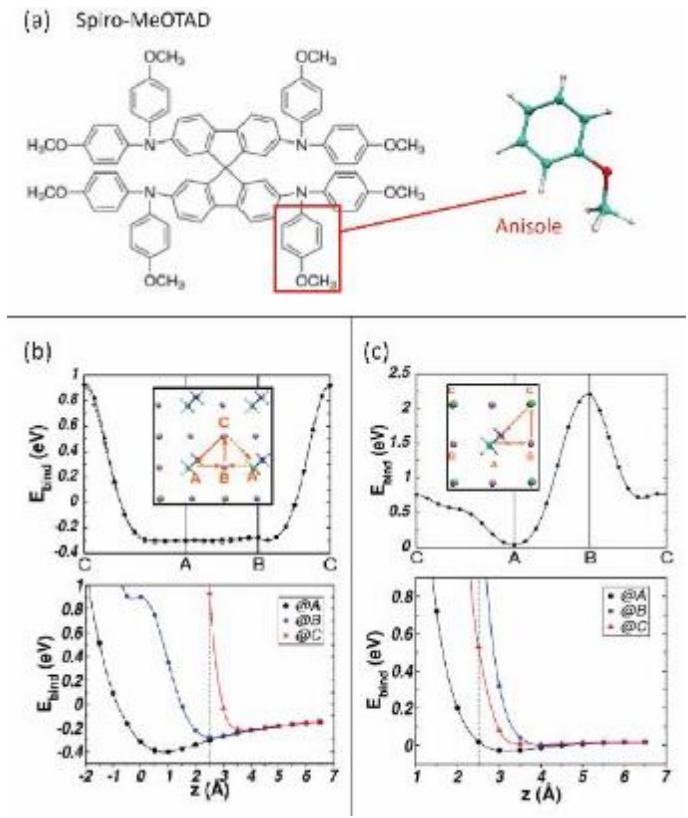

*Figure 9* – (a) Chemical structure of Spiro-MeOTAD. The Methoxybenzene (anisole) unit is marked in red and shown in the inset (C=teal, H=white, O=red).(b),(c) Adsorption of anisole on top of a pseudocubic MAPbI$_3$ slab. On top of the MAI-terminated (b) and PbI$_2$ terminated (c) (001) surface. Calculated binding energy variation for the anisole molecule anchored with the methoxy pointing towards the MAPbI$_3$ surface at a constant height of 2.5 Å along the path C→A →B →C (top view of the cell in inset), and as a function of height in positions A, B and C. Reprinted with the permission from Ref. 91. Copyright 2014 American Chemical Society.

*Oxygen species adsorbed on HaP surfaces:*

Calculations were also performed for the adsorption of water and other oxygen containing species on HaP surfaces. The issue is of high technological relevance, as water ingress into HaP thin-films occurs fast and already on a time scale of seconds, at a relative humidity as low as 10%.[114] The HaP film incorporates water molecules by reversibly forming monohydrates or, in the case of MAPbI$_3$ for instance, transitioning into a bihydrate phase of isolated [PbI$_6$]$^{4-}$ octahedra.[115–117] However, the role of oxygen species in reducing the integrity of the HaP layer is even more pronounced since photocatalytic reactions are readily observed for perovskite films in the presence of oxygen; i.e. in dry air over the course of a couple of hours. For instance, when a MAPbI$_3$ perovskite-based film is exposed to both light and dry air, it rapidly decomposes, yielding the products MA, PbI$_2$, and I$_2$.[118,119]

Deeper insight into the adsorption process and associated chemical reaction was gained from DFT calculations of adsorbed water molecules on HaPs. Koocher *et al.* probed water adsorbates on MAPbI$_3$

(001) surfaces with a variety of surface terminations and polarities (P[+] and P[-]) as depicted in figure 10a.[120]

On the MAI-terminated surface, adsorption was found to be energetically more favorable on the positive polarization (P[+]) than on the negative polarization (P[-]) surface (see figure 10a), while the opposite is true for the PbI$_2$-terminated surfaces. This behavior was attributed to the competition between hydrogen bond interactions between the MA moieties, the Pb-I cage, and the water molecules. For the MAI-terminated P[+] surface, the water molecule can form a hydrogen bond to the exposed $NH_3^+$ surface, while interaction with the $CH_3^+$ group on the MAI-terminated $P^-$ surface is energetically less favorable. The case is reversed for the polarization of the $PbI_2$-terminated surface, where the sub-surface methyl groups and the surface $PbI_2$ lattice exhibit a weaker interaction for the $P^-$ polarization than for the $P^+$ polarization. Hence, the interaction between $H_2O$ and the surface Pb atoms is then dominating the adsorption process.

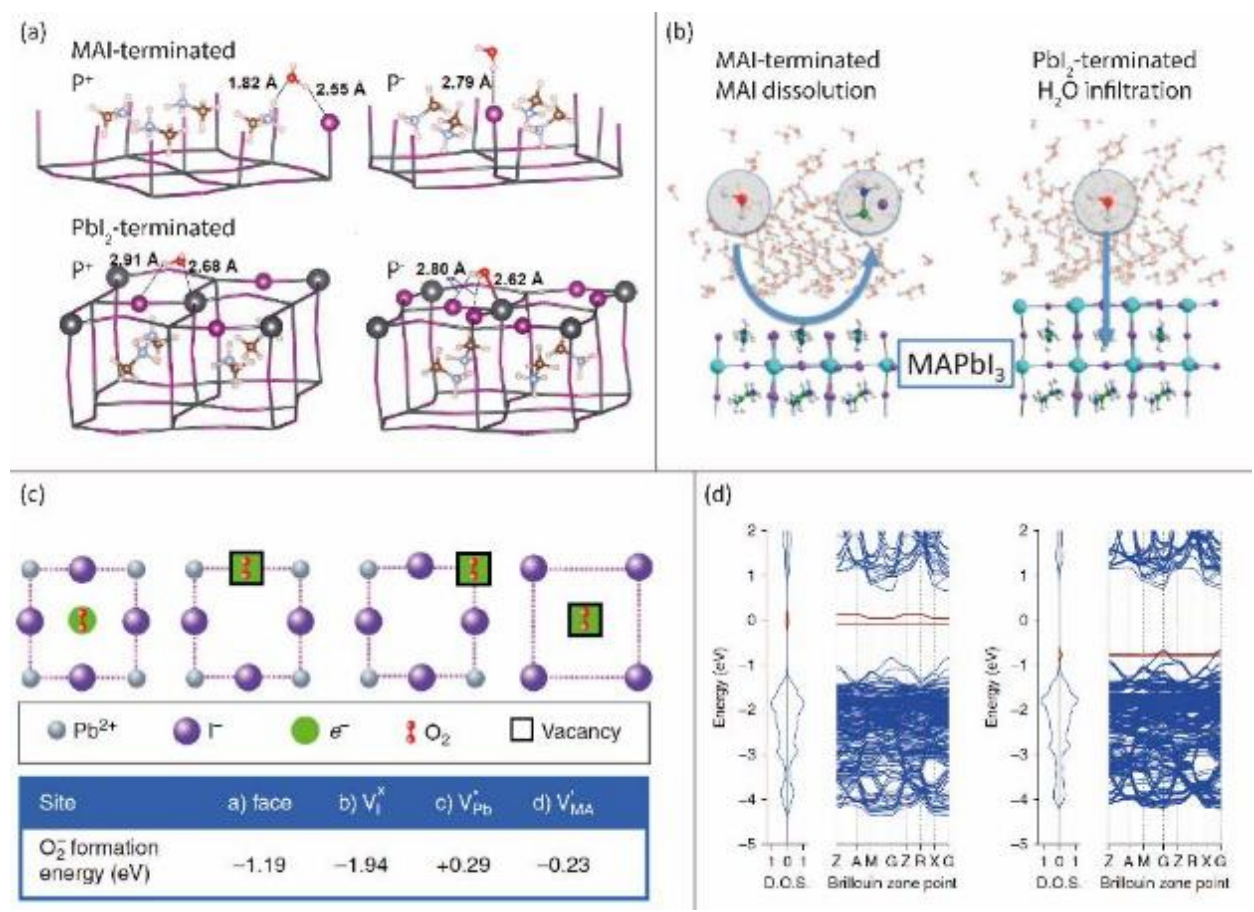

*Figure 10 – Interaction of oxygen species on HaP surfaces (a) Water adsorption on MAI- and PbI$_2$-terminated (001) surfaces of MAPbI$_3$ with different polarities. Reprinted with the permission from Ref. 120. Copyright 2015 American Chemical Society. (b) Dissolution of the MAI termination and percolation of H$_2$O through the PbI$_2$-terminated surface of MAPbI$_3$. Reprinted with the permission from Ref. 121. Copyright 2016 American Chemical Society. (c) Preferred superoxide binding sites in MAPbI$_3$ and respective formation energies. (d) Calculated band structure and DOS for oxygen incorporation into MAPbI$_3$ in defect-free (left) and at an iodine vacancy site (right) with MAPbI$_3$ states in blue and oxygen states in red. Reprinted with permission from Ref. 123. Copyright 2017 by Springer Nature Publishing AG.*

Adding MD simulation to first-principles DFT calculations, Mosconi *et al.* investigated the chemical structure and impact on electronic properties of water molecules adsorbed on $MAPbI_3$ with various surface terminations.[121] On MAI-terminated $MAPbI_3$ surfaces, they found $H_2O$ molecules to interact with the Pb sites, leading to the nucleophilic substitution of I by $H_2O$ and a concomitant desorption of the MA unit. In contrast to this decomposition of the MAI-terminated surface, the $PbI_2$-terminated surface allows for the percolation of $H_2O$ molecules into the layer and their subsequent incorporation into the hydrated phase (figure 10b). Mosconi *et al.* calculate that both hydration processes are exergonic with formation energies of -0.49 and -0.44 eV, respectively, and have only mild effects on the electronic structure of the surface; in the case of the $PbI_2$-terminated surface, the valence band edge is claimed to be stabilized in the interfacial region with the water monolayer. The calculated band gap for a hydrated slab with a 4:1 ratio of $MAPbI_3$ to $H_2O$ increased by 50 meV compared to the non-hydrated unit cell. This change in the electronic properties is rationalized by the minimum impact on the structural properties of the unit cell upon incorporation of one water molecule, as the volume expansion for the hydrated tetragonal slab amounted to only 1%.[121]

Similar results have been reproduced by Zhang *et al.* with the additional observation of photodegradation pathways.[122] Upon attachment of $H_2O$ molecules, the usually more stable MAI-terminated surface undergoes a high degree of disorder, which is implied by a slightly broader radial distribution function and more pronounced fluctuations in the bond lengths (e.g. Pb-Pb) after a 40 ps MD simulation. Likewise, but to a much lesser degree, the defective $PbI_2$-terminated surface exhibits the same dynamics, while the non-defective PbI$_2$-terminated surface appears more stable. Nonetheless, both PbI$_2$-terminated surfaces, defective and non-defective, enable the incorporation of water molecules.

In summary, under humid conditions the MAPbI$_3$ surface reconstructs towards a PbI$_2$-terminated phase, eventually leading to the formation of the monohydrate MAPbI$_3$·H$_2$O. The effects of adsorption of individual water molecules on the MAPbI$_3$ surface on the MAPbI$_3$ surface electronic structure are minor; with adsorption of the water molecule and the associated surface reconstruction, the band gap increases by roughly 50 meV while only shallow surface states localized around the valence band edge are formed. In a separate study, Zhang *et al.* took photoexcitation into account, and found that the change in Pb-I bond strength is especially pronounced at sites adjacent to the adsorbed water molecules. Eventually, the significantly weakened bond enables the formation of the hydrated crystal phases at the defective MAPbI$_3$ surface.[122]

Further details on the role of surface defects for oxygen-related photodegradation processes have been investigated by Aristidou and coworkers .[123] Beside the process of hydrate-induced decomposition, photodegradation also occurs in dry environment under light exposure. They found that molecular oxygen (O$_2$) can be incorporated into the surface at iodine vacancy sites. Subsequently, photoexcited carriers that reach this site can lead to the formation of the reactive superoxide ions (O$_2^-$). In their preceding work, Aristidou *et al.* describe that the superoxide species can then deprotonate the MA$^+$ cation upon photoexcitation, resulting in the decomposition of an MAPbI$_3$ layer into PbI$_2$, water, MA and I$_2$ according to the following reaction scheme:[119]

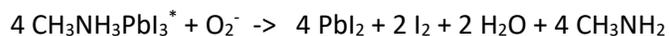

$$4\ CH_3NH_3PbI_3^* + O_2^- \rightarrow 4\ PbI_2 + 2\ I_2 + 2\ H_2O + 4\ CH_3NH_2$$

The rates of oxygen diffusion and oxygen-induced degradation are linked to the structure of crystallites and grain boundaries on the mesoscopic scale, and to the details of the surface defects on the atomic scale.[123] Concerning the mesoscopic scale, HaP films with smaller grains and hence a higher surface to volume ratio tend to allow for a faster diffusion of oxygen and thus superoxide formation with subsequent photodegradation of the HaP film. On the atomic scale, *ab initio* simulations indicate that superoxides form preferentially on defective surfaces, as depicted in figure 10c. However, forming an $O_2^-$ species on either an MA vacancy or a Pb vacancy site is energetically highly unfavorable. Even the formation of $O_2^-$ species on the face center of the cubic surface neighboring four iodine atoms is less likely compared to the low formation energy on the iodine vacancy site; a broader rationalization of which is the matching valence of $O_2^-$ and $I^-$. The implications on the electronic structure are significant and mark the strong contrast between having $O_2^-$ on the face of the unit cell or in an iodine vacancy. If $O_2^-$ is introduced in a defect free unit cell, the oxygen valence level appears as a mid-gap defect state. In comparison, the top most oxygen valence level moves to the top of the valence band of $MAPbI_3$ when the $O_2^-$ moiety is located on the energetically favorable iodine vacancy site (see Figure 10d). Interestingly, the latter case adds to the many defect scenarios for an HaP compound under ambient environmental conditions, which show only minor effects on the HaP's electronic properties.

### B.1.3. Typical surface defects, compensation and passivation

Before giving a comprehensive account of the electronic structure of HaP surfaces and measurements thereof, we summarize this subsection by several remarks on surface defects along with compensation mechanisms and passivation strategies. In general, we learn from the atomic picture, mostly derived from DFT calculations and MD simulations on $MAPbI_3$, that the majority of defect types induced at the surface of the HaP compound are electronically benign compared to similar defects at III-V semiconductor surfaces, which in organic/inorganic hybrid HaPs can be attributed to the previously described organic surface termination.

- For MAPbI$_3$ both MAI- and PbI$_2$-terminated surfaces exist and can form under growth conditions in thermodynamic equilibrium. However, the MAI-surface termination is more stable and hence easier to achieve.[91,100,124]

- Calculations show that, similar to bulk defects,[25] surface defects do not generally introduce electronic states deep in the band gap. However, the PbI$_2$-terminated surfaces exhibit a lower band gap than MAI-terminated surfaces.[91,95,125]

- Surfaces of ABX$_3$ with > 1 halide on the X site or > 1 cation on the A –site, do not exhibit intrinsic mid-gap defect states.[106,108]

- Extrinsic chemical species do not usually dope HaP compounds, i.e. do not function as good electron donors/acceptors, and do not usually introduce recombination centers. However, case-by-case investigations are required, and will be further developed for several examples in the later sections of this review.

- Halide vacancies are common surface defects and can act as nucleation sites for extrinsic species, e.g. in the case of superoxide formation.[119]

- Neither water molecules in monohydrated perovskite films nor superoxides significantly distort the structural or electronic properties of the perovskite layer. However, in conjunction with light exposure, these can lead to a degradation of the HaP film.[120,121,123]

In summary, electronically active (mid-gap) states are generally not expected in HaP-based optoelectronic devices under realistic environmental conditions, and consequently defect surface states are not directly observed. Nevertheless, in spite of their apparent benign nature, the role of defect surface states is still being debated. In particular, surface defects and the energetic disorder they induce do matter when we look into the extraordinarily long carrier lifetimes, reported in HaP single crystals. Even though no clear-cut chemical picture exists on the atomic level, impedance spectroscopy measurements by Duan *et al.* suggest the existence of a defect level 0.16 eV above the valence band maximum.[126] They ascribe this finding to iodine interstitials, which however have been calculated and experimentally corroborated for the bulk only.[25,127,128] Another hypothesis for the existence of trap states is based on local carrier transport properties in HaP thin films, probed by conductive atomic force microscopy measurements. Leblebici *et al.* suggest a pronounced intra-grain heterogeneity of the electrical properties, which they ascribe to different facets for different crystallites, and which could indicate a significant variation of surface defects.[129]

In this regard, multiple passivation strategies have been pursued to improve carrier lifetimes in HaP films, which then approach those of single crystals, and hence, could in principle, enhance performance parameters of devices if those could be made with **ideal interfaces**. Chen *et al.* propose a Type I interface alignment at the $PbI_2$/$MAPbI_3$ interface,[130] which is notably different from the $PbI_2$-termination and respective band gap shrinkage presented earlier. Another common approach is the use of surface modifications through supermolecular halogen bonding donor-acceptor complexation, e.g. iodopentafluorobenzene (IPFB), or ligands such as ethanedithiol (EDT) and tri-*n*-octylphosphine oxide (TOPO), which enhance photoluminescence to lifetimes on the order of multiple microseconds.[131–133] While detailed examples of passivating layers will be discussed in the later sections of this review, the reader is referred to the comprehensive work of Manser *et al.* for a more detailed account on the competing mechanism of bulk and surface recombination.[31]

Passivation to suppress the degradation reactions is another important avenue for improving HaP-based devices. HaP thin films with large grains, such as those grown from nanocrystalline nuclei, exhibit enhanced thermodynamic and phase stability compared to finely grained thin-films,[134] while we note that in certain cases, e.g. $CsPbI_3$, phase stabilization can be achieved by strain relaxation in nanocrystalline form.[78] Alkali halides have been introduced in the fabrication scheme of HaP layers as additives to act as passivation agents in the resultant dry film.[135] Their role on the superoxide degradation pathway has been explored and while oxygen diffusion is not hindered, the excitation-induced formation of $O_2^-$ is suppressed.[123] Furthermore, films that are part of the new generation of mixed cation and anion HaPs are particularly prone to ion migration and can exhibit phase segregation,[103] a potential degradation pathway.[77] In these cases, a specific route employing a significant load of potassium iodide in the precursor leads to an improvement of the optoelectronic properties of the HaP film.[136] This effect is attributed to halide-vacancy management; i.e. a surplus of halide ions is immobilized through complexing with potassium into benign compounds at the grain boundaries and surfaces.

**B.2. Electronic structure and energetics of the HaP thin-film surface**

The electronic properties at the surface of HaP semiconductors are derived from the band structure, which in first approximation is not significantly distorted at the surface as discussed in the preceding section. While DFT calculations, based on the greatest gradient approach (GGA) capture the essence of the band structure, the correct computation of the band gap in the bulk and of the valence band dispersion requires an additional theoretical framework. Using quasi-particle, self-consistent GW (QSGW) with accounting for spin-orbit coupling (SOC), Brivio *et al.* computed band dispersion in various HaP compounds, including that of MAPbI$_3$, pictured in Figure 11a.[137] The formal electronic configuration of the components that make up the PbI$_6$ cage is $5d^{10}6s^26p^0$ for Pb$^{2+}$ and $5p^6$ for I$^-$. The figure provides the orbital character of the individual bands and their contribution to the DOS. The color code of the band structure in Figure 11a attributed the respective electronic states. The valence band maximum consists of a combination of approximately 70% I 5p and 25% Pb 6s orbital character, while the conduction band character is dominated by the Pb 6p orbital. In contrast to this, the molecular MA$^+$ unit exhibits sp$^3$ hybridized σ bonds deep in the valence band. Due to the absence of any strong interaction or hybridization with the inorganic PbI$_6$ cage, they remain without major dispersion.

These QSGW calculation already reveal a trend in the band structure, which was further elaborated using additional relativistic corrections:[26] valence band maximum and conduction band minimum are shifted slightly from the high symmetry point R as a consequence of the strong SOC. The corrections applied in this case already yield a band gap of 1.67 eV and of optical transitions, which is still a bit too large compared to spectroscopic measurements.[138] Finally, the strong dispersion at the band edges indicates a low density of states, a topic that will be discussed in more detail below in section B.2.1.

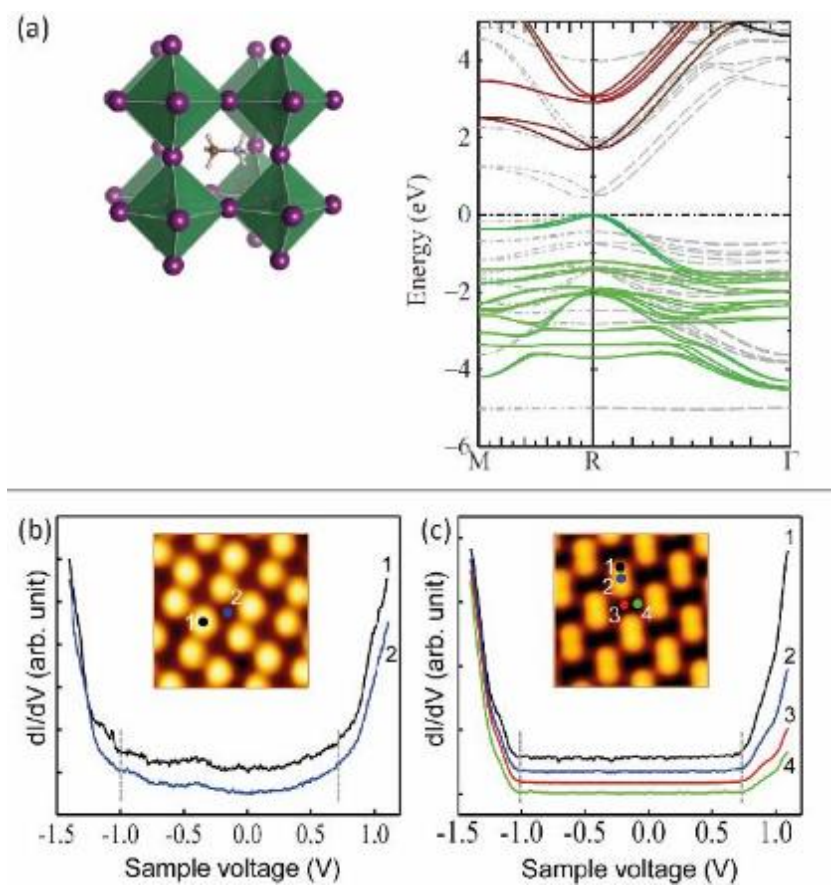

*Figure 11* – (a) Calculated electronic band structure of the cubic phase of MAPbI$_3$ with the valence band maximum set to 0 eV from quasiparticle self-consistent GW computations (QSGW). Color code of the bands according to their orbital character: I 5p = green, Pb 6p = red, Pb 6s = blue. Light-gray dashed lines show corresponding bands in the local density approximation (LDA). VB (CB) are dominated by I 5p (Pb 6p) states in bright green (red), with the darker colors indicating orbital mixing. Reprinted figure with permission from Ref. 137. Copyright 2014 by the American Physical Society. (b,c) STM measurements and differential tunnel current spectra of MAPbI$_3$ acquired from the zigzag (b) and dimer (c) surface reconstruction. Curve intensities are normalized with an identical value at a bias of −1.4 V. Reprinted with the permission from Ref. 100. Copyright 2016 American Chemical Society.

STM measurements of a MAPbI$_3$ layer on top of gold reveal the zigzag and dimer structure on the MAI-terminated (001) face as discussed in section B1.1 (see figure 11b and c).[100] The associated scanning tunneling measurements, i.e. the differential tunneling current d$I$/d$V$ as a function of the bias voltage $V$, yield the onset of the valence and conduction bands, respectively, at the location of the tip and is thus a representation of the local density of states (LDOS). For each tip position, the conduction band minimum is at 0.7 eV above $E_F$, while the valence band maximum is at 1.0 eV below $E_F$. This results in a band gap of 1.7 eV, which is in good agreement with the value computed from QSGW calculations but a bit off from the measured optical gap. Optical gap and transport gap are supposed to coincide in the accuracy limits of these measurements as exciton binding energies in HaPs are low ($E_{B,ex}$ < 2-50 meV).[139,140] The match between the band gap measured at the surface, and the values calculated and measured for the bulk, as well as the invariance of the STM-measured LDOS on the selected surface site corresponds well to the theoretical predictions; the surface electronic band structure and band gap are similar to those in the bulk and are not surface site-specific. We discuss below the experimental determination of these values on larger sample sets and areas, for HaP films used in optoelectronic devices.

**B.2.1. Band edge determination**

Direct and inverse photoemission spectroscopy (PES/IPES) are the methods of choice to measure the energy positions of valence band maximum (VBM) and conduction band minimum (CBM) and the densities of states (DOS) of frontier electronic states at the surface of semiconductors. In addition, with the extraction of the vacuum level ($E_{VAC}$) from these measurements, ionization energy (IE) and electron affinity (EA) can be identified. These quantities are central to the determination of chemical and optoelectronic properties at, and carrier transfer rate across, the interface in a device as laid out in section A.3.

Initially, the determination of the band edge positions proved difficult for HaP films since a precise distinction between bulk, surface or defect states from photoemission spectra is very challenging. This is generally the case for all materials affected by structural disorder or an abundance of defects. Kraut *et al.* offered an approximation of the valence band maximum position by performing a linear regression to the leading edge of the valence band spectrum in X-ray photoemission spectroscopy (XPS) data of crystalline Ge(110) and GaAs(110) surfaces. This approach was found to lead to good agreement with onsets determined from theoretical calculations of the valence band DOS used as reference for these highly crystalline, ordered and pristine surfaces.[141] This fitting procedure was subsequently applied with great success to a broad range of semiconductor systems, including conjugated organic semiconductors whose frontier molecular orbitals are often determined by the delocalized $\pi$ electron system.[142] In many cases however, this *ad hoc* approach can be erroneous for semiconductors that have "soft" energy level onsets, as in the case of non-abrupt band edges. This is the case for materials that exhibit a significant contribution of tail states (e.g. through disorder) or, in case of the HaPs, a particularly low DOS at the band edges.[101] The latter is comparable to what is found for other lead-based compound semiconductors such as PbS.[143] In these cases, extrapolating the perceived linear part of the band onset becomes somewhat subjective and can yield VBM positions that are too deep,[143,144] leading to unphysical results such as the prediction of an excessively large band gap.

In an initial attempt to circumvent this problem, the low DOS at the band edge at the valence band of $MAPbI_3$, $MAPbI_{3-x}Cl_x$ and $MAPbBr_3$ was visualized and approximated in a semilogarithmic representation of the valence band region,[145] similar to the analysis of defect and tail states in organic semiconductors.[146] Together with the evaluation of the conduction band minimum from concomitant IPES measurements, these experiments led to a determination of a single particle gap of 1.7 eV for $MAPbI_3$, and 2.3 eV for $MAPbBr_3$, which are on the same order of the values of experimentally determined optical gaps and DFT-computed band gaps as presented earlier.[137,145] Ionization energies and electron affinities were determined to IE = 5.4 eV and EA = 3.7 eV for MAPbI$_3$, and IE = 5.9 eV and EA = 3.6 eV for MAPbBr$_3$. Similarly, ultraviolet photoemission spectroscopy was used to determine the IE = 5.2 eV of FAPbI$_3$.[147] We will discuss in section B.2.2 that the ionization energy can also depend on the HaP composition and surface termination, which could also cause the difference in IE between MAPbI$_3$ and FAPbI$_3$. Another strategy to determine the band edges involves the fit of the valence band leading edge by a parabolic model. From the case of PbS quantum dots, which exhibit a significant contribution of tail states to the DOS at the band edge, Miller *et al.* adopted and modified the method by Kraut and calculated the DOS of the PbS valence band (DOS$_{VB}$) using GW calculations with SOC.[143] Formally the DOS$_{VB}$ is then expressed by:

$$\mathrm{DOS_{VB}} = A\,(\mathrm{DOS_L} + b\,\mathrm{DOS_\Sigma}) \otimes g,$$

with the density of states at the high-symmetry L and $\Sigma$ points in *k*-space, *A* a global scaling factor, b a weighing factor to account for the DOS contribution at the L and $\Sigma$ points, and *g* the convolution with a Gaussian function to account for linewidth broadening. The DOS in the respective *k* points is then approximated in a parabolic model, which can be expressed as

$$\mathrm{DOS_{L/\Sigma}} = 2(m_e^*)^{3/2}(E - E_\mathrm{VBM})^{1/2},$$

where $m_e^*$ is the electron effective mass at the L or $\Sigma$ point, *E* the electron binding energy, and $E_\mathrm{VBM}$ the energy position of the valence band maximum. A similar parabolic approximation was used by Zhou *et al.* for a set of photovoltaic materials, i.e. Si, CdTe, and several typical HaPs, to corroborate the low density of states for hybrid organic-inorganic HaPs such as MAPbI$_3$.[148] A similar model was also used to fit the VBM from photoemission spectroscopy data of MAPbBr$_3$ single crystals,[149] which will be discussed in more detail in section B.3.2.

In a combined theoretical and experimental study, Endres and coworkers performed direct and inverse photoemission measurements on a set of HaP thin-films (MAPbI$_3$, MAPbBr$_3$ and CsPbBr$_3$) on top of TiO₂/FTO substrates.[101] The measured UPS and IPES valence and conduction band spectra of the three compounds were used to fit DFT calculations including SOC, appropriately shifted and scaled to achieve a good alignment of the spectral features over a broad energy range, as depicted in Figure 12. A summary of the accurate determination of the band gaps along with the value of the work functions and respective ionization energies and electron affinities are indicated in the figure. The results clearly point out the fact that the valence band maximum is marked by a low DOS. The plot on the semi-logarithmic scale further assists in the visualization of the transport gap between valence and conduction band. This low DOS hypothesis has been proposed explicitly for MAPbI$_3$, where it is attributed to the strong coupling between the antibonding orbitals made up of Pb 6s and I 5p levels, with a small contribution to the DOS only at the R-point in reciprocal space.[150] In contrast, the expanded unit cell of PbI$_3^-$ does not exhibit the same type of band dispersion (see figure 13a). The experiments indicate that similar effects prevail for compounds with other HaPs, with different A-site cations and (X) halides HaP. The universality of this finding agrees with further theoretical calculations, which suggest that substitutions of the various ions in the ABX$_3$ structure could well be approximated by an orbital overlap rationalization scheme.[151] The coupling of frontier atomic orbitals determines the band structure, thus the effective masses of electrons and holes, and hence charge carrier transport, charge carrier densities, and the limits for hot carrier dynamics in HaPs.[148,150]

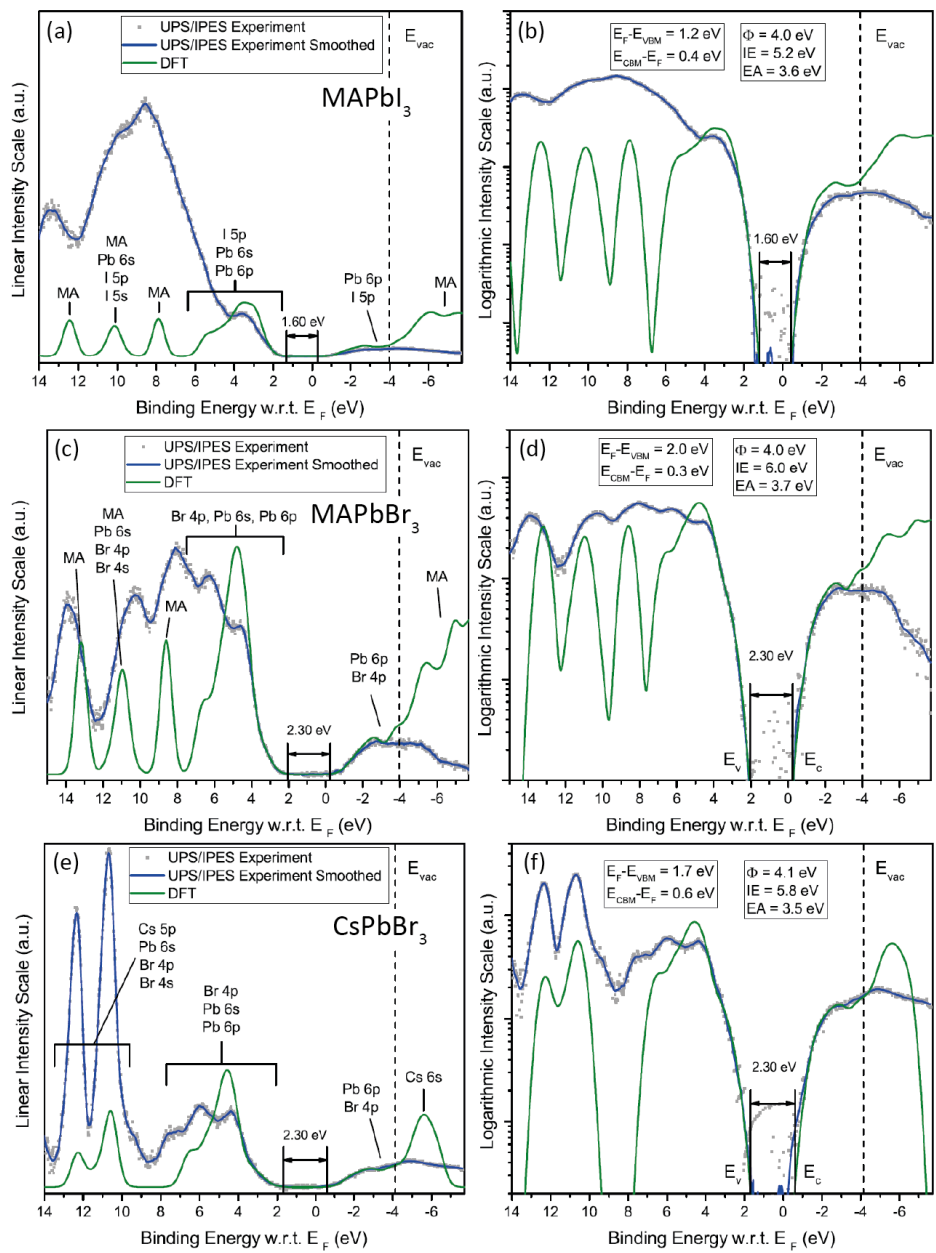

*Figure 12* – Comparison between measured UPS and IPES spectra for approximately 300 nm thick (a,b) MAPbI₃, (c,d) MAPbBr₃ and (e,f) CsPbBr₃ films, grown on TiO₂/FTO, and spectra, derived from DFT calculations. The energy scale is referenced to $E_F$ = 0 eV, while the intensity is plotted on a linear (a,c,e) and logarithmic (b,d,f) scale, respectively. Major atomic orbital contributions are indicated in the spectra and values for IE, EA and the energy gap were extracted from the measured and fitted band onsets in conjunction with the read-out of the work function φ from the secondary electron cut-off in the UPS measurements (not shown here). Reprinted with the permission from Ref. 101. Copyright 2016 American Chemical Society.

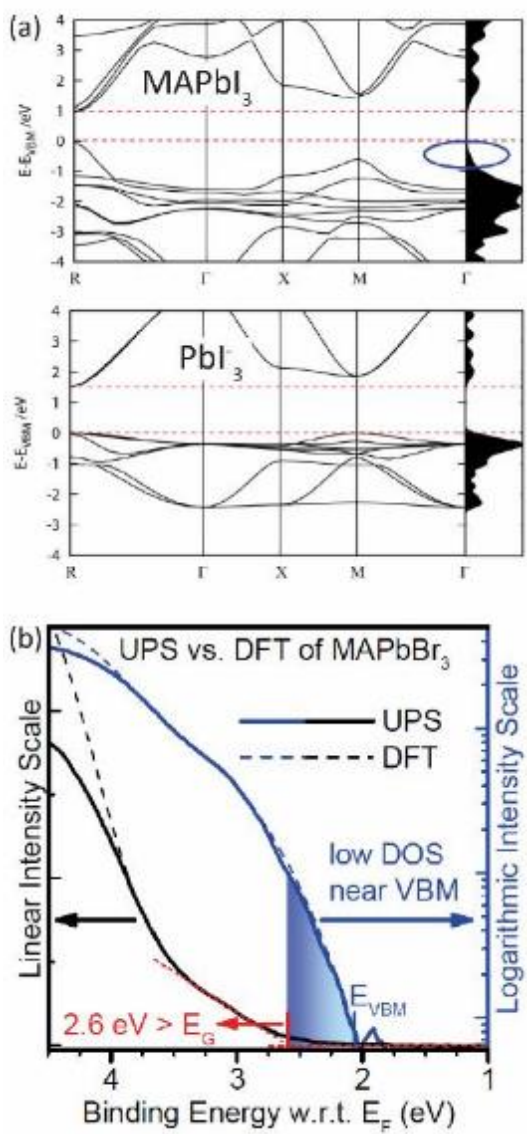

*Figure 13* – (a) Calculated band structure and DOS of MAPbI$_3$ and the expanded unit cell of a PbI$_3^-$ ionic moiety. The MA unit leads to a different overlap in the Pb and I atomic orbitals resulting in a low DOS at the valence band maximum (blue circle in upper panel). The band gaps are underestimated in these LDA-based computations. Reprinted with the permission from Ref. 150. Copyright 2015 American Chemical Society.. (b) Reliable fitting of the low DOS at the VBM, measured in ultraviolet photoemission spectroscopy (UPS), as pictured here for MAPbBr$_3$, requires further evaluation with accurate DFT calculations. Reprinted with the permission from Ref. 101. Copyright 2016 American Chemical Society.

Since the process of a dedicated combined theoretical and experimental assessment is very involved and computationally expensive, reliable approximations to cover the many stoichiometric variations in the HaP composition are highly sought-after. In their orbital overlap approach, Meloni *et al.* link the evolution of the band gap to the overlap of the atomic orbitals.[151] Their approximation can be broken down to a linear correlation between the overlap of the B and X site valence band orbitals and the valence band maximum with respect to the vacuum level, i.e. the ionization energy is correlated to the overlap integral between the B and X atomic valence orbitals. The correlation is poorer for the less localized conduction band orbital overlap and electron affinity. Meloni *et al.* suggest these correlations to work well for pure Pb- and Sn-based HaP systems, while the usability of the scheme is more limited when dealing with more exotic effects.[151] For instance, Hao *et al.* discuss an anomalous band gap opening in ABX$_3$ structures with B = Pb$_{1-x}$Sn$_x$ deviating from Vegard's law.[152] They suggest that such a trend in composition-dependent band gap could have similar origins as in Pb$_{1-x}$Sn$_x$Te chalcogenides, for which band inversion and hence a systematic change in the atomic orbital composition of the conduction and valence band has been reported.[153]

Assessments of the band edge energies, ionization energies and electron affinities have been conducted for many HaP compositions relevant to the current technological interests, including further mixed halide systems,[154–156] and were summarized in previous review articles.[54,157] The above-mentioned difficulty to determine the low-DOS band onsets by common practice fitting procedures has led to significant variations of reported values of the electronic energies, mentioned above. In addition, the composition of the surface is difficult to assess, and non-stoichiometry is expected to lead to variations in the characteristic energy levels and the vacuum level position. In MAPbI$_3$ this could be through variations in the extent of MAI- or PbI$_2$-terminated surfaces.

### B.2.2. Compositional effects on Ionization energy

The most accessible experimental quantity is the ionization energy, and halide excess and deficiency in HaP thin-films has an immediate effect on it. Co-evaporation via a dual-source deposition, with one source for AX and the other for BX'$_2$, enables the fabrication of mixed halide HaP films and affords reasonable control over the stoichiometry and halide content. For films deposited from MAI and PbX$_2$ with X = I, Br, Cl, respectively, incorporation of a significant amount of Br into the mixed compound MAPb(I$_{1-x}$Br$_x$)$_3$ must be contrasted to the negligible amount of Cl incorporated into MAPb(I$_{1-x}$Cl$_x$)$_3$ in a co-evaporation process.[154] Hence, in the case of MAPb(I$_{1-x}$Br$_x$)$_3$, both band gap and valence band position scale with the mixing ratio between the values for pure MAPbI$_3$ and MAPbBr$_3$, whereas the bandgap and VBM position in the nominal MAPb(I$_{1-x}$Cl$_x$)$_3$ films correspond to those of a MAPbI$_3$ film. This is in good agreement with the measured bandgap and IE of solution-processed MAPb(I$_{1-x}$Cl$_x$)$_3$ films, for which, the *de facto* Cl content is negligible.[145]

The analysis of films processed by vacuum deposition reveals another important interdependence. As the ratio between MAI and PbX$_2$ evaporation rates in a co-evaporation experiment tailors the resultant film stoichiometry, one can sample the surface composition from an MAI-rich surface to a PbI$_2$-rich one, which corresponds to the theoretical exploration of film formation as function of the chemical potentials of the individual components in the binary compounds.[25] The trend in the characteristic energy levels is the same for all three compositional gradients, MAPbI$_3$, MAPb(I$_{1-x}$Br$_x$)$_3$ and MAPb(I$_{1-x}$Cl$_x$)$_3$, i.e., IE and EA both decrease with excess MAI and hence an MA$^+$ rich sample (see Figure 14a). In an expanded experimental series, Emara *et al.* demonstrated for a large set of samples and sample processing procedures that the value of IE in MAPbI$_3$ is correlated to the experimentally determined elemental Pb to N ratio, which should reflect the PbI$_2$ / MAI ratio.[144] The plot in figure 14b confirms that the measured values for IE are maximal in the region of PbI$_2$ excess and decrease proportionally with an increasing MAI component. Notably, no peak of PbI$_2$ is found in X-ray diffraction (XRD) measurements even in the high PbI$_2$ excess regime, which suggest that the excess PbI$_2$ is poorly crystalline or does not segregate into large continuous crystallites. This finding leads to the assumption, that even in the high PbI$_2$ excess regime, the films resemble defective MAPbI$_3$. Hence, the measured stoichiometry and the correlation with IE could be confined to the surface region that is probed by PES. In addition to the changed IE, the valence band structure changes with changing MAI/PbI$_2$ mixing ratio. Most prominently, the intensity of the VB leading edge is modulated, which is consistent with a different coordination of the valence electrons and hence average Pb-I bond properties. However, in the case discussed above, no linear correlation is observed between the MAI/PbI$_2$ ratio and the corresponding valence band peak intensity; possibly dedicated DFT calculations, which would need to sample a large range of disorder, can help shed light on what is/are the physical causes for these behaviors.

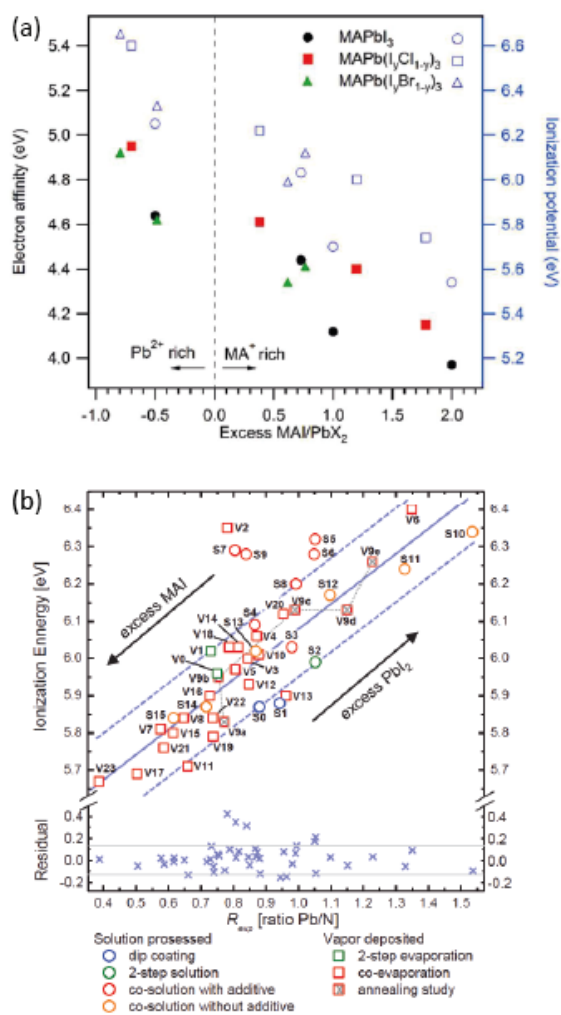

*Figure 14* – a) Measured IE and projected EA (using the optical gap) as a function of MAI to PbX$_2$ excess for MAPbI$_3$, MAPb(I$_{1-x}$Br$_x$)$_3$ and MAPb(I$_{1-x}$Cl$_x$)$_3$ thin films. Reproduced from Ref. 154 with permission from the PCCP Owner Societies. b) Measured IE as a function of elemental Pb to N ratio for MAPbI$_3$ films from various fabrication procedures. Reprinted wih the permission from Ref. 144. Copyright 2016 WILEY-VCH Verlag GmbH & Co. KGaA, Weinheim.

### B.2.3. Defect tolerance and the effects of radiation damage

At this point a question arises as to how to reconcile the proposed defect tolerance and experimentally detected self-healing capabilities of HaPs with the large variation in IE as a function of compositional parameters between mostly PbI$_2$- or MAI-terminated MAPbI$_3$ surfaces. While we see IE and EA as key parameters for estimating the electronic energy level alignment and formation/presence of an energetic barrier at the interfaces, the intrinsic electronic properties of the perovskite alone, i.e. charge carrier transport and recombination, depend more on the modulation of the bulk defect state density and associated shifts of $E_F$ in the gap. However, the surface defect state density and density of interface states become important quantities for the estimation of recombination currents which constitute a main loss factor for photovoltaic performance.

The evolution of defective MAPbI$_3$ (surfaces) and threshold for substantive switching of its electronic properties can be tracked in an experiment, that exploits the limited radiation hardness of HaP compounds. Several early reports indicated that HaP films decompose under harsh radiation conditions: for instance, metallic lead species emerge in the form of nanocrystals in MAPbI$_3$ films upon exposure of the film to X-ray radiation. The effect is due to radiolysis as evidenced for PbX$_2$ compounds already, and limited thermal stability of hybrid HaPs, from which the organic unit A is readily evaporated either as AX or 2A + X$_2$.[158–161] In an XPS experiment one can hence track the decomposition process of HaP as the material is exposed to X-ray radiation used as excitation for photoemission. In MAPbI$_3$ the loss of MAI occurs at appreciable rates at a temperature of around 225°C under atmospheric pressure. Thus, evaporation of MAI happens at even lower temperatures under ultra-high vacuum conditions, under which the XPS experiment is usually conducted.[162]

Steirer *et al.* performed continuous XPS measurements on a solution-processed MAPbI$_3$ film over the course of 42 h.[33] An Avrami phase-change analysis[163] of the decay in the nitrogen signal yielded a single exponent, indicating that the loss of MA follows first-order kinetics, i.e. a steady loss rate of MA species under X-ray exposure in vacuum. The evolution of the nitrogen signal also indicated that, while the MA contribution is decreasing, small amounts of at least two other nitrogen species emerged at slightly lower binding energies. No clear chemical attribution of the peaks was provided in the study, but a likely explanation for these nitrogen components is the formation of molecular fragments of a broken-down MA moiety, which is not further specified in the report but could for instance be adsorbed CH$_3$NH$_2$. During exposure, the iodine content was also shown to drop at the same rate as that of nitrogen, consistent with the loss of MAI and hence suggesting a continuous decomposition of MAPbI$_3$ towards PbI$_2$. Surprisingly, the position of the valence band with respect to $E_F$ was not found to change until an I/Pb ratio of 2.5 was reached (see Figure 15). The origin of this effect lies in the depletion of MAI, which leads to the creation of MA and I vacancy pairs (Schottky defects in purely ionic crystals). This picture is in good agreement with the theoretical description by Yin *et al.*, who locates these defect states close to the conduction and valence band edges, respectively.[25] The defect pairs are self-compensating and hence do not affect the

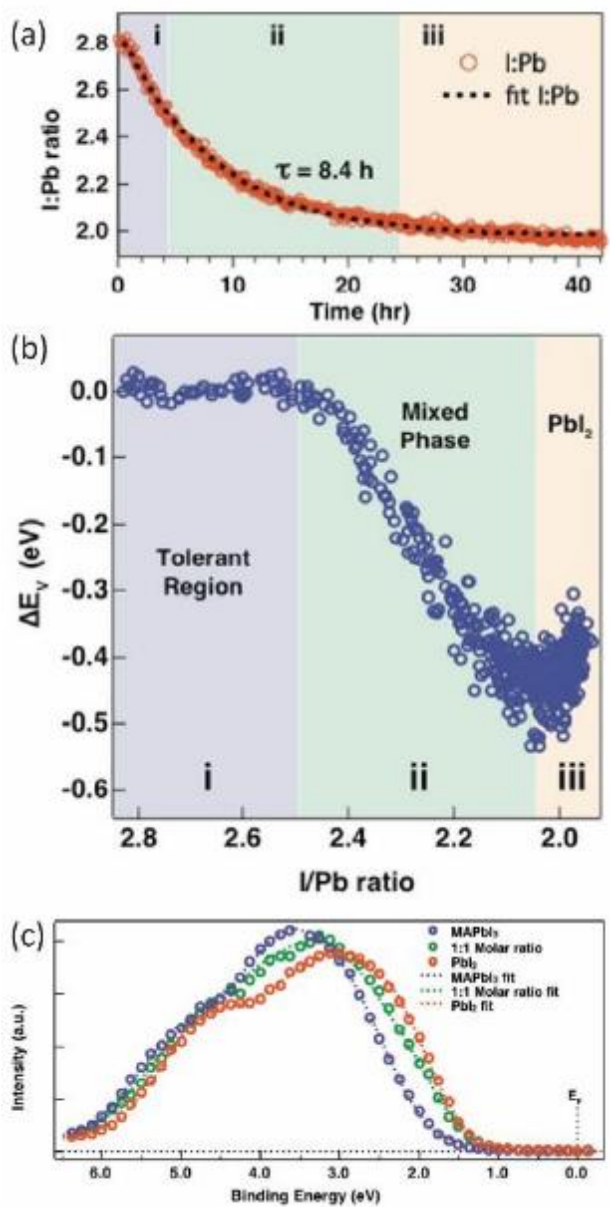

*Figure 15* – *Decomposition of MAPbI$_3$ film on a TiO$_2$ substrate under continuous X-irradiation. (a) decrease of I / Pb ratio with X-ray exposure time. (b) Apparent shift of the leading edge to the valence band maximum as a function of I/Pb ratio. (c) Measured valence band spectra and fits, generated a superposition of the MAPbI$_3$ and PbI$_2$ valence band spectra. Reprinted with the permission from Ref. 33. Copyright 2016 American Chemical Society.*

Fermi level position in the gap or the free carrier density in the material. Only after the I/Pb ratio drops below the threshold value of 2.5 does a $PbI_2$ phase precipitate with significant effect on the electronic properties of the material. The upshift of the Fermi level in the gap can be explained by two concomitant processes. While $MAPbI_3$ deposited on top of n-type $TiO_2$ surfaces is n-type as usually reported for HaP films deposited on $TiO_2$ (further discussion in section C1),[145] $E_F$ is generally found near mid-gap in the $PbI_2$ phase (BG = 2.3 eV). Thus, the steady formation of $PbI_2$ over time will lead to a continuous shift of the leading edge and change in the shape, of the valence band spectrum, which can be reproduced by a superposition of spectra from the initial $MAPbI_3$ valence band and from pure $PbI_2$, as depicted in figure 15c. The process is consistent with an analogous evolution of isosbestic points in the Pb and I core level spectra.[33]

In contrast to earlier reports,[161] no metallic lead ($Pb^0$) species were formed, even after 42 h of continuous X-ray exposure in vacuum. Instead, the process culminated in the complete transformation of the $MAPbI_3$ film into $PbI_2$ under the the X-ray spot (see Figure 16a). In this case, the tendency of the $MAPbI_3$ film to radiolyse with precipitation of $Pb^0$ clusters was inhibited by empirical adjustments of the film preparation methods.[33] In a different example (and for a different HaP sample composition), even *in vacuo* exposure to white light illumination of an $MAPbI_{3-x}Cl_x$ thin-film fabricated at low temperature resulted in a considerable increase in $Pb^0$ content with significant consequences for the optoelectronic properties of the material.[164] The concentration of $Pb^0$ species is largest at the surface, where Pb atoms could even form a metallic film giving rise to a clearly distinguishable Fermi edge from the Pb 6p electrons (see figure 16). The metallic species act as donor defects and thus effectively as trap states on the HaP surface. As a result, $E_F$ is pinned to the conduction band minimum in the case of high density of $Pb^0$ defects. Similar results on this process have been acquired before in an accelerated degradation study, in which XPS measurements were taken on a laser irradiated spot ($\lambda$ = 408 nm, 1000 $W/m^2$) on a $MAPbI_3$ thin-film.[165]

In conclusion, HaPs demonstrate what can be interpreted as remarkable defect tolerance, i.e. an invariance of key energetic parameters such as the position of $E_F$ in the gap, upon the creation of shallow defect pairs (methylamine and iodine vacancies). Nonetheless, metallic defects can pin the Fermi level, potentially act as recombination centers and ultimately affect the energy level alignment process at interfaces with adjacent functional layers. The formation of such defects can be provoked by irradiation and might be substantially accelerated in vacuum conditions. In the presence of oxygen, the zero-valent Pb will be oxidized and, apparently the resulting Pb-O has a surface passivating effect, the nature of which has not yet been elucidated. Further reading on the implications on measurement conditions can be found in literature reports.[159,160]

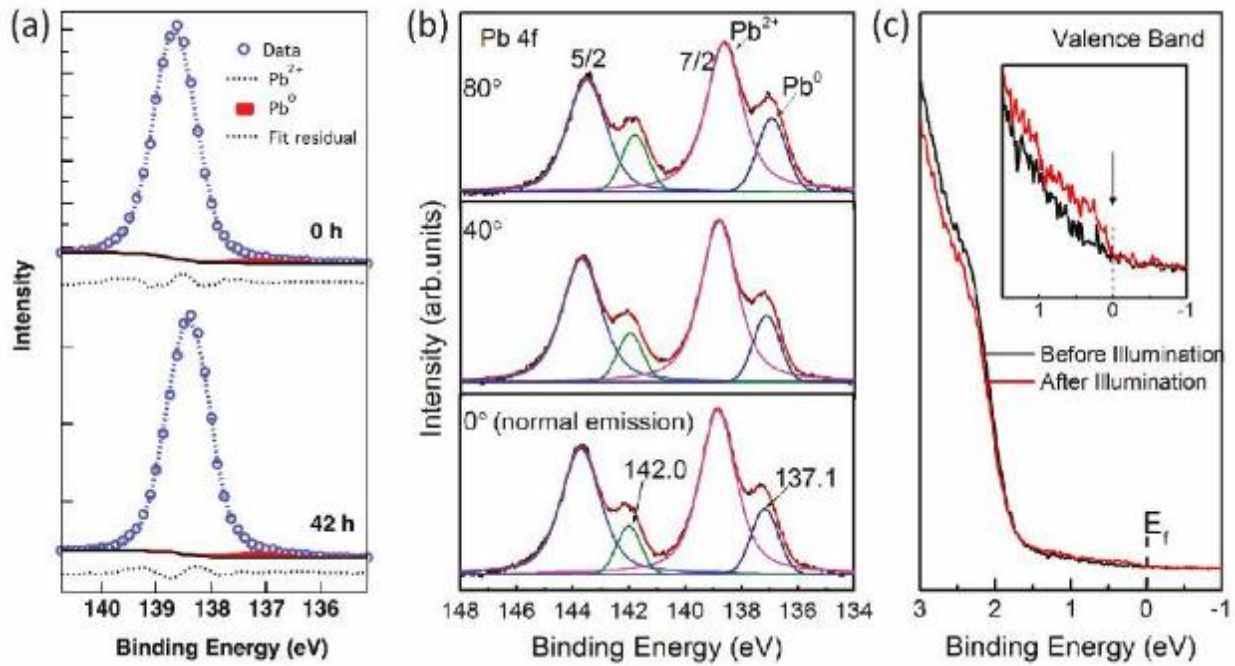

*Figure 16* – (a) XPS Pb $4f_{7/2}$ core level spectra of an MAPbI$_3$ thin-film; initially and after 42 h of X-ray exposure in vacuum, which effectively led to the degradation of the layer to PbI$_2$, with no significant formation of Pb$^0$. Reprint from ref 33. (b,c) PES spectra of MAPbI$_{3-x}$Cl$_x$ film after extended in vacuo illumination with white light bias. (b) XPS Pb 4f core level spectra showing a substantial amount of Pb$^0$ forming a metallic overlayer (c) UPS spectra indicating a distinctive DOS contribution at $E_F$. Reprinted wih the permission from Ref. 164. Copyright 2017 WILEY-VCH Verlag GmbH & Co. KGaA, Weinheim.

**B.3. HaP single crystals**

Further refinement of HaP surface characterization requires a closer look at the sample quality. In this regard, single crystalline specimens fulfill two important roles: first, they can provide high material purity; second, they offer structural order on a macroscopic scale, which allows to extract additional details for the electronic structure determination. Since structural order and defined orientation in real space translates into reciprocal space, PES experiments taken under varying emission angles (angle-resolved photoemission spectroscopy: ARPES) and excitation energies can be used to map out the band structure, and dispersion, *E-k,* relations, of the material. Vacuum-cleaved surfaces of HaP single crystals can serve as a prototypical system with minimal influence of extrinsic defects. Due to the ease of, and recent advances in HaP single crystal synthesis,[166,167] high quality samples have become available as testing platforms to assess intrinsic optoelectronic properties of HaPs. The growth methods, used today, can be grouped into four different categories: solution temperature-lowering, inverse temperature crystallization, anti-solvent vapor-assisted crystallization, and melt crystallization.[167] A potentially important difference between the first three, especially the second and third, and the last method is the ability to grow MAPI crystals below the tetragonal to cubic phase transition.

For many HaP systems, such as $MAPbI_3$ or $MAPbBr_3$, millimeter-sized specimens of high purity and crystalline quality can be grown within a matter of days.[168] This allows for a particularly fast route to perform surface characterization studies and has been exploited to study luminescence and electronic transport phenomena. For instance, $MAPbX_3$ single crystals exhibit very low trap state densities on the order of $10^9$ to $10^{10}$ cm$^{-3}$ and long charge carrier diffusion lengths on the order of 10's of micrometers.[169] In the case of a $MAPbI_3$ single crystal, Dong and coworkers claimed electron and hole drift/diffusion lengths on the order of 100's of micrometers,[170] but the interpretations of the data to arrive at such numbers has been criticized.[171]

However, the long carrier lifetimes compared to the already respectable ones in polycrystalline thin films suggest that surface recombination in single crystals can be effectively suppressed. With respect to surface characterization, we can hence assume the cleaved crystal as close to pristine, with a minimally perturbed surface, and gain insight into the surface termination that results in such. For the various surface planes, the electronic band dispersion is then determined by ARPES experiments. In the following section, we discuss this approach and procedure for MAPbI$_3$ single crystals and demonstrate, that measurements can be used to explore some of the hallmark materials properties in HaPs such as the giant Rashba splitting in MAPbBr$_3$.[149,172]

### B.3.1. Band dispersion by angle-resolved photoemission spectroscopy in MAPbI$_3$

The first hypothesis, i.e. to obtain a pristine surface by cleaving, is not necessarily easy to fulfill; Zu *et al.* took XPS and UPS measurements of a MAPbI$_3$ single crystal, grown by seed-induced nucleation and cleaved in dry N$_2$ prior to introduction in an ultra-high vacuum analysis chamber for PES.[164] In contrast to MAPbI$_3$ thin film, investigated in parallel, the crystal surface exhibited a significant density of Pb$^0$ species, that pinned $E_F$ to the conduction band minimum, prior to any illumination (apart from the X-ray and UV for the PES analyses) was used. A possible explanation for this finding is rooted in the abrupt surface termination resulting from the cleaving process, which can lead to an iodine-deficient surface. Another explanation takes into consideration the fact that, depending on the preparation methods and conditions, HaP crystals can exhibit very different levels of material purity and crystalline quality, even though, based on visual inspection and basic X-ray diffraction experiments, they are indistinguishable from pure single crystals. Compared to MAPbBr$_3$, the sample and surface preparation process is particularly challenging for MAPbI$_3$ and can lead to specimens with defects and solvent remnants. Thus, the evaluation of single crystal PES data must be done with caution, by keeping track of, and reporting potential variations in sample preparation methods and, hence, crystal quality.

Lee and coworkers produced a MAPbI$_3$ single crystal by solvent evaporation from a GBL solution with equivalent parts of PbI$_2$ and MAI.[172] The resultant crystal was mm-sized and its single crystal character was verified by synchrotron-based X-ray diffraction measurements. In the diffraction pattern the tetragonal phase of the crystal (held at 200 K) was identified by distinct diffraction spots in the (*h*0*l*) and (0*kl*) planes. Subsequently, the electronic structure and band dispersion was determined by ARPES measurements, also using synchrotron excitation. In Lee *et al.*'s measurements, a variation of the photon energy corresponds to a change in $k_y$, perpendicular to the sample surface (see Figure 17a) and is chosen such that the high symmetry planes in the Brillouin zone (BZ) can be identified. The periodicity with large *k* values in the $k_y$ scans suggest a cubic structure instead of the tetragonal low temperature phase of MAPbI$_3$, which could have persisted due to the fact that the crystal was grown at 100 °C, i.e. above the cubic-to-tetragonal phase transition temperature (∼ 56 °C at ambient pressure). Constant energy cuts in the ΓMX plane as depicted in figure 17c-f mark where the electronic bands intersect the high symmetry points of the cubic BZ. In the final step the dispersion relations are measured for a constant photon energy as plotted in figure 17g-i along the MΓM direction. As a first result, the ARPES measurements reveal little band dispersion in MAPbI$_3$. While the general position of the band agrees with the theoretical calculations, the measured bands appear broader and are dominated by a non-dispersive background. These findings have a significant impact on the assumed optical transition and transport in MAPbI$_3$ since the bandwidths experimentally determined by Lee *et al.* are smaller than commonly calculated, which implies a lower transfer integral between the atomic sites and point to higher effective masses than expected by them.[172] At this point, the reader is reminded of the low DOS at the band edges in MAPbI$_3$, which presumably originates from the R point in *k*-space.[150] This contribution to the DOS is already hardly observable in angle-integrated PES experiments,[101] and the ARPES experiment presented here would explicitly be "blind" to this direction of the band progression due to the chosen crystal orientation and measurement geometry. Finally, the influence of defect states and disorder in the surface layer cannot be excluded, which could be associated to beam damage, impurities in the MAPbI$_3$ crystal surface or a mixed tetragonal-cubic sample.

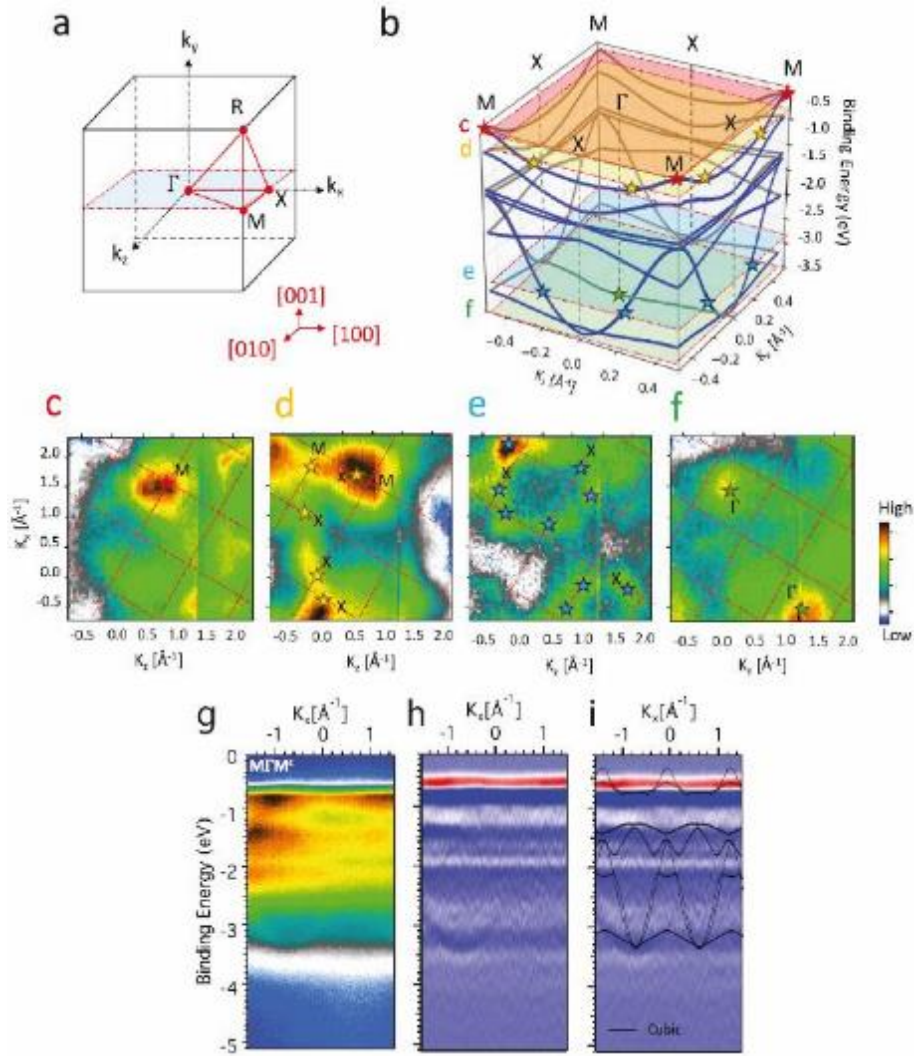

*Figure 17* – *(a,b) Cubic BZ and band structure assumed for the MAPbI$_3$ single crystal with constant energy cuts of the electronic structure in the ΓXM plane at binding energies corresponding to (c) 0 meV, (d) 300 meV, (e) 2440 meV and (f) 2840 meV with respect to the VBM. Stars in each cut indicate the intersection of the bands with the cut planes. (g)Measured band dispersion along the MΓM direction (hv = 142 eV), and (h) second derivative of the data.(i) Superposition of data from (h) with calculated band dispersion. Reprinted figure with permission from Ref. 172. Copyright 2017 by the American Physical Society.*

However, Yang *et al.* independently performed ARUPS measurements of a MAPbI$_3$ single crystal. On the same sample, LEED measurements confirm the cubic phase and a small signal of diffraction spots from a presumably incomplete tetragonal phase, which becomes brighter when the crystal is cooled to lower temperatures.[173] In contrast to the excitation energy in the study of Lee *et al.* (74 eV < $\hbar\omega$ <200 eV)[172] Yang et al. used a He I excitation source (21.22 eV), which yields higher photoionization cross-sections for the frontier molecular orbitals and revealed the expected pronounced dispersion relation. They reported effective masses of holes (m*$_h$ / m$_0$, with the free-electron mass m$_0$) of 0.35 ± 0.15 and 0.24 ± 0.10 along the Γ-X and Γ-M directions in the Brillouin zone, which corresponds well to the theoretical calculations (m*$_h$ / m$_0$ = 0.2 – 0.25, respectively).[173]

**B.3.2. MAPbBr$_3$ and mixed HaP single crystals**

We now turn to the investigations of other HaP single crystals than MAPbI$_3$ to demonstrate, that such studies not only reveal atomistic quantities required to describe charge carrier transport, but could also assist in evaluating fundamental physical mechanisms such as Rashba-splitting.

An initial attempt at visualizing the band dispersion in MAPbBr$_3$ was reported from a laboratory source-based UPS measurement. The data were acquired at sample-to-analyzer take-off angles θ between 0 and 28.5° to map $k_\parallel$ along the ΓX and ΓM directions in the BZ of the cubic lattice.[174] The valence band features were coarsely fitted with Gaussian functions and attributed to individual bands, for which the shifts of the centroid peak positions serve as a measured approximation of the band dispersions along the respective $k_\parallel$ values, with

$$hk_\parallel = [2m(\hbar\omega - \Phi - |E_B|)]^{1/2} \sin\theta,$$

where $\Phi$ is the work function, $\hbar\omega$ the excitation energy, $E_B$ the binding energy; here we assume a lifetime-broadened free electron-like final state dispersion.[175] From this measurement the lower limit of the effective mass of holes is derived as the inverse of the curvature at the valence band maximum which yields a mobility of $\mu_h > 33.90$ cm$_2$V$^{-1}$s$^{-1}$ for the MAPbBr$_3$ single crystal, compatible to values determined in optical and electrical measurements.[169,176]

Komesu *et al.* performed synchrotron-based ARPES measurements of MAPbBr$_3$ single crystals and compared the results to those from electronic structure calculations at the DFT level, to yield a refined exploration of the band dispersion.[177] ARPES measurements were performed of the (001) surface of a cm-sized MAPbBr$_3$ crystal, as depicted in Figure 18a, and compared to the calculated band structure along the Γ−M direction of a MAPbBr$_3$ surface Brillouin zone with either MABr or PbBr$_2$ surface termination. Beside a strong non-dispersive contribution at 3 eV below $E_F$, which is particularly intense at the Γ point, a dispersive band with a much lower intensity range, was seen up to 1.8 eV below $E_F$ at the M point. Komesu *et al.* attribute this band to a mixed bulk Br 4p and Pb 6s orbital in line with independent calculations, which also show minor contributions of the Pb 6p orbitals to the valence band.[101] Two main observations are made for this set of experiments: First, the MABr surface termination leads to a better fit of the experimental data than the PbBr$_2$ termination. This is in line with earlier reports of a preferential MABr (or more generally AX)-terminated surface for MAPbX$_3$ compounds. Second, with the valence band located at 1.8 eV below $E_F$, the crystal appears n-doped, in good agreement with combined PES/IPES measurements of MAPbBr$_3$ thin-films on top of TiO$_2$ substrates.[145] Here, Komesu *et al.* suggest that the n-type doping behavior originates from a self-doping mechanism of the halide-deficient surface (in the ionic point defect model V$_I$ is a donor defect) and to metallic lead (Pb$^0$) species, similar to the case of the cleaved MAPbI$_3$ crystal. We note that Pb$^0$ can charge balance charged V$_I$ defects. Notably, the inclusion of SOC in the DFT calculations led to changes for the band dispersion and positions, on the order of 10s of meV.[177]

In a separate study, Niesner et al. demonstrated that ARPES on a MAPbBr$_3$ single crystal can aid in elucidating strong prevalent relativistic effects in cubic HaP compounds.[149]

First, they showed that the phase transition from the orthorhombic (low temperature) to the cubic phase (room temperature) directly impacts the surface electronic structure and is clearly visible in the symmetry of the isoenergetic cuts through the BZ (see Figure 18). The VBM is fitted by a parabolic model to account for the low DOS.[101,143]

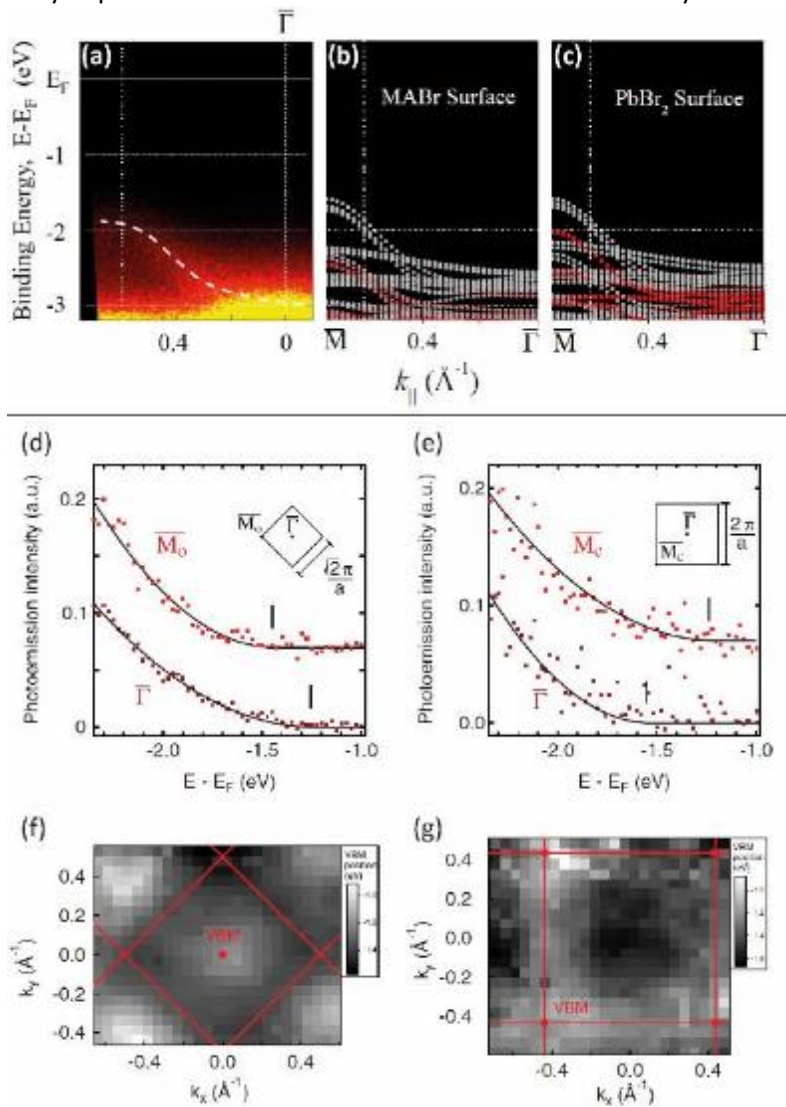

*Figure 18* – *Electronic band structure of the (001) surface of MAPbBr$_3$. (a) ARPES data acquired at a photon energy of 34 eV. (b,c) Calculated band structure with MABr (b) and PbBr$_2$ surface termination (c) along the Γ–M direction. Surface states are indicated in red and bulk states in light grey in (b) and (c). Reprinted with the permission from Ref. 177. Copyright 2016 American Chemical Society. (d-g) ARPES data from the low-temperature temperature orthorhombic (d,f) and the room-temperature cubic (e,g) phase. (d,e) VBM region of the ARPES spectra at the centers (Γ) and the corners M$_o$ and M$_c$ of the orthorhombic and the cubic surface BZ (shown in inset), respectively.(f,g) Constant energy cut through the VB as a function of parallel momentum. Reprinted figure with permission from Ref. 149. Copyright 2016 by the American Physical Society.*

Second, the valence band scans revealed an absence of energy levels at mid-gap, suggesting low densities of surface trap and/or recombination states, which is consistent with the determination of (shallow) bulk defects in MAPbBr$_3$ by Laplace deep level transient spectroscopy (DLTS).[178] Furthermore, the symmetry observed from the isoenergetic cuts in both phases, room temperature cubic and low temperature orthorhombic, suggests that no long-range surface reconstruction occurs. This local surface structure has been confirmed in a dedicated STM experiment of an MAPbBr$_3$ single crystal, where Ohmann *et al.* reported a non-centrosymmetric zigzag (ferroelectric) and centrosymmetric dimer (antiferroelectric) arrangement of the Br anions in the top-most MABr-terminated layer.[179] The non-centrosymmetric pattern exhibits a smaller band gap and, hence, Niesner *et al.* suggest that this scenario matches best their measured band onset and assume it to be the predominant surface termination.[149]

In the next step, Niesner *et al.* used their laser-based ARPES experiment with circular polarized light ($\hbar\omega$ = 6.2 eV) to reveal a pronounced dichroism map, i.e. the difference ($\sigma^+$ - $\sigma^-$) photoemission intensity from the VB under excitation by right- and left-handed circularly polarized light, respectively. The split bands resemble parabolas off-set in *k*-space by a Rashba parameter $\alpha$, and can be expressed in a quasi-free electron picture with the energy dispersion relation

$$E^\pm = \frac{\hbar k^2}{2m_h^*} \pm \alpha|k|.$$

While the limited experimental resolution did not allow for the extraction of the effective mass m*$_h$, $\alpha$ was determined as 7 ± 1 eVÅ for the orthorhombic and 11 ± 4 eVÅ for the cubic phases, respectively. This is at the higher end of the range of predicted SOC effects in HaPs,[124] and corresponds to an offset of the VBM by k$_\parallel$ = 0.043 Å$^{-1}$ giving rise to a substructure of the top of the valence band and a local minimum at the high symmetry points 0.16 and 0.24 eV below the VBM for the orthorhombic and cubic phases, respectively. The exact origin of this surprisingly large effect is hard to pinpoint; in contrast, a separate study gives a detailed theoretical inspection of the effect of the 2 x 2 surface reconstruction (zigzag and dimer) in MAPbBr$_3$ and predicts only a slight Rashba effect one order of magnitude smaller than the values measured in Niesner's experiment.[102] Hence, point and step-edge defects or mesoscale polar reconstruction rather than the dimer surface reconstruction could be the root cause of the observed Rashba splitting. Irrespective of these uncertainties of the SOC-induced phenomena, the example demonstrates that the HaP surface can denote a complex energetic landscape. However, further corroborating measurements are needed to arrive at a comprehensive picture and quantify the impact of the measured effect on a device-relevant situation, i.e., at an interface instead of a surface under ultra-high vacuum conditions.

We conclude the section on HaP single crystals by noting that the measurement of band dispersion yields an important data point for the determination of macroscopic transport properties, but is only a first step to determine charge carrier dynamics at the interface. It is in the nature of the experiments that the measurement conditions are quite unique and strictly picture only the exposed surface in UHV, while the formation of an interface – as it is the case in a device – would disrupt those conditions. In addition, the effect of alloying and phase segregation in multi-cation or mixed halide perovskites is very relevant to HaP-based semiconductor devices. Yet, aside from the examples presented for MAPbI$_3$ and MAPbBr$_3$ single crystals, to date no detailed experimental and theoretical studies exist for HaPs single crystals with mixed A, B and/or X species in the ABX$_3$ compound. However, recent reports demonstrate a successful synthesis of mixed HaP single crystals,[167] which bring the exploration of band dispersion for additional systems within reach. Particularly, the readily available fabrication route to mixed halide HaP single crystals has the potential to lay bare their dominant surface termination and preferential surface cell composition.[180]

**C. Interfaces to adjacent functional layers**

After having discussed the electronic properties of the HaP surface and the interaction between the exposed surface and isolated adsorbates, e.g. prototypical organic molecules and oxygen species in section B.1.2., we turn to the fully assembled interfaces between HaP semiconductors and functional layers. As in the other sections of this review, the focus remains on energy level alignment and electronic properties at the interface. We recognize that the range of HaP compositions is large, in particular when considering the compositional variants that originate with different sample preparation methods and lead to different surface terminations. Still, mostly the family of methylammonium lead halide ($MAPbX_3$) compounds has been explored in detail by surface science methods, and only few data points exist for the wide range of HaPs that are not based on methylammonium and lead. The compositional space of HaPs is then considerably multiplied for interfaces with transport/passivation layer. We give here a broad survey of the various potential HaP/layer combinations, but pick the most relevant ones to derive what we consider to be the most prevalent interaction phenomena. The main goal of this exercise is to present the many electronic level alignment processes and connect these to the role of the interface,[16,54,157,181] as it is becoming increasingly clear that the interfaces dominate PV device performance of HaP-based cells.[55–57]

Interfaces of HaPs can generally be categorized into two subdivisions, which differ from each other by their formation mechanisms as well as accessibility in surface science experiments:

- The buried interface between substrate and HaP layer that is deposited on top of the substrate (often called "bottom" interface)
- The interface between HaP layer and transport/passivation layer that is deposited on top of the HaP (often called "top" interface)

The formation processes of these two types of interfaces are fundamentally different and hence the resulting interfacial electronic alignment and prevalent defect structure are expected to be different as well. In the first case, i.e. the bottom substrate/HaP interface, a solid substrate sets the foundation for the HaP film formation. Hence, the substrate and the conditions under which it is held (e.g. temperature, -OH surface termination) template the growth, morphology and structure of the HaP layer,[182,183] which can have wide-ranging implications for device operation.[184] In terms of film morphology, the reader is referred to a review by Saparov and Mitzi for details on the adoption of various growth modes of HaP thin films and the correlation with the performance of photovoltaic and other HaP-based semiconductor devices.[185] Importantly, the substrate surface interacts with the precursor solution and determines the chemical environment, thermodynamic driving force and formation kinetics for the HaP precursor phase.[186,187] In the final product, the solid layer on the substrate, the structural properties and energetics of the film at the immediate interface could be significantly different from those of the bulk or the exposed surface.

In the case of the top interface, i.e. the HaP layer with a coating deposited on top, the solid HaP layer is already formed and can be assumed to be in (at least thermal) equilibrium, which to a first approximation allows to discard any effects due to a precursor phase. However, depending on the deposition conditions of the coating, the HaP film can undergo structural and chemical, and resulting energetic rearrangement, e.g. due to non-orthogonal solvents or reactive overlayer materials. Again, in this scenario the electronic properties of the HaP film at the interface would differ from those of the bulk or of the originally exposed HaP surface. Thus, while understanding of the exposed surface is an important starting point to rationalize potential chemical reactions with the coating, for instance through excess MAI or $Pb^0$ as effective reactants,[188–190] its energy levels may well differ from those, relevant for (top) interface energy level alignment.

We note that the resulting HaP film-interface systems are usually not in thermodynamic equilibrium. The non-negligible mobility of ions in the HaP film can lead to structural and compositional changes, which can be particularly pronounced at interfaces and strongly impact the energy level alignment at, and electronic transport properties across these boundary regions, especially if the material that contacts the HaP can accept halide ions (e.g., polymers). For example, many reports discuss interface halide depletion or enrichment as a result of light or electrical bias-induced ion migration in HaP.[71,191,192] Even though these transient effects are difficult to capture and quantify in many of the surface science experiments presented here, they clearly affect the operation, and at times the degradation behavior, of HaP-based devices. The exact role of ion migration on the stability and I-V hysteresis in PSCs, and especially in top-performing ones, remains a challenging academic question and technological conundrum.[70,73]

**C.1. Substrate typology and interactions with HaP thin-films**

*Conventional PSC device architecture:*

A straightforward way for the determination of the most relevant substrate typologies is to look back at the application that led to the recent breakthrough and surge in HaP research: the perovskite solar cell. At first, PSCs were thought of as dye-sensitized solar cells (DSSC) with the halide perovskite as the dye, first in an electrolyte-based architecture[6] and then in the more stable solid-state DSSC.[8,193] In both cases the substrate configuration comprised a transparent conductive oxide (TCO)-covered glass slide, covered with a mesoporous titanium oxide film acting as an electron transport layer. The HaP/substrate interface was therefore between the n-doped $TiO_2$ and the HaP. Interestingly, the mesoporous $TiO_2$ films, which were infiltrated with the perovskite light absorber, could be substituted by insulating mesoporous oxides (e.g. $Al_2O_3$) or be omitted altogether to form a planar heterojunction without (much) decreasing the PV device performance.[8,194] This finding demonstrated that the HaP film in the device configuration would facilitate ambipolar transport, unlike the commonly applied dyes in DSSCs. However, in all these scenarios, the bottom electron transport layer (ETL) for charge extraction in contact with the light absorber film was still *n*-doped $TiO_2$ (also in case of the mesoporous $Al_2O_3$). Eventually, the device would be completed by the deposition of a hole transport layer (HTL) – usually an organic semiconductor – on top of the HaP film, followed by the subsequent deposition of a metal top anode, which we will discuss in a later section. Much effort has been spent to substitute the bottom $TiO_2$ oxide in this configuration using other oxides (ZnO, ZnO:Al, $SnO_2$), other inorganic compound semiconductors like CdS or organic ETLs.[54,195,196]

The requirements for the ETL layer are stringent and can be summarized as follows:

(i) compactness without pinholes, to prevent leakage currents between HaP layer and the TCO cathode;

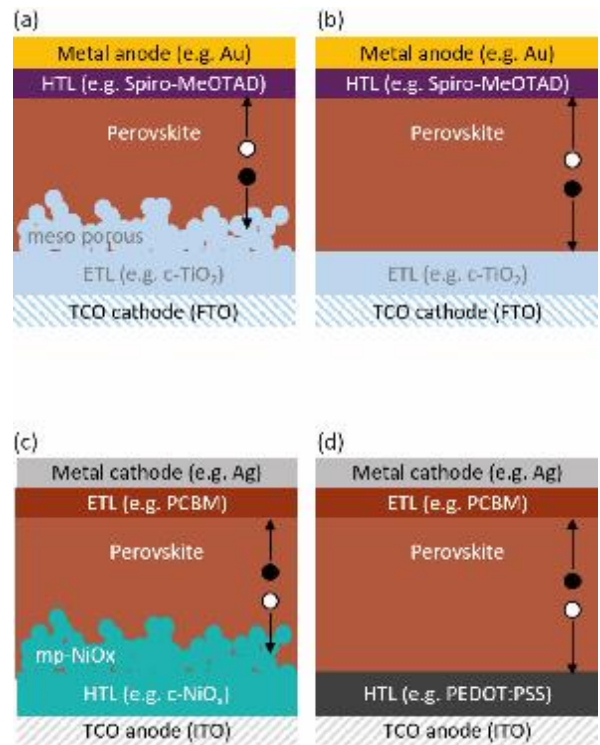

*Figure 19* – *Substrate choices for various HaP -based PSC configurations. (a) Conventional cell stack with mesoporous oxide and ETL as substrate. (b) Planar configuration with oxide ETL. (c) "Inverted" device architecture with mesoporous p-type oxide HTL substrate. (d) Inverted planar structure with organic HTL substrate. Graphic, courtesy of C. Dindault.*

(ii) good optical transmittance, i.e., wide band gap and minimal reflectivity over the solar spectrum range, to minimize optical losses of incoming light (also needed for light outcoupling in LED devices),

(iii) high electrical conductivity to minimize electrical losses for charge carrier transport in the layer,

(iv) energy level alignment, particularly vacuum level and conduction band minimum for minimal energy loss and efficient electronic coupling to, and hence charge transfer to the adjacent HaP film.

This latter point, i.e. the energy level alignment between ETL and HaP, is difficult to assess as the suitability of the ETL/HaP pairing does not systematically derive from the equilibrium positions of the respective energy levels in the separate phases. While the absence of an ETL layer between HaP film and TCO cathode is detrimental to the solar cell device characteristics due to the aforementioned hole leakage currents and recombination losses, not every nominally charge-selective contact can be used for efficient electron extraction and hence improved device functionality. Comparing $TiO_2$ and ZnO, both ETL materials generally enable reasonably high initial power conversion efficiencies, whereas CdS, which exhibits similar charge selectivity promoting electron extraction, leads to a significantly reduced device efficiency.[195] The reason is its lower band gap (than $TiO_2$ and ZnO) resulting in significant light absorption, which does not contribute to the generated photocurrent, as becomes apparent in incident photon-to-electron conversion efficiency (IPCE) measurements. In addition to the effects of energy level alignment, chemical reactions upon formation of the interface or under device operation play a major role in the integrity of the interface and their study requires dedicated analyses. With a ZnO ETL layer, the stability of the top HaP film is significantly reduced compared to that of an HaP film deposited on $TiO_2$. This instability is generally ascribed to the higher acidity of the ZnO surface, but the stability of the ZnO/HaP system can be improved by annealing of the ZnO film in air prior to HaP deposition or even more rigorously by adding Al as a dopant in ZnO (ZnO:Al, AZO, also used in CIGS PSCs).[197]

Suppressing chemical reactions at the oxide/HaP interface has been attempted and enabled by either a hybrid typology, where the bottom oxide layer is capped with an organic self-assembled monolayer (e.g. $PC_{60}BM$) on top of the $TiO_2$ ETL,[198] or by using a more inert organic ETL altogether.[199] This latter approach features the additional advantage that an organic ETL on the bottom interface would more easily allow for integrating HaP layers into layer architectures for flexible devices.

Moreover, the choice of the ETL has a direct effect on charge transfer from or to the HaP layer and consequently charge transport in the adjacent HaP film itself. Often the current-voltage measurements of HaP-based solar cells exhibit a transient behavior (or hysteresis effect).[200] This effect has been attributed to various mechanisms, such as ion migration in the HaP film, (re-)polarization of ferroelectric domains in the HaP film, the build-up of space charge regions at the interfaces with charge transport layers (or trap-filling or emptying at these interfaces), or a combination of these cases. Due to the dominant time scale of this effect (in the range of few seconds) as well as its sensitivity to the contact materials, the most prevalent scenario is the formation of space charges at interfaces occurring with the photogeneration of charge carriers in the HaP bulk, both through purely capacitive effects and ion migration.[68,201,202] While neither the thick PCBM-only ETL nor most $TiO_2$ ETLs produce hysteresis-free PSCs, the thin PCBM modification of a compact $TiO_2$ ETL succeeds in making the interface less prone to charge accumulation and hence polarization.[203] Therefore, the correct choice of the ETL, its chemical inertness, and the respective energetic alignment to the HaP are prerequisites to minimize transient behavior and capacitive effects in HaP-based semiconductor devices.[204]

*Inverted PSC device architecture:*

An alternative configuration to the conventional *n-i-p* layer stack for the PSC layout is derived from the most common cell *p-n* architecture in conventional organic photovoltaic devices.[205] In an "inverted" *p-i-n* structure, the perovskite film is deposited on top of a p-type hole conductor, which can be a planar oxide layer such as NiO$_x$ or V$_2$O$_5$, or a hole-conducting layer of small molecules or polymers (e.g. PEDOT:PSS). An organic electron transport layer based on fullerene compounds (C$_{60}$, PCBM) or bathocuproine (BCP) deposited on top of the perovskite absorber followed by a metal cathode generally completes the device stack. Initial inverted PSC devices based on a PEDOT:PSS bottom electrode yielded respectable power conversion efficiencies on the order of 6%, while the analogous devices comprising planar metal oxide hole conductors as a bottom HTL fell short of achieving high PCEs despite all interface systems resulting in similar PL quenching efficiencies.[205]

A related approach is to combine the layer sequence (*p-i-n*) with film fabrication routines that are applied in the DSSC production scheme, one of the parent technologies for HaP-based PSCs. Mesoporous p-type metal oxide layer (e.g. mesoporous NiO$_x$) then emulate the scaffolded morphology of the DSSC while maintaining the direction of hole extraction from the HaP absorber layer into a bottom HTL.[206,207] The same requirements established for the ETL substrates in the conventional PSC apply to the development of improved or modified bottom hole transport layers: (i) improved structural integrity, (ii) optical properties that assist in- and out-coupling of photons in the device, (iii) chemical inertness and electronic properties that enable trap passivation and efficient charge transfer. In particular, the electronic properties of these bottom HTLs have been improved, based on surface treatments and variations in the doping level, which eventually led to inverted PSCs with PCEs around 20%.[204,208–210]

*Impact of substrate on charge carrier profile in the layer stack:*

In this context the most pressing question remaining for all the various substrate typologies is how the substrate/HaP interface impacts the profile of mobile charge carriers (electrons and ions) in equilibrium condition, but more importantly under device operation with either electrical bias in the dark, or under illumination. Alternatively, the issue boils down to understanding to what degree the response of the carrier distribution in the HaP film is modulated by its interfaces with adjacent layers (here: the substrate), with an external perturbation such as photons or an electric field. A sensitive way to visualize and measure the charge carrier distribution in the layer stack is given by cross-sectional Kelvin probe force microscopy (KPFM) on a cross-section of the layer stack obtained by cleaving, ion milling, and/or polishing, bearing in mind the damage (changes from the pristine state) that forming the cross-section can cause. The cross section is then analyzed via scanning force microscopy. Aside from AFM-derived topographical images, KPFM provides a spatially-resolved measure of the contact potential difference (CPD), that is the difference between the work functions of the probe tip, $\Phi_{tip}(x)$, and the sample surface with electrostatic potential $V_E(x)$. The electrical field $E(x)$ across the layer stack is then given by the spatial derivative of the CPD by

$$-\frac{d}{dx}\text{CPD}(x) = -\frac{d}{dx}\left(V_E(x) - \frac{1}{e}\Phi_{tip}(x)\right) = -\frac{d}{dx}V_E(x) = E(x).$$

Under the assumptions that, (a) the field in the bulk does not deviate strongly from the field at the surface of the exposed cross section and (b) sample preparation leads to negligible changes of the surface (where cleaving is the method with the highest probability for this) we can extract the progression of the electron electrochemical potential (=WF) and hence charge distribution, $\rho(x)$, of resident and photogenerated charges in the device, according to Poisson's equation

$$\rho(x) = -\varepsilon_r\varepsilon_0\frac{d}{dx}E(x) = -\varepsilon_r\varepsilon_0\frac{d^2}{d^2x}V_E(x),$$

where $\varepsilon_0$ is the vacuum permittivity and $\varepsilon_r$ the local dielectric constant of the probed material. While the latter 's value is not agreed on, this will not affect the spatial variation, but only the absolute value of $\rho$.

With this type of analysis, the cross-section of a standard glass/FTO/c-TiO$_2$/mp-TiO$_2$/MAPbI$_3$/Spiro-MeOTAD/Au (*n-i-p*) device layer stack (figure 20a) reveals that, under illumination, the electrical field distribution points to unbalanced charge-carrier extraction under short-circuit conditions. In the case depicted in Figure 20, one can distinguish between two regions in the active HaP (MAPbI$_3$) absorber layer: a region of HaP-infiltrated mp-TiO$_2$ and a bulk HaP capping layer on top. However, the main issue arises from the HTL rather than the substrate layer: during illumination holes accumulate in the layer and lead to a potential barrier for charge carrier extraction.[211] We note that KPFM measurements can be ambiguous when it comes to defining whether a junction is *p-n* or *p-i-n*, which depends on the density of dopants and free carriers in the layers. For instance the results from Jiang *et al.* suggest a scenario that slightly deviates from an immediate *p-n* junction at the TiO$_2$/MAPbI$_3$ interface.[212] However, Bergmann *et al.*'s finding agrees with earlier electron beam-induced current (EBIC) measurements, which point to a higher electron extraction efficiency through the ETL than hole extraction through the HTL (Figure 20b).[213] EBIC measurements also enable a more direct probe of the extraction of carriers injected locally into a specific spot at the layer cross section. Notably, Edri's EBIC measurements on device structures that feature mesoporous Al$_2$O$_3$ films at the ETL side demonstrate that the device operation is feasible even with an insulating scaffold. Hence, a PSC, based on this HaP absorber (MAPbI$_3$) operates in *n-i-p* mode.[213]

### C.1.1. Substrate/HaP Interface Chemistry

*The HaP/substrate interface under bias and illumination: Transient and reversible electrochemistry*

The barrier formation at the interface between substrate and HaP film, proposed above, implies the movement (and trapping) of charge carriers – electronic and ionic ones – at the interface, which then results in an additional measurable component to the capacitance. This behavior can be observed to some extent in a broad range of HaP compounds (e.g. MAPbI$_3$, FASnI$_3$, etc.) and surfaces in the low frequency modulation of the capacitance under light bias, where the effect is in general more pronounced for the oxide/HaP interface compared to an organic semiconductor/HaP interface.[214] These observations suggest that at the least the interface exhibits different degrees of trapping charge carriers. In a more quantitative picture, the order of magnitude of the capacitance cannot be explained any longer with only the accumulation of photogenerated electronic charge carriers (electrons and holes). The theoretical upper bound for a purely electronic capacitance without ion movement and chemical reaction at a planar interface is limited to around 0.05 mF/cm,[215] which is almost two orders of magnitude below the capacitance (10 mF/cm) measured for MAPbI$_3$-based PSCs with either mesoporous or planar TiO$_2$ ETL substrates.[214] A major hypothesis to describe the large capacitive effect in HaP-based devices is stoichiometric polarization, which is defined as ion accumulation and the resulting space charge created at the respective carrier selective electrodes, leaving the overall stoichiometry and defect density of the material unchanged.[216] While this effect would be sufficient to describe the magnitude of hysteresis in a PSC, it would be independent on the type of contact. This stands in contrast to the varying degree of hysteresis as a function of the interface (formation) with the charge transport layer or electrode. Hence, in addition to the stochiometric polarization, Faradaic processes similar to the chemistry occurring at electrode interfaces in electrochemical cells, can contribute to the discrepancy between expected and observed capacitances in the case of planar oxide/HaP interfaces.[216]

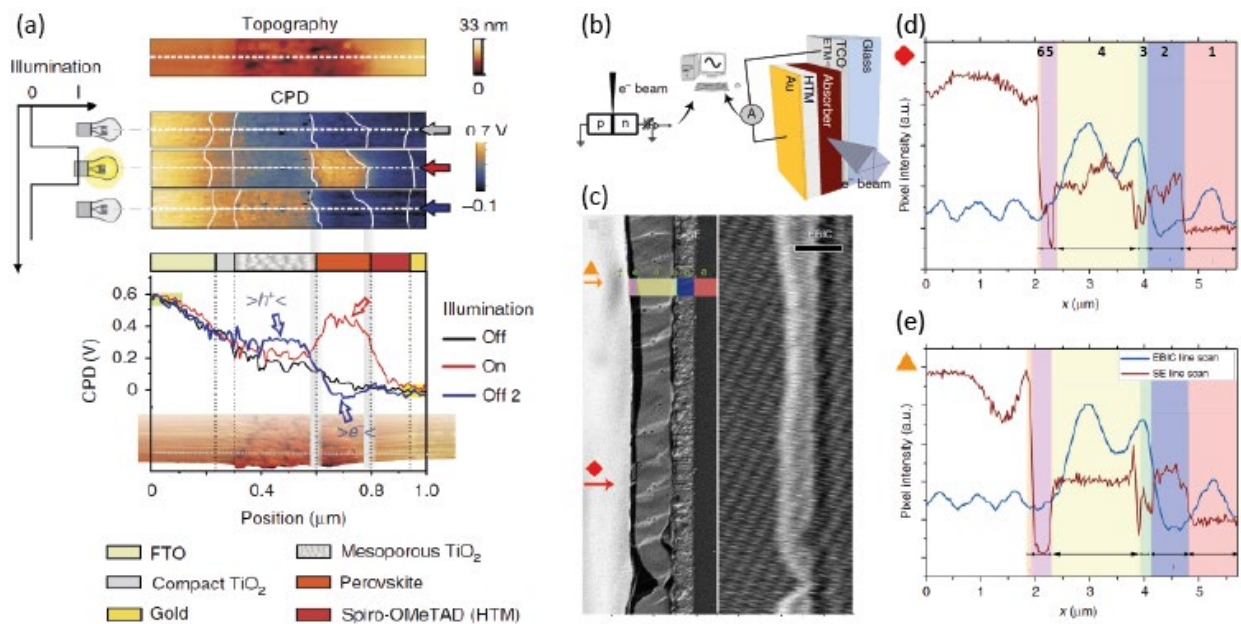

*Figure 20* – *(a) KPFM images of topography and CPD before, with, and after illumination of a PSC stack cross-section in short circuit conditions. The increase in potential in the perovskite capping layer upon illumination corresponds to hole accumulation. Post light exposure, the unbalanced charge extraction leads to holes trapped in the mesoporous layer (>h$^+$<) and electrons in the HaP capping layer (>e$^-$<). Reprinted with permission from Ref. 211. Copyright 2014 by Springer Nature Publishing AG. (b) EBIC measurement configuration, (c) secondary electron (SE), and electron beam current (EBC) image of a PSC stack cross-section. (d-e) SE and EBC line profiles for lines marked in (c) with numbers designating the layers: 1, glass; 2, FTO; 3, TiO$_2$; 4, MAPbI$_{3-x}$Cl$_x$; 5, HTM; 6, Au. Reprinted with permission from Ref. 213. Copyright 2016 by Springer Nature Publishing AG.*

One such transient and reversible chemical process occurs readily for the prototypical interface between MAPbI$_3$ and TiO$_2$, and can be inferred to from the reaction of iodine gas with the isolated surfaces (figure 21a):

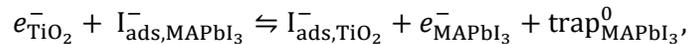

$$e^-_{TiO_2} + I^-_{ads,MAPbI_3} \leftrightharpoons I^-_{ads,TiO_2} + e^-_{MAPbI_3} + trap^0_{MAPbI_3},$$

where an iodine atom from MAPbI$_3$ can move across the interface, leave a surface trap on the MAPbI$_3$ side, adsorbs onto the TiO$_2$ layer, and depletes the oxide of an electron.[216] The corresponding reaction kinetics are subject to an extended set of parameters. While the structure at the interface defines the reaction front, the availability of reactants along with their replenishment, i.e. through field-driven ion migration from the bulk, play a major role in the quantification of this effect. Consequently, the timescale for the built-up of this (partially) reversible chemical capacitance can span tens of seconds and is hence compatible with reported hysteresis parameters.

The interfacial chemistry described above is not limited to the MAPbI$_3$/TiO$_2$ system, but manifest in similar reaction pathways at other interface to the HaP film. HaP compounds are prone to further dissociation under external stresses (light exposure, temperature),[217] and one needs to consider that the degradation products of the HaP layer can serve as reaction partners. In the case of MAPbI$_3$, exposure to light and elevated temperatures can lead to the formation of PbI$_2$ and further NH$_3$, MAI, HI, MA, I$_2$ and Pb$^0$ (figure 21b). In turn, this leads to compositional variations of the HaP film, which induce changes in the energetic landscape of the interface. The energy offset between the conduction band minimum of the TiO$_2$ layer and that of MAPbI$_3$ is calculated to be smaller for the PbI$_2$-terminated surface than for the MAI-terminated one (see section B). An MA deficiency at the interface would therefore lead to an energetic landscape that is beneficial for charge transfer but would also increase quenching rates at the interface.[218]

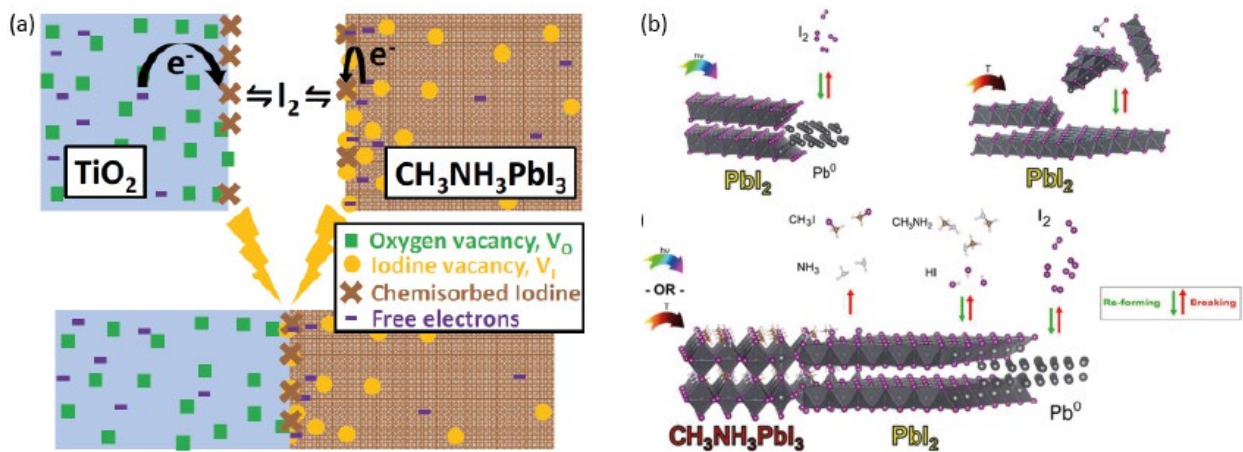

*Figure 21* – *(a) Reversible surface chemistry for TiO$_2$ and MAPbI$_3$ surfaces leading to solid state interfacial chemistry that dominates transient optoelectronic behavior. Reprinted with the permission from Ref. 2016. Copyright 2017 American Chemical Society. (b) Upon light exposure and with elevated temperatures MAPbI$_3$ and PbI$_2$ undergo various degradation mechanisms resulting in potential reaction partners for interfacial chemistry. Reproduced from Ref. 217 with permission from The Royal Society of Chemistry.*

*Chemistry-dominated HaP layer growth:*

In addition to the transient electrochemical processes taking place at an existing interface, reactions can change the formation process of the growing interface – and eventually the bulk of the film – during the formation of HaP compounds on top of reactive substrates. In a reverse example, a $PbI_2$ film deposited on a $TiO_2$ scaffold can be converted to $MAPbI_3$ by post-deposition exposure to an MAI-containing solution (e.g. dip-coating), in which case the chemical activation of the reaction front is strongly modulated by additional illumination. Ummadisingu *et al.* performed electrochemical impedance spectroscopy of the $PbI_2/TiO_2$ system and suggest that their data show the presence of surface traps in the $PbI_2$ film.[219] They propose these traps to be populated with photo-generated holes under illumination, which increases the amount of surface charges. Iodide anions in the MAI solution migrate to the $MAI/PbI_2$ interface to compensate the positive charges, which Ummadisingu *et al.* suggest to happen not only in this two-step deposition process but could also occur in a one-step HaP crystallization process.[219]

Generally, these formation processes are difficult to monitor, but tracking the fate of ultra-thin evaporated HaP films can be an important first step to deconstruct the interface and infer on the nature of the interfacial chemistry. Hence, we turn to the deposition via an evaporation process to closer examine the interface formation. Condensation and crystallization of high-quality HaP films from evaporation are primarily governed by the precursor materials in the sources, deposition rates and partial pressures during deposition.[220] However, these processes are also strongly influenced by the substrate used for the layer growth. In the early stages of the HaP film formation, the evaporated precursors interact with the surface. The complexity of this process is substantially increased as the precursors can break up into volatile compounds that can induce further reaction schemes. As a result, different substrates can catalyze the formation of intermediate compounds that differ significantly from the stoichiometric ratio of the desired HaP compound. The effect is very clear for the evaporation of MAI and $PbI_2$ precursors to grow $MAPbI_3$ on top of a variety of oxide and organic substrates. XPS measurements performed at various stages of the growth provide a chemical analysis of intermediate phases and confirm that the formation of MAPbI$_3$ is preceded by the growth of non-stoichiometric compounds.[187] The spectra in figure 22 give evidence for surfactant species that are specific to the respective HaP/substrate combination: formation of CH$_3$-O, In-I and NH$_3$-O bonds on indium-doped tin oxide (ITO); PbSO$_4$ on (sulfur-containing) PEDOT:PSS; O-CH$_2$-O and CH$_3$-O on MoO$_3$; and Pb-O on PEIE-covered ITO. In general, reactivity and break-down of the precursor phase into non-HaP components are more pronounced on top of oxide surfaces compared to organic semiconductor surfaces. In the case of evaporated MAPbI$_3$ on MoO$_3$ and ITO, the measured thickness of the film strongly deviates from the nominal evaporated thickness (below a nominal thickness of 10 nm) due to the formation and re-evaporation of volatile compounds. The effect is particularly clear with the observed deficiency in deposited lead, presumably due to the formation of volatile tetramethyl lead. In contrast to the decomposition of the precursor phases on top of the oxide surfaces, growth of stoichiometric HaP film sets in earlier on top of organic substrates after deposition of a nominal 3 nm HaP film.[187]

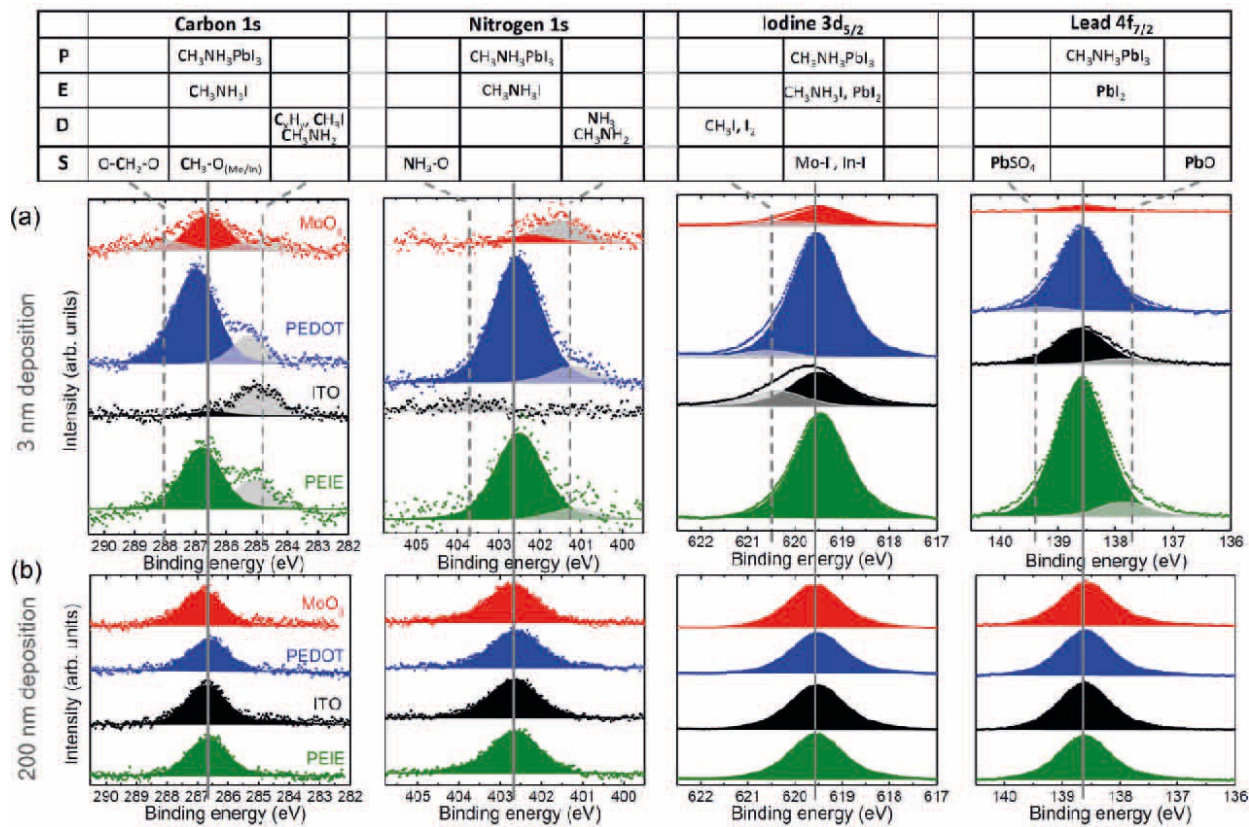

*Figure 22* – *XPS core level spectra of nominally (a) 3 nm and (b) 200 nm thick MAPbI$_3$ films from co-evaporation of MAI and PbI$_2$ precursor materials on top of various substrates. The table on top lists possible components of interfacial chemical reactions, divided into products (P), educts (E), decomposition products (D), and surfactants (S). Binding energies of the peaks were shifted in some cases to an overlay of the Pb 4f$_{7/2}$ level of the Pb$^{2+}$ component to enhance comparability. Reprinted with permission from Ref. 187. Copyright 2017 by Springer Nature Publishing AG.*

It is noteworthy that the decomposition can be both simply catalyzed by the substrate or resulting in a strong reaction of the substrate components themselves. Remarkably, even though the formation of stoichiometric HaP starts at an earlier stage in case of growth on $MoO_3$ compared to ITO, the $MoO_3$ substrate exhibits a more pronounced reaction by decomposition into sub-stoichiometric $MoO_{3-x}$ species (see figure 23). In general, however, the trend in compositional variations of the HaP film stoichiometric is confined to the interfacial region to the substrate, and the surface of a 200 nm thick evaporated $MAPbI_3$ film shows the expected elemental ratios of the top-most layers far away from the interface. Nonetheless, the film formation and surface morphology can be severely affected by interface chemistry, as shown in the SEM micrographs of 200 nm thick $MAPbI_3$ films (figure 23).[187]

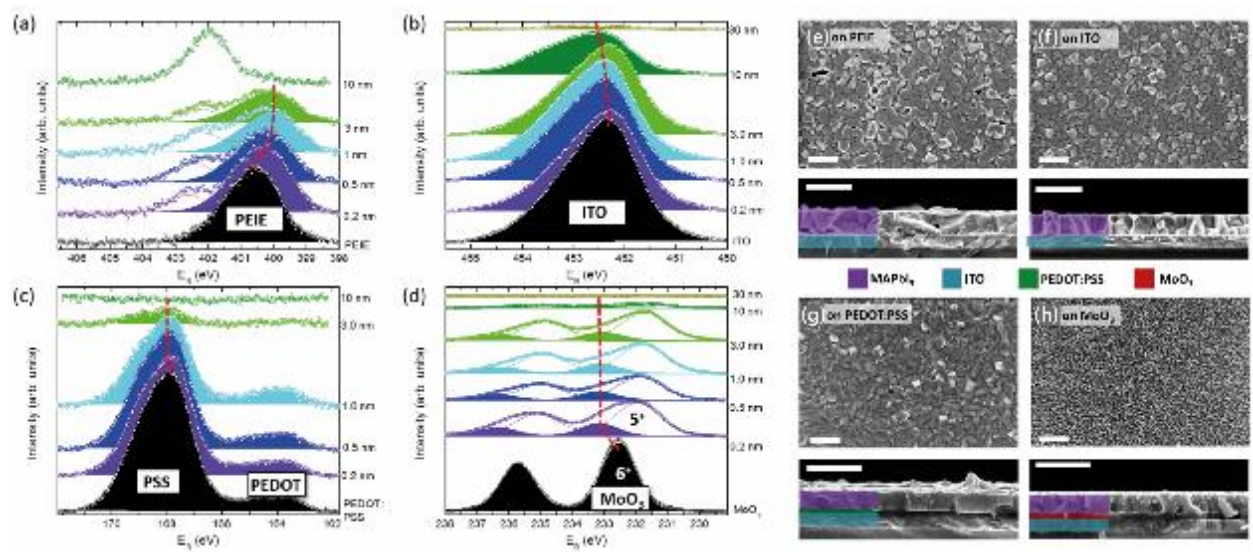

*Figure 23* – *XPS spectra of substrate specific core level peaks during incremental deposition of $MAPbI_3$ on top, with the evolution of (a) N 1s originating from PEIE, (b) In $3d_{3/2}$ from ITO, (c) S 2p from PEDOT:PSS, and (d) Mo 3d peaks from $MoO_3$. The morphology of the various 200 nm thick $MAPbI_3$ layers is captured in the SEM top view and cross section images for $MAPbI_3$ on top of (e) PEIE covered ITO, (f) ITO, (g) PEDOT:PSS, and (h) $MoO_3$. Reprinted with permission from Ref. 187. Copyright 2017 by Springer Nature Publishing AG.*

In the case of MAI and PbI$_2$ precursor evaporation on a TiO$_2$ surface, again with the intent to grow MAPbI$_3$ thin-films, the initial nucleation and subsequent growth of the HaP are strongly impeded. The initial stage of growth yields island-like aggregates on the bare TiO$_2$ surface and does not result in the nucleation of the pure HaP phase until approximately 15 nm of precursor material are deposited.[221] Another report concludes that the initial stage of growth (∼ 5 nm) of this interface system mostly leads to the formation of PbI$_2$,[222] which stands in contrast to the substantial lead deficiency observed in the corresponding experiment of MAPbI$_3$ deposited on MoO$_3$, ITO, PEIE and PEDOT:PSS.[187] However, the occurrence of a PbI$_2$-rich interfacial layer was described for other HaP/oxide systems, such as MAPbI$_3$/ZnO, where the PbI$_2$ layer forms a barrier for electron extraction and could be the origin of a nucleation front for the degradation of MAPbI$_3$ films in ZnO-containing MAPbI$_3$ devices.[197]

In summary, we note that discrepancies in reported intermediate layer composition and evolution with increasing thickness can be attributed to: (i) Variations in the evaporation process and precursor integrity, which can sensitively fluctuate with minor alterations to the experimental setup,[220] (ii) variations in the substrate preparation, cleaning and pre-conditioning (various chemical passivation strategies for oxide surfaces will be discussed in section C.1.4), (iii) transient changes in the composition and chemical constituents of the intermediate phase induced by the exposure of the interfacial layer to illumination and X-ray radiation in vacuum.

We described documented changes in the film composition earlier in section B.2.3, but need to underline here that changes could occur quasi-instantaneously, in particular in the vicinity of a catalyzing oxide surface. Particularly, the radiolysis of nitrogen-containing fragments of the MAI precursor and reactive Pb$^0$ in lead-based HaP materials has been well documented.[33,158,160,164]

In spite of these considerations, one should question the degree to which the observations made for ultra-thin evaporated HaP films are representative of the interface between any given substrate and thick HaP films, grown on top without specific perturbation or stimulus. Accordingly, the evaluation of peak level position, valence band onsets, and work function changes for the determination of energetic alignment and band bending in both substrate and HaP film at the interface would be subject to significant disruption in incremental growth studies, particularly for the early stage of growth, for which we find intermediate phase formation.

### C.1.2. Substrate-dependent HaP doping characteristics

The limited access to the HaP/substrate interface becomes a major shortcoming for the description and understanding of the electronic interactions at the buried interface. Beside assuming a central role in interfacial electrochemistry as well as in templating the initial stages of growth, the substrate also exerts immediate influence on the apparent doping characteristics of the HaP film. The effect is illustrated for MAPbI$_3$ films, deposited by spin-coating on top of substrates with different doping characteristics. XPS measurements performed on these films give the position of the VBM (here approximated by a linear fit to the leading edge, see discussion in section B.2.1.) with respect to the Fermi level $E_F$ in the gap.[37] If one coarsely groups the substrates into two different categories, i.e. n-type FTO, Al$_2$O$_3$/ TiO$_2$, TiO$_2$, ZnO, ZrO, and p-type PEDOT:PSS, NiO, Cu$_2$O, one observes that the Fermi level position at the perovskite surface (and presumably in the bulk) follows the doping characteristics of the substrate: the Fermi level appears very close to the CBM for MAPbI$_3$ on the former group, and much closer to mid-gap on the latter group (see Figure 24). Note that the reported values of the absolute position of $E_F$ depend sensitively on the determination of the valence band edge, which in turn should take the low density of states into account as described in section B.2.1.

The behavior was confirmed in an extensive combined UPS/IPES study that probed the surface of MAPbI₃ films spin-coated on top of heavily n-doped TiO₂ and p-doped NiO substrates.[38] While the MAPbI₃ films exhibited the same fundamental structural, morphological, and optical properties, as measured by XRD, AFM, and PL, the direct and inverse photoemission spectra point out clear changes in the electronic properties of the surface (figure 25): the work function increases by 0.7 eV, the valence band features and VBM shift by ~0.6-0.7 eV to lower binding energy (toward $E_F$), and the conduction band onset shifts by ~0.7 eV away from $E_F$, when changing from the TiO₂ substrate to the NiO substrate. Overall, the direction and magnitude of the change in work function and shift in the frontier electronic energy levels is evidence of a true shift of $E_F$ in the gap of the HaP film, which is hence n-doped on TiO₂ and slightly p-doped on NiO. The lack of any marked difference in the structural and optical properties of the HaP bulk film indicate that the change of the HaP's doping character must originate from the interface region to the underlying substrate rather than from any variation of the bulk. The remarkable aspect of this observation lies in the fact that the electronic properties were determined at the sample surface: given the MAPbI$_3$ film thickness, the change in Fermi level position at the surface was propagated from the interface, located 250 – 400 nm underneath.

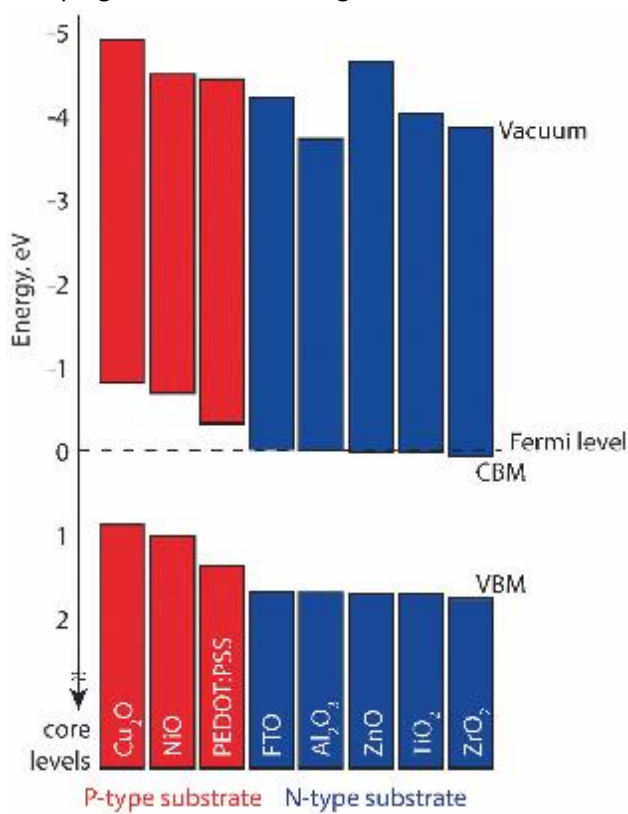

*Figure 24 – Band offset of MAPbI₃ film on various substrates with values for the VBM (derived from linear fit to band onset) and vacuum level determined from XPS and values for the CBM projected from a calculated band gap of 1.7 eV. Reproduced from Ref. 37 with permission from the PCCP Owner Societies.*

The origin of the effect can be threefold:

- The different substrate surfaces could induce the formation of non-stoichiometric intermediate HaP phases. The reaction products and surfactants then act as "self-dopants" in the interfacial region, change the chemical potential for the growth conditions in the short window of the film growth, and tip the doping range depending on a *de facto* PbI$_2$ or MAI rich precursor phase.[25]
- The work function of the substrate could dictate the alignment of the energy levels in the HaP semiconductor through vacuum level alignment, a mechanism that has been recognized for the majority of organic semiconductor interfaces.

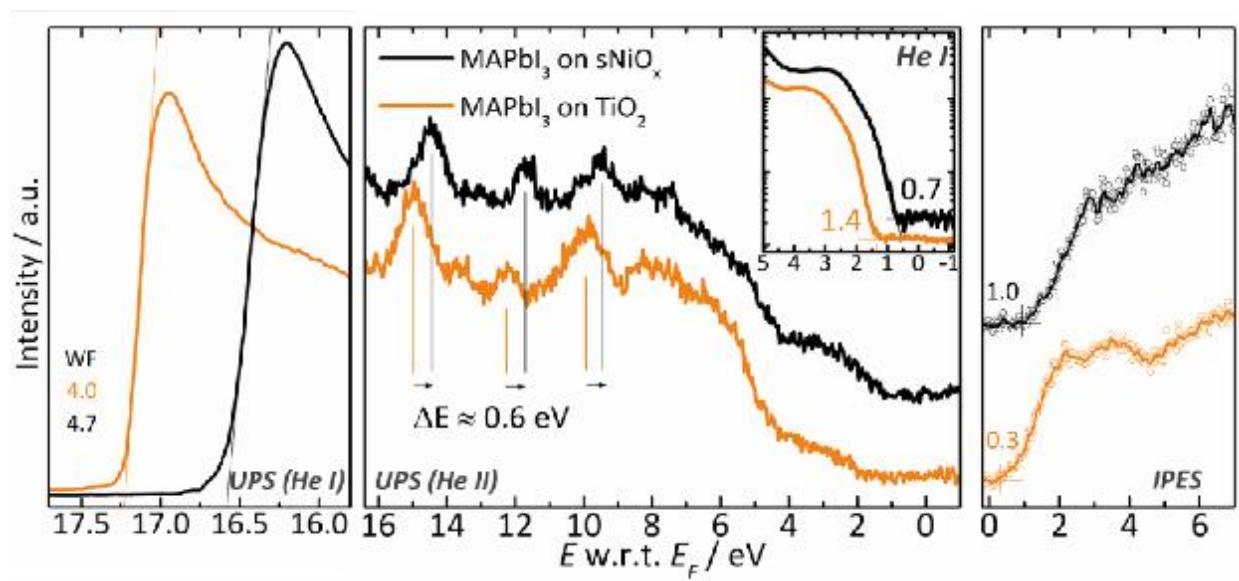

*Figure 25* – UPS and IPES spectra of MAPbI$_3$ films spin-coated on top of TiO$_2$ (orange curves) and NiO from a sol-gel process (sNiO$_x$, black curves). Change in work function and a concomitant shift of all frontier energy levels indicate a shift of E$_F$ in the gap. Reprinted wih the permission from Ref 38. Copyright 2015 WILEY-VCH Verlag GmbH & Co. KGaA, Weinheim

- The highly doped substrate could pin the Fermi level of the adjacent HaP film. In this latter case of remote doping, no chemical species would be exchanged between the two material systems.

The correlation between substrate work function, or doping type, and Fermi level position in the HaP film suggests that one of the latter two cases describes the scenario more consistently, while no such correlation has been found for the reactivity of the substrates, which ranges from inert (PEDOT:PSS, $Al_2O_3$) to rather reactive or photocatalytic (NiO, ZnO, $TiO_2$).[37] In contrast to bulk chemical doping, the observed remote doping at the interface lacks a clear origin or rationale for screening lengths. In this context, we remind the reader that the term remote doping (or modulation doping) was coined to describe the physical separation of the dopants, placed in one layer, from the carriers, released in an adjacent layer, thereby decreasing ionized impurity scattering and increasing mobility for these carriers.[223] The effect of remote doping is important in intrinsic (low density of free charge carriers) and high purity (low density of deep or mid gap defect states) semiconductors, in particular in terms of the temperature dependence of carrier mobility. In a uniformly doped semiconductor, mobility generally scales with temperature according to $\mu_{e,h}(T) \propto T^{3/2}$ at low temperature (ionized impurity scattering) and $T^{-3/2}$ (or stronger, i.e. more negative) at higher temperature (phonon scattering). In a remotely doped semiconductor, the absence of ionized impurity scattering should result in a more metallic-like behavior at low temperature, i.e. a smooth increase of mobility with decreasing temperature.[223] Yet, experimental verification of carrier mobility has proven to be a complicated task in HaPs, requiring careful navigation around measurement artifacts and extrinsic factors. As a result, it has been noted that for any given HaP stoichiometry, reported carrier mobility values encompass a considerable range depending on the probed time scale.[224] While such conclusion (as others) is an interpretative one, there is some consensus that Fröhlich interactions (i.e. scattering with polar optical phonons, where the electron density polarizes in response to lattice fluctuations)[225] can explain much of the carrier scattering.[226]

With respect to device fabrication, the exploitation of remote doping could give rise to substantial tailor-made adjustments in device functionality. In particular, HaP films on top of p-doped substrates do not exhibit a strong p-type character, as seen from mostly mid-gap $E_F$ positions. Various reports on improved solar cell device parameters due to additional doping of the underlying NiO HTL substrate (e.g. through Cu impurities or oxygen doping) could hence in part be related to better transport properties in the HaP films.[208,227,228] However, it is generally not possible to test the above hypothesis alone, nor feasible to deconvolute the various contributions to the performance of the device. For instance, the fabrication of MAPbI$_3$ films on top of heavily Li-doped NiO leads to an increase in HaP film crystallinity, reduced trap densities in the bulk, and potentially a more favorable interfacial energy level alignment compared to similar films produced on top of PEDOT:PSS substrates. For completed PSC devices, the interchange of the HTLs even led to the improvement of the device characteristics with a remarkable increase of $V_{OC}$ from 0.9 V to 1.1 V, in case of a PEDOT:PSS and a Li:NiO HTL, respectively, but the exact contribution of each abovementioned factor to the device functionality remains to be quantified.[228]

### C.1.3. Buried interface analysis via Hard X-ray photoemission spectroscopy

As seen above, the choice of the bottom transport layer affects the position of $E_F$ at the top of the HaP film. However, the exact evolution of the position of $E_F$ throughout the HaP film, including the possibility of band bending, has not been clearly established by surface sensitive measurements, which probe regions limited to the top most layers of the HaP film. Hard X-ray photoemission spectroscopy (HAXPES) studies offer the same electronic and chemical insight as conventional XPS measurements, but the higher photon energies result in higher kinetic energies $E_{kin}$ of ejected electrons. Consequently, these photoemitted electrons escape from deeper inside the sample, according to a Lambert-Beer law with a mean free path $\lambda$ of electrons in solids derived from the empirical relation:[229]

$$\lambda = \frac{A}{E_{\text{kin}}^2} + B\sqrt{E_{\text{kin}}} \approx B\sqrt{E_{\text{kin}}} \quad \text{for} \quad E > 150 \text{ eV},$$

where A and B are material constants.

At very high excitation energies (on the order of 10 keV), the technique can be used for direct assessment of the electronic properties and elemental composition hundreds of nm below the surface, and down to the interface with the substrate. The technique was thus successfully employed for a broad range of semiconductor heterostructures such as oxides and thin-film silicon for solar cell applications.[230,231]

HAXPES measurements have been performed on several HaP/substrate systems but often without capturing the buried interfacial region.[232–238] Nonetheless, spectra taken at various excitation energies enable us to derive a first assessment of chemical and electronic structure gradients in the material, even without reaching down to the substrate layer. In an approach to probe the effect of non-stoichiometric ionic distribution, Jacobsson *et al.* measured core level spectra of $FA_{0.85}MA_{0.15}PbBr_{0.45}I_{2.55}$ films on $SnO_2$ substrates, made from precursors that were either in a stoichiometric ratio or had 10% excess or deficiency in $PbI_2$.[237] Spectra of the perovskite core levels (Br 3d, I 4d and Pb 5d) and valence band were taken at excitation energies of 758 eV, 2100 eV and 4000 eV, which correspond to probing depths of $t_p$ ~5, ~11, and ~15 nm, respectively. The spectra for 758 eV and 4000 eV excitation energy are shown in figure 26. The measurements confirm that the I/Pb ratio (a) increases with $PbI_2$ deficiency and (b) decreases at higher excitation energies, i.e. with the signal originating from further down in the bulk. With their compositional analysis, along with a generally lower I/Br ratio at the surface, Jacobsson *et al.* propose a model in which the HaP film is terminated with an overlayer of unreacted FAI, followed by a Br-rich perovskite phase modeled in figure 26c. In this picture, however, the FA/MA ratio should be different at the surface than in the sub-surface region, which can be evaluated from the N 1s core level spectra with clearly different signatures for MA ($E_B$ = 402.5 eV) and FA ($E_B$ = 400.9 eV).[237] The absence of any trend in this metric as a function of probing depth is somewhat surprising and could be explained by a concomitant

FA deficiency in the top-most perovskite layer. Overall, the model remains coarse and points to the difficulty to derive exact layer models for these complex quinternary compounds from HAXPES measurements alone. In a complementary set of PES measurements, the evaluation of the valence band onsets suggested the absence of any measurable band bending in the top 15 nm of the HaP films, despite the previously determined compositional gradient.[235] We add as a further remark, that the PES analysis can be subject to beam-induced alterations, particularly for intense synchrotron radiation as excitation source (see section B.2.3.).

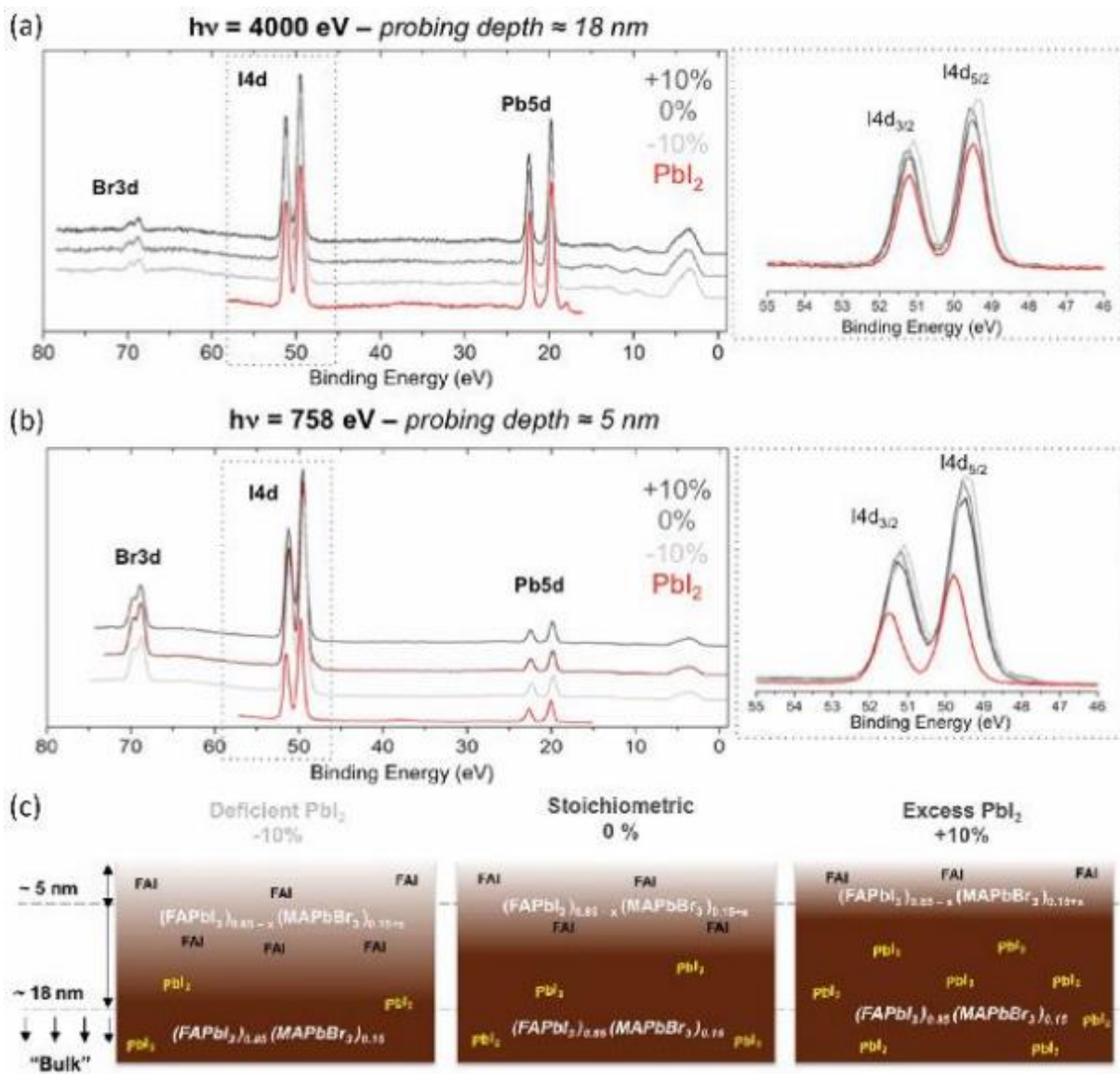

*Figure 26* – HAXPES measurements of $FA_{0.85}MA_{0.15}PbBr_{0.45}I_{2.55}$ at excitation energies of (a) 4000 eV and (b) 758 eV. The spectra show the regions of the Br 3d, I 4d and Pb 5d core level peaks for films that were prepared with either stoichiometric (0%), excess (+10%), or a deficiency (-10%) in $PbI_2$. Reference spectra of a $PbI_2$ film are plotted for the sake of comparison. (c) Species distribution in the surface and sub-surface region from the HAXPES measurements indicating $PbI_2$-rich bulk (dark brown) and a surface FAI layer (light brown). Reprinted with the permission from Ref. 237. Copyright 2016 American Chemical Society.

Nonetheless, the method opens an avenue to qualitatively determine compositional heterogeneity in HaPs. Analogous HAXPES measurements of HaPs exhibiting even more components (quadruple cation: $FA^+$, $MA^+$, $Cs^+$, $Rb^+$) by Philippe *et al.* demonstrate not only that their incorporation could be inhomogeneous, but also that the degree of inhomogeneity can vary strongly with the type and combination of cations employed.[234] For instance, very little Rb is found in the surface layers in RbFAMA based systems, which could be consistent with Rb segregation to the bulk/bottom of the sample. In contrast to that, Cs in CsFAMA seems to be more homogeneously dispersed and, in case of the CsRbFAMA, lead to a better dispersion of Rb in the film with an equally stoichiometric concentration of Rb in the surface region as well.[234] We note that further verification of this compositional gradient could be envisioned by employing complementary experimental tools to investigate the sub-surface region. For instance, time of flight secondary ion mass spectrometry (ToF-SIMS) could be used to generate a chemical profile down to the substrate layer.[239]

The HAXPES approach can also be enriched if combined with other more bulk sensitive spectroscopic techniques that yield additional chemical information. Fluorescence yield X-ray absorption spectroscopy (FY-XAS) is a complementary technique that can be applied in tandem to the HAXPES investigation of HaP compounds, for instance, to better analyze phase segregation. In the case of $MAPbI_{3-x}Cl_x$, no stable compound with stoichiometric alloying could be achieved and a concomitant XPS analysis of the surface layer yields only negligible amounts of Cl even when the film is produced in a wet process from MAI and $PbCl_2$ precursors, which should yield a I:Cl ratio of 1:2. In the HaP chlorine concentrations only as small 2% are reported.[240,241] The actual amount of chlorine incorporation – as measured by the concentration of surface Cl – is strongly dependent on the preparation technique. For instance, post-annealing a film composed from MAI and PbCl$_2$ precursors to 140°C under UHV conditions leads to no detectable surface Cl in the accuracy limit of XPS (< 1%), and the *in situ* monitored decrease in Cl content measured at the surface sets in at temperatures around 50°C already.[238] This experiment alone, however, does not reveal if the loss of Cl from the initial $PbCl_2$ precursor, tracked via XPS and leading to substantially chlorine-deficient $MAPbI_{3-x}Cl_x$ films, is homogenous. HAXPES data of a sample, depicted in figure 27a, do not show any detectable trace of chlorine for excitation at 2 keV nor 6 keV photon energy, which correspond to probing depths of 10 nm and 26 nm, respectively. Since the relative photoionization cross sections of the I and Cl core level orbitals change with photon energy, Starr *et al.* calculate the minimum amount of detectable Cl with respect to I as Cl:I < 0.024 for 2 keV and Cl:I < 0.15 for 6 keV. Thus, for the (3-x)/x concentration profile of Cl with respect to I in a $MAPbI_{3-x}Cl_x$ sample, the upper limit for the Cl-deficient stoichiometry would be $MAPbI_{2.93}Cl_{0.07}$ for the top-most 10 nm and $MAPbI_{2.6}Cl_{0.4}$ for the top-most 26 nm of the film. However, the FY-XAS measurements performed in the same spot (Figure 27b) reveal an extra component at 2827.4 eV binding energy, tentatively attributed to chlorine in PbCl-coordination.[238] Given the calculated detection limit in the FY-XAS measurement, which are averaged over the entire film thickness, the Cl concentration in the lower half of the ~60 nm thick sample would at least be $MAPbI_{2.6}Cl_{0.4}$. This suggests therefore that the layer is not completely depleted of chlorine but exhibits a non-negligible Cl content at the buried $TiO_2$/HaP interface, which could potentially be linked to persistent $PbCl_2$, i.e., the most insoluble of the precursors and thus the first to precipitate in a solution deposition process. This scenario has direct implications in terms of barrier formation and charge carrier transfer across the bottom interface: Introducing chlorine substitutions of iodine at a $TiO_2/MAPbI_3$ interface, DFT slab calculations, which are reminiscent of those presented in section B on the exposed surfaces, indicate a redistribution of charges (figure 27c). Mosconi *et al.* calculated that in this case the coupling between the Ti and Pb conduction band states is increased and project that the alignment between the energy levels of $TiO_2$ and HaP is slightly more favorable than without Cl at the interface, due to an upshift of the CBM on the $TiO_2$ side (besides further changed properties such as modified growth and film stability of the HaP layer).[105]

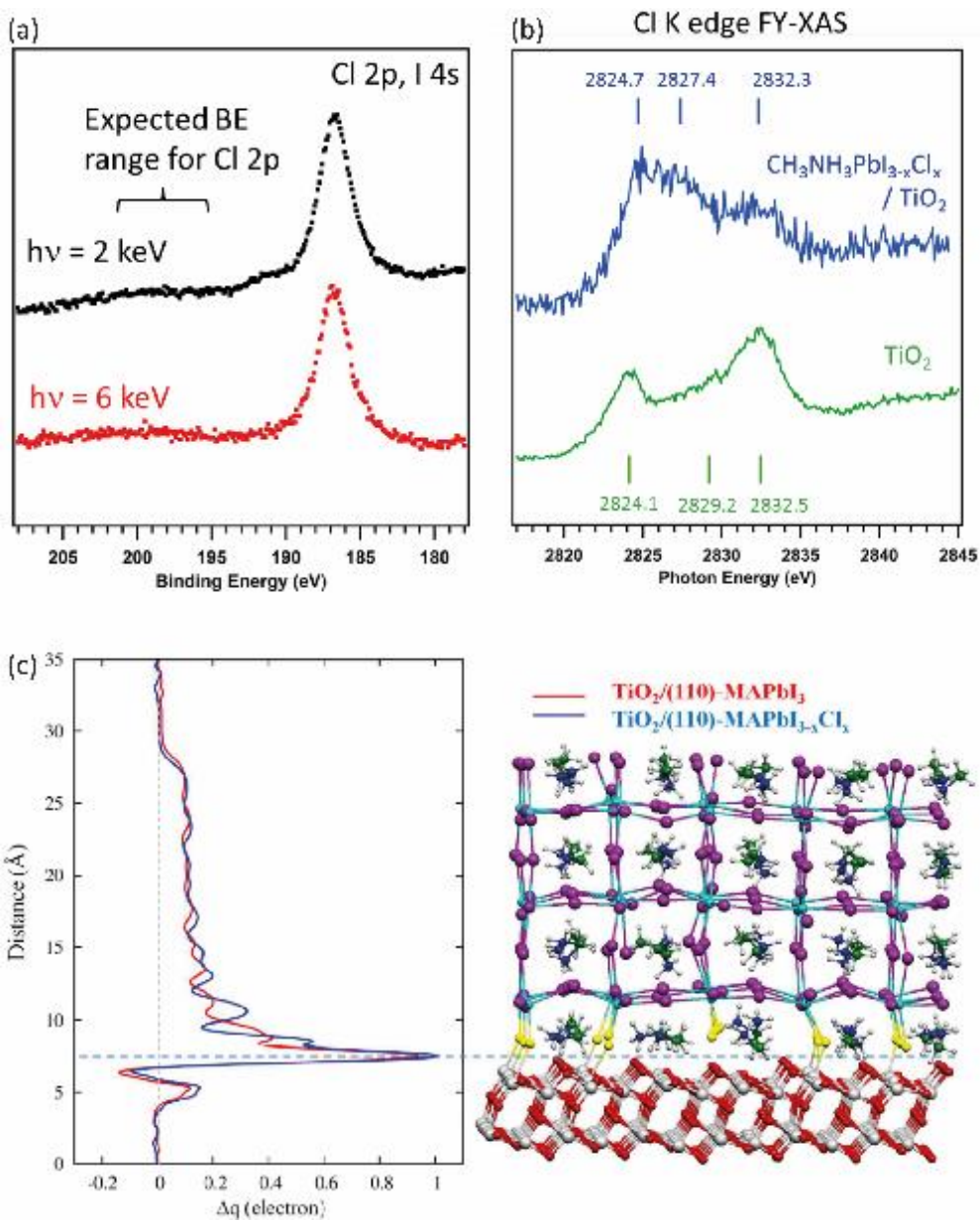

***Figure 27*** *– (a) HAXPES and (b) FY-XAS data of a MAPbI$_{3-x}$Cl$_x$ thin film. (a) Cl 2p and I 4s region with no detectable trace of Cl in the first 26 nm. (b) The comparison of the Cl K edge for a plain substrate (compact TiO$_2$) and substrate + HaP layer system gives a lower bound for the overall chlorine concentration in the sample. Reproduced from Ref. 238 with permission from The Royal Society of Chemistry. (c) Charge displacement analysis and simulated slab structure comparing TiO$_2$/MAPbI$_3$ and TiO$_2$/MAPbI$_{3-x}$Cl$_x$ interfaces. Reprinted with the permission from Ref. 105. Copyright 2014 American Chemical Society.*

### C.1.4. Examples for oxide substrate modifications and passivation

In the previous sections we described that, while the electronic energy level alignment between oxide substrate and HaP layer has hardly been accessible, the interface chemical reactivity has been identified, e.g. through the incremental growth studies, as a major obstacle for the control of junction formation. In the conventional PSC device architecture, a set of strategies has emerged to either substitute the bottom oxide layer by a less reactive ETL (e.g. replacing $TiO_2$ by $SnO_2$),[196,242] employ a surface modification such as an ultra-thin interlayer on the oxide ETL prior to HaP film deposition, or a combination of both. Various interlayer approaches have been tested, ranging from polymer films over organic self-assembled monolayers (SAMs) to alkali halide salts, which could undergo (partial) incorporation into the HaP film.[56] It is important to keep in mind that the effect of these buffer layers may not only be to suppress interfacial chemistry, but also to passivate trap states in the oxide that could otherwise act as recombination centers, and improve the energy level alignment between oxide and HaP film to reduce barriers for carrier extraction.

The case of ZnO ETL surfaces provides a good example of a particularly reactive substrate for the formation of a top HaP film. PSCs fabricated using ZnO ETLs generally suffer from a strong thermal instability and HaP film degradation which originates from the ZnO surface. The *basic* ZnO surface deprotonates the methylammonium cation, which leads to a loss of methylamine and the formation of $PbI_2$.[243] Like for the evaporation of $MAPbI_3$ on top of other oxide surfaces, discussed in section C.1.2., XPS measurements reveal the formation of a $PbI_2$ phase at the immediate interface with ZnO substrate for $MAPbI_3$ films deposited on top by co-evaporation.[244] Yang *et al.* compute that this decomposition process should be accelerated by the presence of surface hydroxyl groups and residual acetate ligands from the HaP precursors.[243] Hence, they describe the removal of these moieties by calcination as a strategy to achieve a thermally more robust HaP film. This could be the mechanism that improves the stability of the HaP film when a ZnO substrate from a nanoparticle solution is substituted by Al-doped ZnO (ZnO:Al) from a magnetron sputtering process as the amount of surface hydroxyls would be reduced substantially.[197] Similar outcomes are partially reached in alternative fabrication methods of the ZnO layer by electrodeposition or sputter deposition, which is usually done to improve the morphology and optoelectronic properties of the ETL itself (e.g. conductivity, optical transmittivity, etc.).[245,246] However, in both examples, neither hysteresis in the PSC device, which presumably stems from interfacial chemistry, nor ageing, i.e. the decline of the device performance parameters over days of testing, could be entirely averted. A different approach thus seeks out to mitigate the detrimental interface chemistry by introducing buffer layers between ETL and HaP film. The case was demonstrated for ZnO ETLs to a first order by employing a substantial amount of $PbI_2$ presumably forming an interlayer,[247] or in a more advanced approach using PCBM or PEI coatings.[248,249]

*Organic molecular (mono)layers as buffer layers:*

The concept of employing thin organic interlayers or self-assembled monolayers (SAMs) to improve metal oxide transport layers was introduced well before their application in PSCs mentioned above. As the idea has been widely explored in the field of molecular electronics, a thorough review gives an in-depth account on the modification of chemistry and electronic properties of semiconductor surfaces,[250] and more precisely oxide semiconductor surfaces.[251] SAMs based on fullerene derivatives, e.g. PCBM, have found use early in the development of solid state PSCs to decorate the mesoporous $TiO_2$ scaffold (Figure 28a).[252] Interestingly, the initial conjecture by Abrusci *et al.* had been that while the PCBM acts as a very effective electron acceptor from the $MAPbI_3$, it would inhibit further electron transfer into the nanocrystalline $TiO_2$ due to unfavorable energy level alignment. This approach was hence thought to be beneficial for the mesoporous structure, for which Abrusci *et al.* suggest that electron transport could be maintained in the perovskite film via multiple trap-and-release events at the fullerene-based SAM, instead of electron transport via the TiO$_2$ nanocrystalline film. Nonetheless, the application of a fullerene SAM was eventually translated to planar PSC device structures as well, where significant improvement in the device characteristics and reduction of interfacial capacitance were observed (Figure 28b).[198] In contrast to the earlier assumption that electron transfer to the metal oxide would be impeded, time-resolved photoluminescence measurements indicate a faster quenching rate of photo-excited charge carriers at the interface than without the fullerene surface modifier. Additional electroluminescence measurements and time-resolved microwave conductivity measurements show that the SAM surface modification does reduce non-radiative recombination and indeed promotes electron transfer at the interface. In this scenario, the SAM would also passivate localized traps, i.e. under-coordinated surface $Ti^{4+}$ ions. Furthermore, it was suggested that trap states in the perovskite film could be passivated by the fullerene,[253] which in itself could aid to the reduction of recombination at the interface region. Such a passivation mechanism for the HaP film is still under intense debate; under-coordinated halogen such as $I^-$ as well as $Pb^{2+}$ sites could form trap states (halogen anions as hole traps and lead cations as electron traps), which can act as recombination centers.

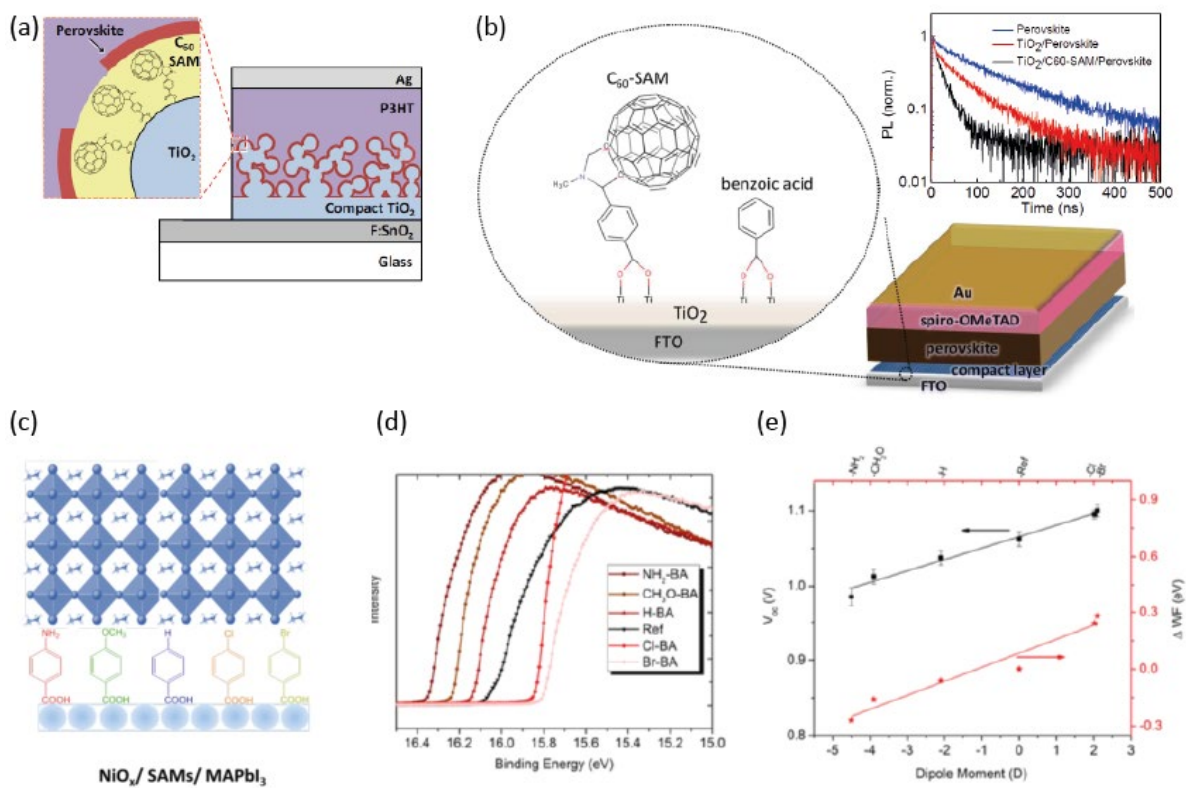

*Figure 28* – *(a) Schematic of a PSC device with a fullerene-based SAM covering the mesoporous TiO$_2$ ETL. Reprinted with the permission from Ref. 252. Copyright 2013 American Chemical Society. (b) fullerene-based (C-60) and benzoic acid SAMs on a compact TiO$_2$ ETL in planar PSC cell geometry. The inset depicts PL decay transients of the MAPbI$_3$ emission band. Reprinted with the permission from Ref. 198. Copyright 2014 American Chemical Society. (c-e) Benzoic acid SAM functionalization of NiO surfaces in inverted MAPbI$_3$ based PSCs with (d) the work function determination of the modified NiO surfaces from the UPS spectra and (e) a correlation between substrate work function and device open circuit voltage. Reprinted wih the permission from Ref. 258. Copyright 2017 WILEY-VCH Verlag GmbH & Co. KGaA, Weinheim.*

We will discuss in a later section (C4.1), that Lewis acids can be employed to bond halogen anions and passivate the surface defects of a HaP film, while Lewis bases can be used to passivate under-coordinated $Pb^{2+}$ sites,[254] like in well-established molecular passivation routines for II-VI and III-V semiconductor surfaces.[255,256] Here, in a similar fashion the interface modifier at the bottom substrate interface of the HaP film could take over an analogous role. Despite the rather preliminary explanations of interfacial properties between oxide, organic SAM and perovskite film, as well as only rudimentary stability data, many research groups reported that the power conversion efficiency, taken as a significant parameter of the PSC, was generally improved after surface modification. Qiao *et al.* summarized these cell enhancements for HaP-based semiconductor devices in their survey for many different substrate oxide layers, such as $TiO_2$, ZnO and $SnO_2$.[257]

In a similar fashion, SAMs can be employed in inverted perovskite solar cell structures where the molecular layer (e.g. based on para-substituted benzoic acids) functionalizes the HTL NiO substrate as shown in Figure 28c.[258] The various substitutions on the para-position of the benzoic acid change the dipole moment of the molecule and thus the dipole of the resultant SAM, for which the average molecular orientation is well-defined and aligned to the substrate surface normal, as has been extensively reported before for a wide range of semiconductor surfaces.[256] As a result, the work function of the NiO surface is either lowered (negative dipole) or increased (positive dipole), leading to a change in energy level alignment with the subsequently deposited $MAPbI_3$ film. Wang *et al.* measured the work function change ($\Delta$WF) by UPS (see figure 28d) and compared these values to the change in $V_{oc}$ for devices produced with the same SAM interlayers between NiO and HaP absorber (Figure 28e).[258] We note that while the two values change parallel, the excursion of the work function is significantly larger than that of $V_{oc}$ ($\Delta WF_{max}$ = 600 meV, $\Delta V_{oc,max}$ = 100 meV). Hence, the work function affects the energy level alignment at the interface and could induce band bending in the HaP film but does not fully determine the attainable photovoltage. The mechanism is complex, as tr-PL measurements indicate that carrier extraction is enhanced with the application of a benzoic acid SAM on top of the NiO, which can also be attributed to a reduction in trap state densities as observed in the case of the fullerene-based SAM on top of $TiO_2$. However, a clear assignment of these effects is further complicated by the concomitant changes in HaP film morphology, estimated from the apparent grain size in SEM images of the sample surface.[258] More generally speaking, only a precise determination of the interfacial chemistry and subsequent HaP layer growth would allow to determine the specific effect of the energy level alignment on the transfer of charge carriers across the interface.

*Chemistry and HaP growth on passivated oxide surfaces:*

As laid out in section C.1.1. the growth modes of (vacuum evaporated) HaPs, particularly $MAPbI_3$, is strongly affected by the chemical reactivity of the substrate, which leads to the formation of a $PbI_2$-rich phase at a reactive oxide interface and a conformal $MAPbI_3$ phase on top of an organic semiconductor substrate. These results translate well into HaP growth on a SAM-modified $TiO_2$ substrate compared to a bare $TiO_2$ surface. Shallcross *et al.* investigated the growth of $MAPbI_3$ from co-evaporated MAI and $PbI_2$ on top of both bare $TiO_2$ and $TiO_2$ coated with an amine buffer layer. Their XPS and UPS results reveal marked differences between the initial nucleation and subsequent growth of the $MAPbI_3$ film on the two different surfaces. Film growth appears to follow a layer-by-layer mode on the amine-terminated $TiO_2$, promoting the formation of the stoichiometric MAPbI$_3$ phase within the first 5 nm of HaP deposition, while an island-like growth of the PbI$_2$-rich phase is observed on top of the unmodified TiO$_2$. In the same study, control over the morphology and crystallinity of solution-processed HaP films was achieved by these modifications of the TiO$_2$ surface.[221] It is important to keep in mind that the exact nature of the interface chemistry is still being debated, and seems to depend on the details of the HaP processing. In a similar study, Will *et al.* investigated MAPbI$_3$ films grown from solution of a 1:1 mixture of PbI$_2$ and MAI in DMF and DMSO on top of TiO$_2$ surfaces with and without modification with a C$_{60}$-SAM.[259] Their fit to X-ray reflectometry (XRR) data suggests a PbI$_2$-rich phase (not excluding the possibility of forming PbO$_x$ as well) at the interface between HaP and SAM, and an MAI-rich phase at the TiO$_2$/MAPbI$_3$ interface (Figure 29), which appears to be in stark contrast to the expectation that the $C_{60}$-SAM termination of the $TiO_2$ should suppress decomposition. However, here the claim is made that the interfacial phases of the HaP material extend to only a few Å, well in the sensitivity regime of the XRR measurement. These changes have a pronounced impact on the carrier distribution in the layer stack under light bias. Cross-sectional KPFM measurements depicted in Figure 29c-e indicate that the difference in local CPD between dark and illuminated state in the $MAPbI_3$ film is increased for the $C_{60}$-SAM modified $TiO_2$ substrate in comparison to the reference $TiO_2$ substrate, which corresponds to a higher photoinduced voltage.[259]

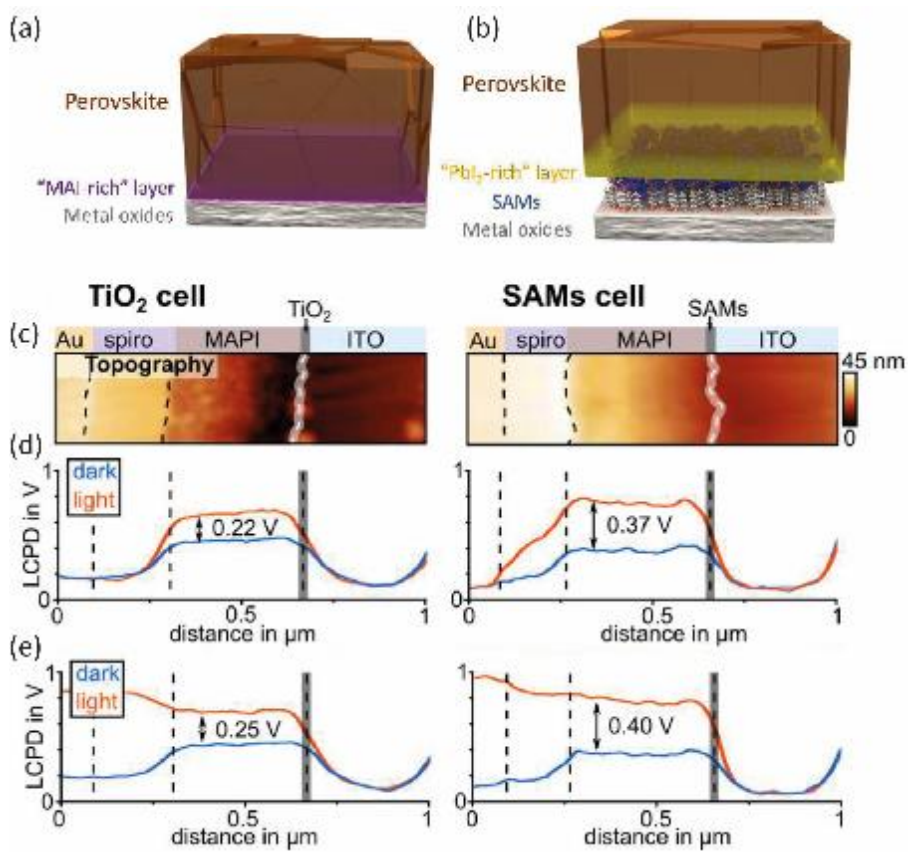

***Figure 29*** – *MAPbI$_3$ based PSC with (a) bare TiO$_2$ ETL and (b) C$_{60}$-SAM modified TiO$_2$ ETL forming thin interlayer phases at the immediate ETL/HaP interface. (c) Cross-sectional KPFM measurements and local CPD und (d) short circuit and (e) open circuit conditions. Reprinted with the permission from Ref. 29. Copyright 2018 American Chemical Society.*

Finally, we stress that the community is exploring numerous further routes to improve the oxide CTL / HaP interface, among which we find comprehensive approaches that seek to improve the electrical properties of the $TiO_2$ while reducing surface trap state densities. Singh and co-workers investigated the incorporation of alkali metal dopants into mesoporous $TiO_2$ scaffolds by applying various alkali bis-(trifluoromethanesulfonyl) imide (TFSI) salts in acetonitrile solutions.[260] They suggest that in this configuration the electronic conductivity of the $TiO_2$ is enhanced while oxygen vacancies are effectively being counterbalanced. In addition, the TFSI treatment grafts sulfate bridges on the $TiO_2$ surface and could lead to the formation of $PbSO_4$ anchors to the HaP film, which has been suggested before for the PEDOT:PSS/$MAPbI_3$ interface from a XPS multi peak fit of the Pb 4f core level.[187] The exploration of these design rules for the bottom CTL has important consequences for improving the functionality and reliability of HaP-based devices.

Particularly, double-layer architectures, comprising an oxide CTL and an organic CTL, could mark an interesting strategy to mitigate photocurrent hysteresis rooted in the interfacial chemistry. Kegelmann *et al.* found that the magnitude of hysteresis could be reduced by applying a $TiO_2$/PCBM double-layer ETL instead of only a single $TiO_2$ or PCBM film. They attribute improved hole blocking by the additional wide band-gap metal oxide and decreased loss for charge transport, due to an energetically more favorable contact, to a reduction of shunt paths through the fullerene to the ITO layer.[202]

**C.2. Semiconducting charge transport layers on top of HaP films**

We now turn to the other side of the HaP thin-film in the device, i.e. the top contact layer. In the conventional PSC geometry, this top contact layer is a semiconducting HTL, whereas in the inverted PSC geometry, a semiconducting ETL is deposited on top of the HaP. Here, we will differentiate between the use of (i) organic semiconductors, (ii) transparent conductive oxides, and (iii) conductive carbon contacts used in a range of HaP-based devices. These material classes vary widely in their expected energy level alignment, chemical reactivity with the constituents of the HaP film, and inherent optoelectronic properties that define key parameters, such as trap state densities, electrical resistivity and optical transparency.

Historically, organic semiconductors have by far been the most commonly employed transport layer materials deposited *on top of* the HaP film. The true merit of using organic CTLs lies in the ease with which they are introduced in the device structure. This includes processing conditions compatible with maintaining the HaP film and surface integrity, and upscaling potential for large area devices. In this context, the processing conditions refer either to solution-based fabrication routes, in which the organic CTL is dissolved in a solvent orthogonal to the HaP precursors, or to low temperature evaporation-based deposition of the organic CTL. In both cases the perturbation of the HaP surface can be expected to be minimal (but not necessarily negligible as will be discussed later). The fabrication procedure has been so successful that numerous organic transport materials and dopants have been used to form high-efficiency devices.[261,262] Eventually, the major promise of a HaP device employing all organic CTLs, as substrate and top layer, is that the fabrication scheme could rely on low temperature processes for cost reduction and integration in mechanically flexible device structures.[263]

In contrast to their organic counterparts, transparent conductive oxide (TCO) semiconductors are less commonly used as CTLs directly deposited on top of a HaP film. While TCOs generally promise better protection and encapsulation for the underlying perovskite,[264] the reasons for the avoidance are twofold: First, the most relevant deposition methods for oxides (e.g. sputter deposition, pulsed laser deposition, etc.) are harsh; i.e. they involve the formation of highly energetic metal and oxygen species that impinge on the HaP film surface, thereby damaging it. Second, as discussed in the previous section, oxide/HaP interfaces are prone to chemical reactions that lead to many reaction products but generally comprise a partial reduction of the metal oxide layer.

We will also discuss top layer conductive carbon contacts as an alternative configuration for HaP devices. This material class comprises carbon nanotubes (CNT), as well as graphene, reduced graphene oxide (rGO), or HTM-free micro-graphite carbon counter electrodes, which have been employed successfully in PSCs.[265–267] The integration of these nanomaterials and related carbon-based nanocomposites could mark a distinct technological advancement as they combine the facile applicability of organic semiconductors with the robustness of inorganic semiconductors. Yet, the approach remains the least tested, with only a limited number of study cases that explore the interaction between HaP films and conductive carbon overlayers.

While the following sections focus on the energy level alignment and the chemical reactivity at the interface to the HaP, it is important to note that the inherent optical and electronic properties of the CTLs themselves bear requirements similar to those described in the previous section for the substrate films (e.g. high carrier mobility, good optical transmittance). Oftentimes these properties are not entirely disconnected from the interfacial chemistry and energy alignment processes. As an example, for Spiro-MeOTAD, the most ubiquitously used organic HTL,[268] the addition of dopants (e.g. through ambient oxygen) leads to a change in conductivity and a concomitant change in energy level position with respect to $E_F$, which affects carrier extraction from the adjacent HaP film.[269] Additionally, the case demonstrates that we now need to take additional chemical reaction pathways into account, such as the formation of superoxides at the (doped) Spiro-MeOTAD/HaP interface (see section B.1.2). It thus becomes virtually impossible to deconvolve the sum of these effects into specific contributions by only performing electrical and optical measurements.

### C.2.1. Energy level alignment to organic semiconductor films

In the following section, we describe the electronic energy level alignment processes at the interface between the HaP film and organic semiconductors. Data from measurements and calculations then serve to substantiate whether an improved energy level alignment at the HaP / transport layer interface has a distinct impact on the transport characteristics across the interface and ultimately on the device performance. Such a correlation has been proposed early in the literature and follows well-known approaches in semiconductor physics in general,[270] and injection/extraction into organic semiconductors in particular.[271] For HaP semiconductors, we want to determine whether the energy level alignment with the many different doped and undoped organic transport materials that have found use in HaP-based devices,[261,262,268] follow the established rules or exhibit unique character.

*Energy level alignment and IE matching between HaP and HTL in PSC:*

With the rise of the first solid state perovskite solar cell in 2012, the energy level alignment between MAPbI$_3$ and the organic hole conductor, spiro-MeOTAD, was estimated by comparing the IE of the individual components. Those values were determined independently from each other (5.4 eV for MAPbI$_3$ and 5.2 eV for spiro-MeOTAD) and served as a first guide for drawing the energy level diagram.[193] The formation of the interface was subsequently tracked in a dedicated PES experiment, in which thin spiro-MeOTAD films were evaporated on top of three different HaP layers, MAPbI$_3$, MAPbI$_{3-x}$Cl$_x$, and MAPbBr$_3$.[145] A set of direct and inverse photoemission (UPS/IPES) measurements on an incrementally deposited (undoped) spiro-MeOTAD layer grown on top of a MAPbI$_3$/TiO$_2$/glass layer stack is presented in figure 30a. Several points can be noted in this plot. First, no work function change occurs when the first thin layer of spiro-MeOTAD is deposited, indicating that no interface dipole is formed between the two materials. Second, the offset between the VBM of the MAPbI$_3$ film and the spiro-MeOTAD HOMO level amounts to 0.4 eV, with the spiro-MeOTAD HOMO being closer to $E_F$. Hence, no barrier for hole extraction from the HaP onto the HTM is expected, although a barrier exists for hole injection in the other direction. Third, the HOMO level onset shifts toward higher binding energies (i.e. away from $E_F$) with increasing spiro-MeOTAD thickness. The work function changes accordingly, as the IE of the organic layer remains equal to 5.0 eV. The absence of interface dipole, equivalent to vacuum alignment, between spiro-MeOTAD and the HaP is confirmed for MAPbI$_{3-x}$Cl$_x$ and MAPbBr$_3$, as depicted in figure 30b.[145] It is important to note, that for all the different iterations (i.e. thickness of the spiro-MeOTAD overlayer) XPS measurements found the Pb and I core level positions to be constant, indicating that no band bending occurs in the HaP film as a result of the deposition of the spiro-MeOTAD layer. This finding is somewhat unexpected when recalling the variations in Fermi level position in a HaP film as a function of the underlying substrate (section C.1.2.). In the present case, the organic CTL on top of the HaP does not change the carrier density in the perovskite.

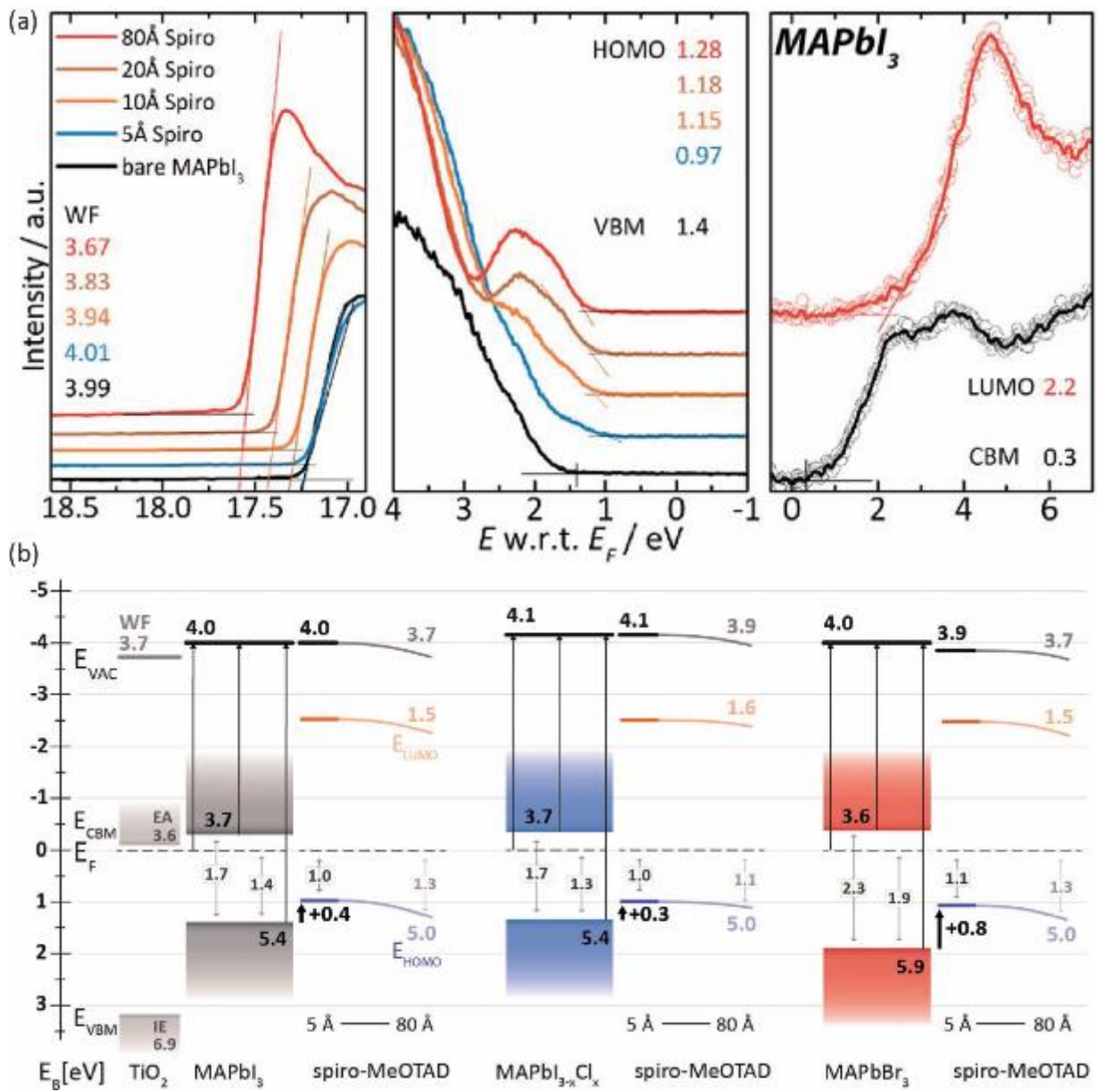

*Figure 30* – Energy level alignment between the HTM spiro-MeOTAD and HaP thin films. (a) UPS (left and middle panel) and IPES (right panel) of incrementally evaporated spiro-MeOTAD films on top of a MAPbI$_3$/TiO$_2$/glass layer stack. (b) Energy level diagram derived from UPS/IPES measurements of incrementally grown spiro-MeOTAD films on top of MAPbI$_3$, MAPbI$_{3-x}$Cl$_x$, and MAPbBr$_3$. Reproduced from Ref. 145 with permission from The Royal Society of Chemistry.

In this picture, it was initially proposed that the energy level misalignment between the charge transport levels of the organic semiconductor and the HaP could limit the attainable solar cell device parameters. The most direct impact is expected to be on the achievable photovoltage (open-circuit voltage), as the quasi-Fermi level splitting (QFLS) in the HaP could be limited by pinning of the electron Fermi level ($E_{F,n}$) close to the TiO$_2$ CBM at the bottom interface and the pinning of the hole Fermi level ($E_{F,p}$) close to the organic HTL HOMO on the other side of the cell. A straightforward solution to this problem would be to engineer the energy level alignment at the HaP / HTL interface by changing the IE of the HTM and make the HOMO onset resonant with the HaP VBM. Since indications are that vacuum level alignment exists between the two materials (i.e. no Fermi level pinning and no interface dipole, figure 30b), a change in the HTL IE would translate into a minimization of the energy offset between the HaP VBM and the HTM

HOMO. In MAPbBr$_3$ based PSC in particular, a hole transport material with an IE ~6.2 eV, N,N'-Dimethyl-3,4,9,10-perylenedicarboximide (PTCDI-C$_1$), exhibits a HOMO level about 0.3 eV deeper than the VBM of MAPbBr$_3$.[145] Edri *et al.* projected that, while this situation results in an extraction barrier for holes, generally higher V$_{oc}$ values would be achievable. This hypothesis, however, could not be decoupled from the need of heavily doping the HTL for achieving high photovoltages above 1.5 V.[272]

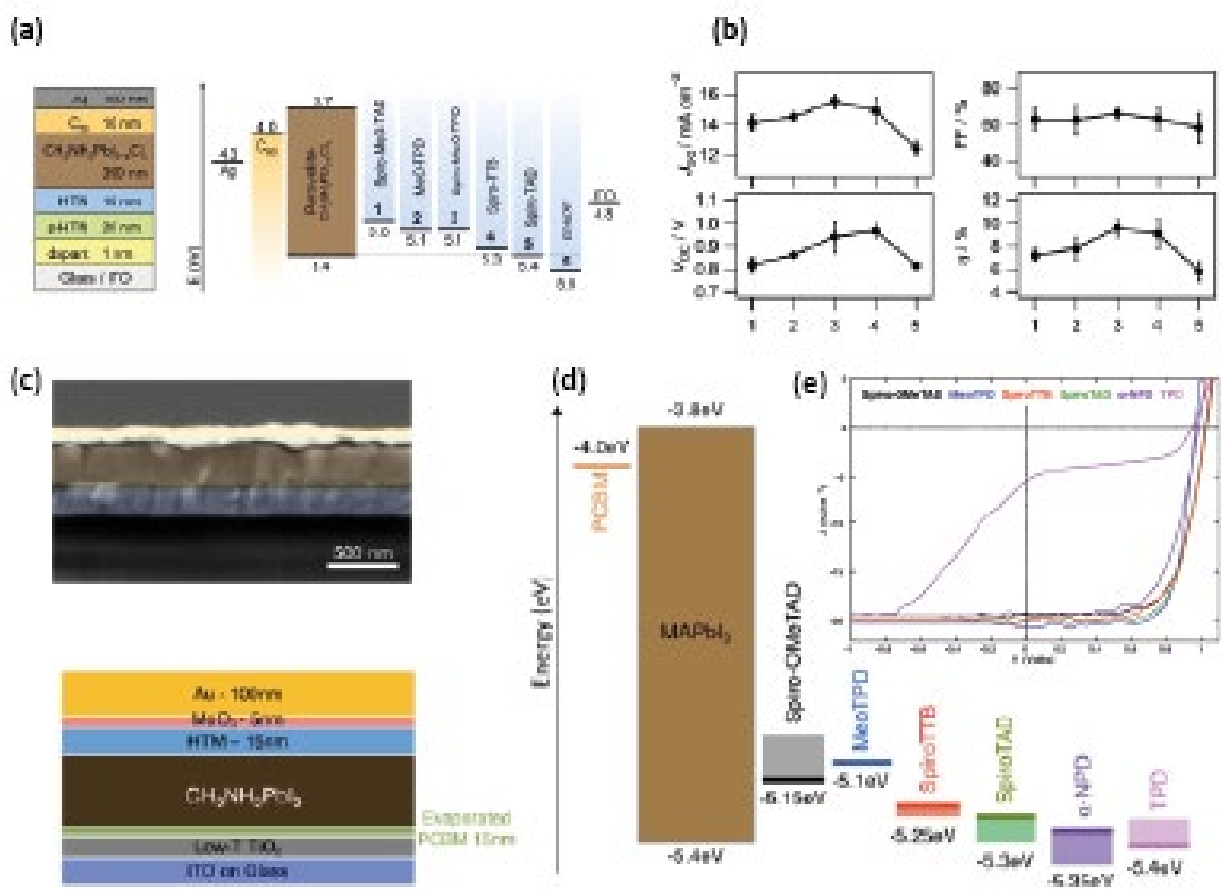

*Figure 31* – *(a) Energy level diagram of a set of PSCs (all layers from evaporation) with various triphenylamine-based organic HTMs with IE ranging from 5.0 - 5.5 eV. (b) PV parameters (FF = fill factor, η = PCE) for PSCs fabricated with HTL layers. Numbers on the x-axis indicate the molecular compounds from panel a and are described in the text. Reprinted figure with permission from Ref. 273. Copyright 2014 by the American Physical Society. (c) SEM image and schematic of PSC device layout and (d) energy level diagram with different organic HTLs. (e) j-V curves of PSCs fabricated with the various HTLs, showing no apparent correlation of device characteristics with IE mismatch. Reprinted with the permission from Ref. 274. Copyright 2016 American Chemical Society.*

Polander *et al.* investigated the effect of matching the IEs of the organic HTM and HaP on the open voltage.[273] They find that for PSCs fabricated with various HTMs with IE ranging from 5.0 eV to 5.5. eV, the maximum $V_{oc}$ and PCE are achieved when the IE of the organic semiconductor (5.3 – 5.4 eV) corresponds to that of the evaporated $MAPbI_{3-x}Cl_x$ film (5.4 eV) used in this study (see Figure 31a,b). In this case the $MAPbI_{3-x}Cl_x$ film is evaporated on top of the organic HTL (referred to as HTM in Fig. 31 a) in an inverted structure, but the example still illustrates the impact of mismatched IEs. The HTMs (referred to by number in figure 31a,b) used in this study are based on triphenylamine moieties and are spiranes with the exception of compound 2, methoxy-N,N'-diphenylbenzidine (MeO-TPD), and hence have a molecular structure related to spiro-OMeTAD. In a similar study, Belisle and coworkers investigated $MAPbI_3$-based PSCs in the conventional structure with several HTMs with IEs varying from 5.1 to 5.4 eV.[274] For each device configuration, a 15 nm thin HTL of the organic semiconductor was evaporated on top of the perovskite film. The device was then completed by applying a 5 nm thin $MoO_x$ film and a Au top electrode (see Figure 31c,d). In contrast to the previous results, the device parameters, represented by the respective j-V curves in Figure 31e, show no correlation with the IE mismatch between HTM and HaP. On the contrary, the $V_{oc}$ of the PSCs remains unaffected by the choice of the HTL and, hence, the adjustment of the IE. A similar lack of sensitivity of the device performance with respect to the IE of the organic HTL was further corroborated by Park *et al.* who employed a wide range of HTMs, looking into a new group of molecular compounds based on triarylamino-ethynylsilyl derivatives with IEs varying from 4.8 to 5.8 eV as prospective HTLs.[275] Again, the device parameters remained unchanged within the margin of error for HTMs with IE between 4.8 and 5.4 eV. However, devices using HTLs with larger IEs were found to lead to a significant decline in $J_{sc}$ and fill factor of the cells, which is consistent with the findings of Polander *et al.* and Belisle *et al.*, for whom the HTMs at the upper end of the IE range (5.6 eV and 5.35 eV, respectively), produced cells with lower $J_{sc}$ and in the latter case a pronounced S-shape of the j-V curve.[273,274] The S-shape diode characteristics indicate the presence of a barrier for hole extraction, as the HTL's IE is higher (lies deeper) than that of the HaP.

While the results from the different groups regarding the energy level mismatch between HTL and HaP seem contradictory at first glance, i.e. in the one case impacting device efficiency but not doing so in the other case, we can derive a few seemingly general coarse rules:

(i) If the IE of the (undoped) organic HTM ($IE_{HTM}$) is smaller than the IE of the MAPb-based HaP ($IE_{HaP}$), vacuum level alignment occurs.[145]

(ii) In a PSC, for the case that $IE_{HTM} < IE_{HaP}$, the quasi-Fermi level splitting, and hence attainable photovoltage, does **not** seem to be limited by pinning $E_{F,p}$ at the HTL HOMO level.

(iii) If $IE_{HTM} > IE_{HaP}$, the formation of a hole extraction barrier can limit charge transport across the HTL / HaP interface, which ultimately leads to a decline in $J_{sc}$ and fill factor of PSCs that constitute such an interface.[273–275]

*Generalized assessment of the energy level alignment between organic CTL and HaP:*

The general conclusions given above depend sensitively on the details of the experiments. First, one should note that the IE of the HaP film is not directly determined in each of the device studies presented above, but instead reproduced from the literature. Yet, as described earlier in section B.2.2., the HaP IE can vary substantially as a function of stoichiometry even for plain $MAPbI_3$, i.e. the MAI to $PbI_2$ ratio during film preparation.[144] This case is further laid out by Hawash *et al.*, who produced MAI-rich surfaces of $MAPbI_3$ by evaporating a thin MAI layer on top of $MAPbI_3$ films before the deposition of a spiro-OMeTAD HTL.[276] In that study, the decrease of the IE of $MAPbI_3$ by excess MAI was determined directly by PES and, assuming no changes upon spiro-OMeTAD deposition, could be linked to changes in device performance.

Hawash *et al.* suggest that the change of the IE through the ultrathin MAI layer would enhance hole extraction due to "staircase energy level alignment", reducing the offset between MAPbI$_3$ VBM and the HOMO level of spiro-OMeTAD. In addition, they suspect that recombination at this interface could be reduced due to the decreased number of electrons at the interface because of slight band bending at the top of the MAPbI$_3$/MAI layer.[276]

More generally, however, the energy level alignment at the HaP / HTL interface is not explicitly measured, mostly due to the inaccessibility of the buried interface. The energy level diagram instead relies on literature values of the isolated compounds, which could differ, depending on the actual interface formation. For instance, the close vicinity to a strong p-dopant (e.g. MoO$_3$) or direct doping through additives in the HTM would lead to a different, unknown energy level alignment to the HaP film.

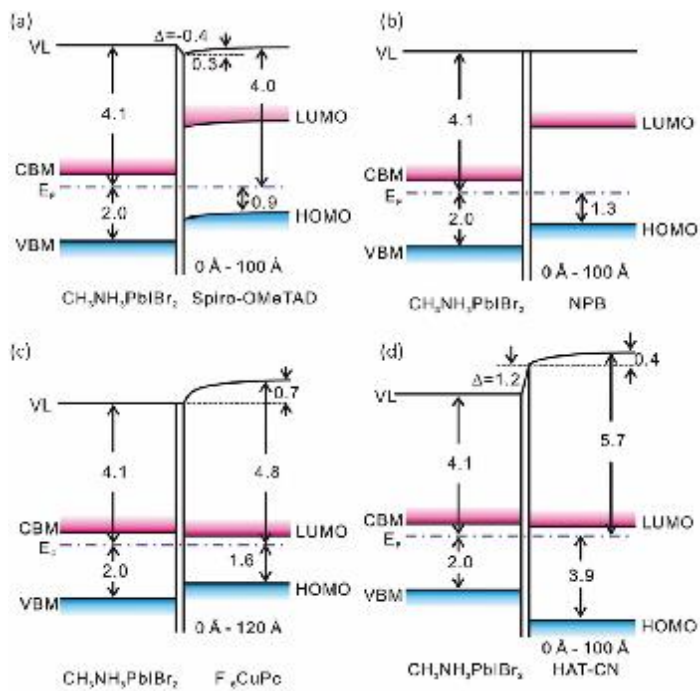

*Figure 32* – *Energy level alignment determined by UPS measurements of incrementally deposited HTLs on top of MAPbIBr$_2$. (a) Interface dipole formation and HOMO level shift for spiro-OMeTAD, (b) vacuum level alignment for NPB, (c) vacuum level alignment with subsequent vacuum level shift for F$_{16}$CuPC, and (d) interface dipole formation and vacuum level shift for HAT-CN. Reprinted wih the permission from Ref. 277. Copyright 2015 WILEY-VCH Verlag GmbH & Co. KGaA, Weinheim.*

Under this premise, it is worth examining cases of energy level alignment between HaPs and organic CTLs that deviate from the above-mentioned guidelines (i) – (iii). In that regard, another set of PES measurements for sequentially grown organic semiconductor layers, which cover a large range of IE and EA values, on top of MAPbIBr$_2$ conducted by Wang *et al.* extend the initial picture.[277] In a first example, the energy level alignment (pictured in figure 32a) between evaporated spiro-OMeTAD and MAPbIBr$_2$ exhibits a different trend compared to the results reported by Schulz *et al.* for MAPbI$_3$.[145] The first 0.2 nm of spiro-OMeTAD reduces the work function down to 3.7 eV from a value of 4.1 eV for plain MAPbIBr$_2$, which Wang *et al.* assume to be related to charge transfer, chemical bonding, surface rearrangement, interface states, image effect, or permanent dipole orientation. The spiro-OMeTAD layer subsequently exhibits an upshift of the HOMO level as the thickness is increased.[277] In a second example, Wang *et al.* investigated the interface between the HTM N,N'-di(naphthalene-1-yl)-N,N'-diphenylbenzidine (NPB) and MAPbIBr$_2$ (Figure 32b), and observed the expected vacuum level alignment and absence of band bending in both layers. In Wang's third example, the growth of a copper-hexadecafluorophthalocyanine (F$_{16}$CuPc) film on MAPbIBr$_2$ was probed. Given the heavy fluorination of the molecule, its EA (~4.6 eV) is about 0.5 eV larger than the work function of the HaP layer (deposited on TiO$_2$). Hence, Wang *et al.* expect a net electron transfer from the perovskite to the organic layer and the LUMO level of the F$_{16}$CuPc to be pinned to the CBM of the HaP. As the thickness of the F$_{16}$CuPc film is increased, they observe no band bending on either side of the interface (Figure 32c). Instead they find a gradual change of the vacuum level position as the thickness of the F$_{16}$CuPC layer increases until the equilibrium EA of F$_{16}$CuPC is attained at a sufficient distance from the interface to the HaP film. This interface formation is similar to that between MAPbBr$_3$ and PTCDI-C$_1$, for which the EA of the organics (4.2 eV) is also larger than the work function of the perovskite deposited on a TiO$_2$ substrate (4.0 eV).[145] Finally, Wang *et al.* probed the interface between MAPbIBr$_2$ and 1,4,5,8,9,11-hexaazatriphenylene hexacarbonitrile (HAT-CN). In this series of organic charge transport materials, HAT-CN marks the most extreme case with the largest EA (~5.6 eV), a high work function electron acceptor material, which is commonly used at the hole extraction (or hole injection for light emitting devices, respectively) contact in organic electronics.[278] As a consequence, a significant vacuum level offset (1.2 eV) is measured at the MAPbIBr$_2$ / HAT-CN interface, indicating the formation of a pronounced interface dipole (Figure 32d). As in the case of F$_{16}$CuPC, Wang *et al.* observe signs of Fermi level pinning to the LUMO level of HAT-CN. Beyond the interface, the incremental deposition of HAT-CN shows no shift of either the HOMO level onset, which is now positioned at higher binding energies about 2 eV below the VBM of the HaP, or of the core levels, measured by XPS. However, the vacuum level is gradually shifted by an additional 0.4 eV for a HAT-CN layer thickness of 10 nm until thermodynamic equilibrium of the free HAT-CN surface is re-established. In summary Wang *et al.* conjecture that, if accompanied with a deep-lying HOMO level (e.g. for HAT-CN) the functionality for hole extraction would be impaired due to the formation of an interfacial barrier and hence a region with increased electron–hole recombination.[277]

Overall, we note that the observed trends for the energy level alignment at the HaP/organic CTL interface is not conclusive and hence setting up general guidelines remains elusive. We note, that the differing accounts on the energy level alignment between various studies could be linked to differences in the interface reactivity and related chemical processes. For instance, in the study of Wang *et al.* the fact that band bending in the HaP layer is found upon interface formation with spiro-OMeTAD, but not with other CTLs, could be related to differences in the surface conditions of the MAPbIBr$_2$ film.[277] An indication of such variations is given by the slightly different shapes of the VB DOS, similar to deviations observed for MAI- or PbI$_2$-rich MAPbI$_3$,[144] as well as by different contributions of metallic lead to the XPS Pb 4f core level signals. This Pb$^0$ component could be related to faintly visible interface states at low binding energies when the first organic layers are deposited. But before we turn to a closer examination of possible interactions with surface states, we first look into additional examples of energy level alignment processes for fullerene-based ETL's as commonly used in inverted PSCs.

*Energy level alignment to fullerene-based ETLs:*

As introduced in section C.1. the inverted PSC structure is often realized by depositing a fullerene-based ETL on the HaP absorber film,[205] a configuration which has been suggested to assist in passivating trap states, and hence to reduce recombination.[253] Fullerene-based organic semiconductors usually function as acceptor molecules. A first snapshot of the electronic structure at the interface was obtained for solution-processed PCBM layers on MAPbI$_3$ films on an ITO/glass substrate. By diluting the PCBM solution, Lo and coworkers were able to realize PCBM films with approximate layer thickness of 5 Å and 100 Å (see Figure 33).[279] The PES measurements of this interface reveal that, while no Fermi level pinning occurs (the EA of PCBM is smaller than the work function of MAPbI$_3$ on ITO), the vacuum level is shifted from ~4.9 eV to 4.6 eV upon deposition of the PCBM layer. In the same study, C$_{60}$ films were evaporated on MAPbI$_{3-x}$Cl$_x$/ITO/glass samples, yielding a similar offset between the vacuum levels. We note that in both cases the VBM of the HaP film was derived from a linear extrapolation of the leading edge in the UPS valence band data, which means that the position of $E_F$ could be closer to a mid-gap position if the low DOS were to be accounted for.[101]

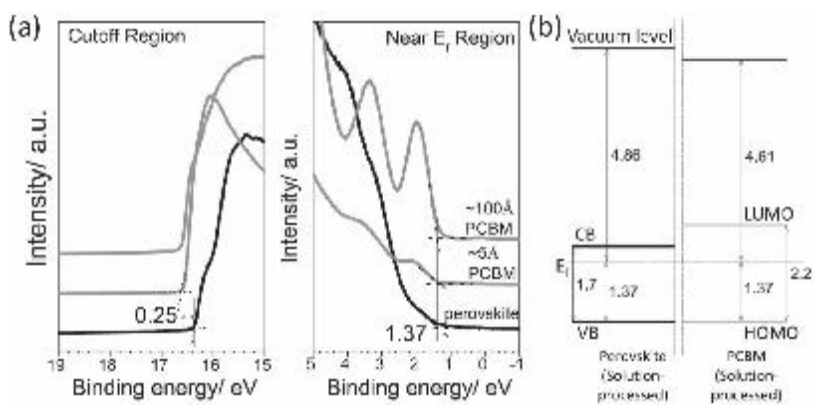

***Figure 33*** *– (a) UPS spectra with secondary electron cutoff region and valence band region, and (b) energy level diagram for the MAPbI$_3$ / PCBM interface. VBM of MAPbI$_3$ is estimated by a linear fit and unoccupied levels are projected by the band gap's found in literature. Reprinted wih the permission from Ref. 279. Copyright 2015 WILEY-VCH Verlag GmbH & Co. KGaA, Weinheim.*

In a different set of experiments, Schulz *et al.* performed a thorough PES/IPES investigation of the interface of an incrementally evaporated $C_{60}$ film on a MAPbI$_3$/NiO$_x$/ITO/glass sample (Figure 34).[38] Again, a small vacuum level offset (0.1 eV) is observed between HaP and fullerenes, indicating the presence of a small interface dipole without significant charge transfer. The onset of the LUMO level of $C_{60}$ is closely aligned with the MAPbI$_3$ CBM and remains so for thicker $C_{60}$ overlayers. With increasing $C_{60}$ layer thickness, the work function increases gradually until the equilibrium values for EA (4.1 eV) and IE (6.4 eV) of the $C_{60}$ surface are achieved. This could correspond to polarization effects as we move further from the interface to the MAPbI$_3$ film with the higher dielectric constant or originate from changes in the $C_{60}$ film morphology at different stages of growth. In line with all earlier examples of organic films grown on HaP specimens, the XPS measurements of the HaP-related core levels (Pb 4f and I 3d) show no shift in the peak centroid positions, which suggests that the HaP film does not undergo any band bending.

Wang *et al.*, who also used PES/IPES on $C_{60}$ films evaporated on MAPbI$_3$/PEDOT:PSS/ITO/glass samples, offer a differing account, in which they observe that the energy levels of the MAPbI$_3$ layer undergo a drastic change when the first 2 Å of $C_{60}$ are deposited, shifting by up to almost 1 eV to lower binding energies.[280] At the same time a shift of the $C_{60}$ HOMO level with respect to $E_F$ is observed with increasing $C_{60}$ layer thickness. They suggest that a pronounced electron transfer occurs at the interface, and propose this process could originate from the passivation of a hypothetically high density of trap states at the MAPbI$_3$ surface. Nonetheless, the discrepancy with other organic CTM/HaP interfaces, which did not show any band bending, remains and likely requires a re-evaluation considering the peculiarities and challenges of measuring the position of $E_F$ in HaP compounds, i.e. the sensitivity to X-ray radiation.[160]

*Impact of HaP surface states and trap states on the energy level alignment to an organic CTL:*

Before further conclusions on the energy level alignment at the organic CTL/HaP interface can be drawn, one should be reminded of numerous additional factors that could vary from experiment to experiment and between two nominally similar sample systems. One of the factors that pertains to all the studies presented in this section is the actual surface termination of the HaP layer, which serves as the *de facto* substrate for the growth of the organic film. Presumably the HaP films presented earlier in this section comprise a $MA^+$ moiety on the A-site and tend to be MAX-terminated, where X is the halogen species used for the HaP film fabrication (compare to section B.1.). Shin *et al.* shed more light on the energy level alignment between $MAPbI_3$ and a $C_{60}$ ETL by conducting a comprehensive PES/IPES investigation of the interface,[281] similar to the studies of Lo *et al.*, Schulz *et al.* and Wang *et al.*.[38,279,280] In their approach, however, Shin *et al.* systematically changed the surface electronic structure of the $MAPbI_3$ film: the HaP film preparation from solution is based on a two-step process, in which a $PbI_2$ film is deposited first, followed by a conversion step into $MAPbI_3$ through a subsequent treatment with a MAI solution. Changing the concentration of MAI in the second step led to a variation of the vacuum level and Fermi level positions of the resulting $MAPbI_3$ layers. Counter to common expectations, the N/Pb ratio was found to decrease for high MAI concentrations, yielding a Pb-rich surface, which is corroborated by the appearance of a small metallic lead component in the Pb 4f core level spectra,[281] apart from the $PbI_2$ precursor phase, which is observed in XRD patterns. While UV-Vis absorption measurements indicate the formation of $MAPbI_3$ in the bulk ($E_G$ = 1.6 eV), the combined UPS/IPES measurements suggest an opening of the band gap at the surface. At the same time, the $PbI_2$-rich films exhibit a shift of $E_F$ in the gap similar to earlier observations made by Steirer and coworkers.[33] Despite the suggested correlation between IE and surface stoichiometry,[144] the IE did not change for the different $MAPbI_3$ films produced in this two-step process. We remind the reader that discrepancies between the studies, but also observed trends within one study, such as the band gap opening, could be affected by the challenging determination of band edges in the PES measurements and, even more so, in the IPES measurements (see section B.2.1.). Here, Shin *et al.* used the differently prepared $MAPbI_3$ surfaces with different Fermi level position in the gap (VBM onset with respect to $E_F$ at 1.4, 1.3 and 1.0 eV, respectively) to grow $C_{60}$ layers on top and investigate the interface electronic structure (summarized in Figure 34c).[281] The data seem to indicate that only for the sample with the intermediate concentration of MAI, i.e. VBM at 1.3 eV with respect to $E_F$, the LUMO level of the $C_{60}$ is well aligned to the CBM of the MAPbI$_3$. In the other cases either a small energy level mismatch and/or interface dipole formation on the order of up to 0.2 eV is found, which is accompanied by band bending in the $C_{60}$ layer to compensate the charge transfer at the interface. In the same study, PSC devices were fabricated in this inverted device architecture and with the various MAI concentrations used in the interface investigation. In fact, the device parameters, i.e. PCE and hysteresis index, were best when no mismatch of energy levels at the interface was observed in the concomitant PES analysis. Shin *et al.* attribute this finding to charge transfer due to the band offsets between HaP photoabsorber and the organic CTL,[281] which we note, points to a variation in the interface state densities at the interface. This in turn would affect recombination and drive interfacial chemistry similarly to the case described for the interface between HaP layer and TCO substrate in section C.1.

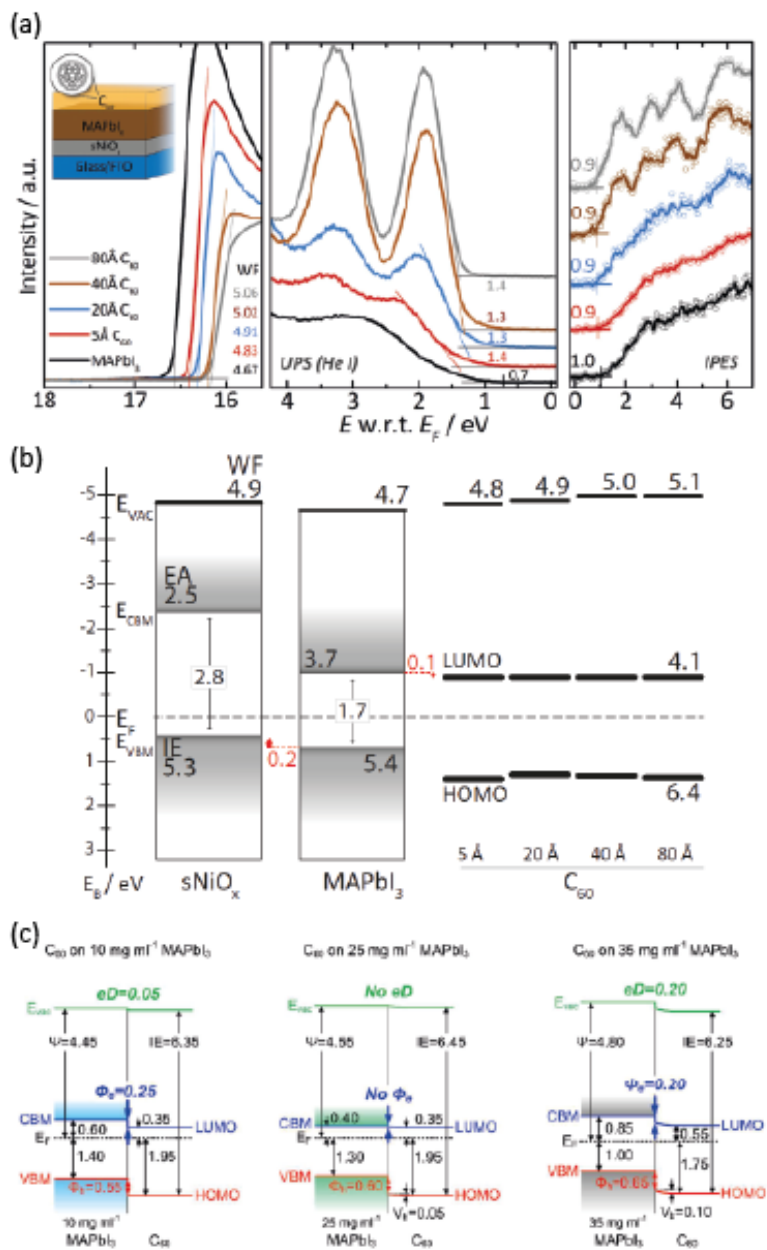

*Figure 34* – (a) UPS and IPES spectra and (b) energy level diagram of incrementally grown $C_{60}$ films on a MAPbI$_3$/NiO$_x$/FTO/glass sample. Reprinted wih the permission from Ref. 38. Copyright 2015 WILEY-VCH Verlag GmbH & Co. KGaA, Weinheim. (c) Energy level diagram from PES/IPES measurements of $C_{60}$ films that are incrementally evaporated on MAPbI$_3$ films from different two-step solution processes. For each MAPbI$_3$ sample, the concentration of MAI used in the second step to convert PbI$_2$ into MAPbI$_3$ was varied. Reprinted with the permission from Ref. 281. Copyright 2017 American Chemical Society.

A targeted approach to evaluate the effect of surface states on the HaP semiconductor and trap state densities on the energy level alignment with adjacent organic CTLs has been pursued by Zu and coworkers.[190] As noted earlier, the group had described the formation of high surface state densities and the formation of substantial amounts of Pb$^0$ as they illuminated their MAPbI$_{3-x}$Cl$_x$ samples in the vacuum chamber.[164] In the next step, they used this treatment to precondition the surface as pristine, i.e. MAI-terminated, or highly defective after illumination, i.e. comprising a significant amount of metallic lead. The surfaces are hence denoted as low density of state surfaces (low DoSS) or high density of states surfaces (high DoSS), respectively. Subsequently, and analogous to the study by Wang *et al.*,[277] the acceptor molecule HAT-CN was deposited in incremental steps to monitor the energy level alignment by PES.[190] In contrast to the findings by Wang *et al.*, Zu *et al.* report that the core levels of the HaP immediately shift to lower binding energy upon formation of a very thin (4 Å) overlayer of HAT-CN on top of the MAPbI$_{3-x}$Cl$_x$ sample. This corresponds to a shift of $E_F$ from close to the CBM to a mid-gap position at the HaP surface.

The deviation from Wang's study, where the Fermi level in the perovskite layer remained close to the CBM even after deposition of the HAT-CN, can be explained by the different substrates used in the two cases and the respective position of the Fermi level in the HaP gap, as described in section C.1.2..[38] Wang et al. produced their MAPbIBr$_2$ film on top of ZnO, bringing the Fermi level close to the CBM throughout the HaP layer,[277] while the MAPbI$_{3-x}$Cl$_x$ film in Zu's study was produced on a PEDOT:PSS film, which presumably leads to a mid-gap position of $E_F$. At the surface, however, the presence of Pb$^0$-related donor states in the upper part of the gap pins the Fermi level close to the CBM, leading to a downward band bending (Figure 35b).[164] In the low DoSS case, the layer of HAT-CN acceptors compensates the donors, empty these states and unpins the Fermi level, restoring the flat-band condition.[190] The case is slightly more complex for the illuminated high-DoSS surface, as tracked in the PES data in Figure 35a: while the amount of $Pb^0$ compared to $Pb^{2+}$ is decreased with increasing HAT-CN film thickness, i.e. the reduced Pb is re-oxidized, the shift of the $Pb^{2+}$ core level is minimal. This means that the donor state density in the HaP section of the interface is too large and pins $E_F$, hence not returning the layer back to flat band condition. The charge transfer mechanism between the $Pb^0$ donor states and the HAT-CN molecule is further evidenced by the appearance of an additional state in the valence band spectra centered around 1 eV below $E_F$ and denoted L*, which corresponds to the partially filled LUMO state. The two cases are summarized in the energy level diagrams in Figure 35b: For a low-DoSS HaP surface a strong organic acceptor molecule can induce upward band bending which results from the saturation of donor like defects, while a high-DoSS HaP surface maintains its n-type character as $E_F$ is pinned to the CBM.[190]

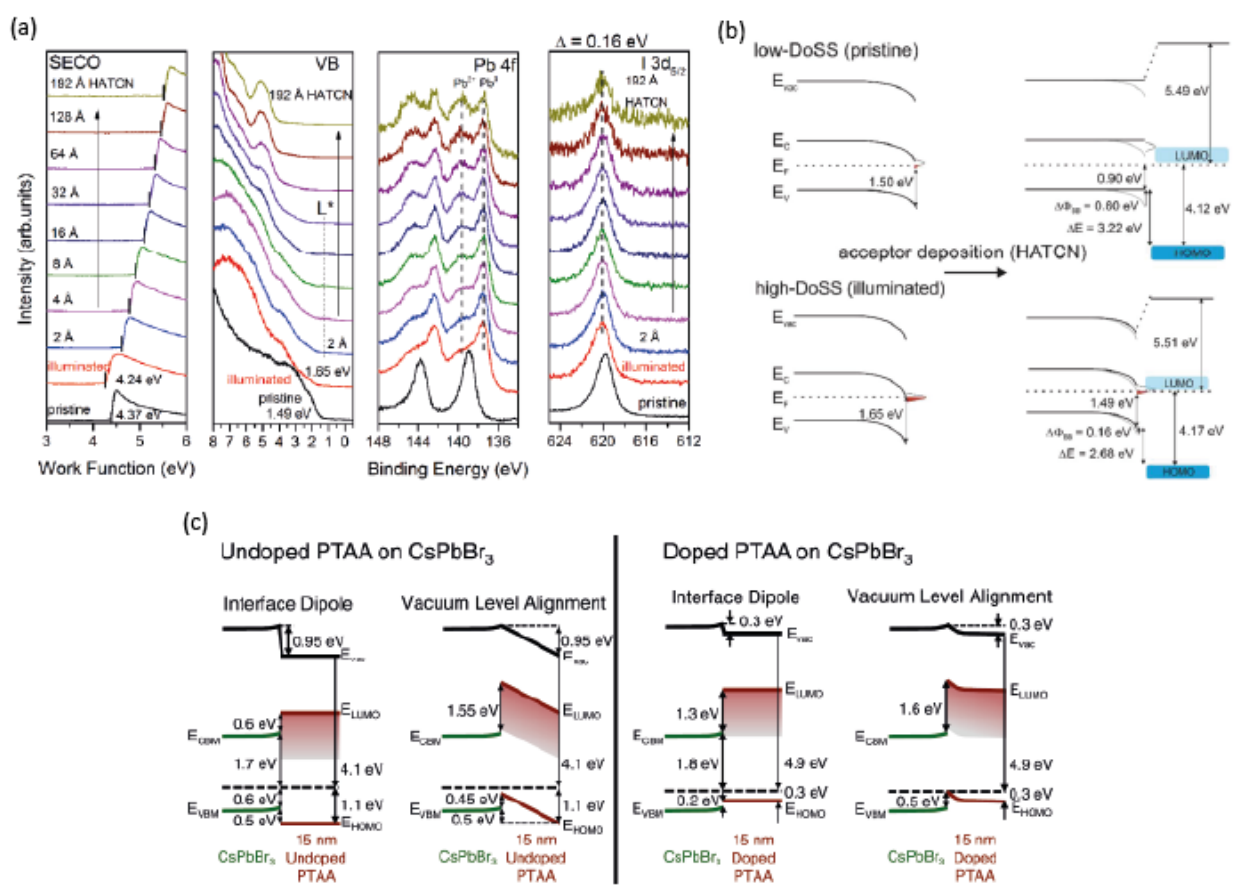

Figure 35 – (a) PES spectra of incrementally evaporated layers of HAT-CN on an illuminated MAPbI$_3$ sample, showing the secondary electron cutoff region, the valence band region of the UPS spectra and the Pb 4f and I 3d core levels of the XPS spectra. (b) Interfacial energy level diagrams, derived from the PES data for HAT-CN films grown on pristine (low-DoSS) and defective (high-DoSS) MAPbI$_{3-x}$Cl$_x$ samples. Reprinted with the permission from Ref. 190. Copyright 2017 American Chemical Society. (c) Proposed energy level alignment between undoped (left) and doped (right) PTAA on CsPbBr$_3$ assuming either vacuum level alignment or the formation of an interface dipole. Reprinted figure with permission from Ref. 282. Copyright 2017 by the American Physical Society.

Finally, we need to highlight the fact that all interfaces discussed above were formed on MA-based compounds, which are supposedly more unreactive than those of inorganic semiconductors, due to the predominant organic surface termination of the HaP film, except for the high-DoSS cases pictured above. Hence, generalization of the findings for all-inorganic HaPs require some additional scrutiny. For instance, $CsPbBr_3$ has been reported to exhibit a non-negligible density of states in the band gap tailing below the CBM, in contrast to its hybrid organic inorganic counterpart, $MAPbBr_3$.[101] In a follow-up investigation, Endres *et al.* produced $CsPbBr_3$ films on ITO substrates on which they deposited a 15 nm HTL of undoped as well as p-doped poly[bis(4-phenyl)(2,4,6-trimethylphenyl)amine] (PTAA).[282] A shift in the Br 3d and Cs 3d core level positions by approximately 0.1 eV after the deposition of either HTL on the $CsPbBr_3$ film indicates a slight upshift of $E_F$ in the gap for the HaP. The VBM of the perovskite shifts from 0.7 eV to 0.6 eV below $E_F$ with the deposition of the undoped PTAA layer and to 0.5 eV below $E_F$ when brought into contact with the p-doped PTAA film.

The PTAA HOMO is found at 1.1 eV below $E_F$ for the undoped layer and 0.3 eV for the doped layer. However, in this experimental configuration, the buried HaP/polymer interface was not directly assessed as a sufficiently thin HTL layer could not be produced. Thus, the energy level alignment could only be proposed as either the formation of an interface dipole or vacuum level alignment as illustrated in Figure 35c. Most likely, it is a combination of band bending and interface dipole formation that leads to a barrier for hole extraction (0.5 eV) from the $CsPbBr_3$ film in the case of the undoped PTAA, whereas doping of the PTAA film removes this barrier and raises the VBM in the HaP layer.[282] In this example of a HaP/PTAA interface, it cannot be ruled out that the band bending observed in the perovskite film is related to the passivation of the $CsPbBr_3$ surface states.

In conclusion we can formulate two key rules concerning the energy level alignment at the interface between an HaP film and an organic CTL on top:

- For a low-DoSS HaP layer and in the absence of surface band bending due to surface states, the interface formation to most organic layers (but not for strong electron acceptors) occurs under vacuum level alignment or by the formation of only a small interface dipole. If the EA of the organic film is larger than the WF of the HaP or the IE smaller than the WF of the HaP, Fermi level pinning can occur. Strong acceptor molecules however can cause a large interface dipole.
- For an HaP layer with band bending due to surface states (observed here: downwards due to $Pb^0$ donor states), the interface formation can be accompanied by a neutralization of these states, e.g. through the deposition of strong acceptor molecules in case of donor state defects. However, for a high-DoSS HaP layer, even the deposition of a strong acceptor molecule cannot alleviate the Fermi level pinning at the HaP/CTL interface.

These broad rules leave a large window of potential organic CTL materials that do not need to fulfill narrow requirements for an exact matching of the HaP's IE or EA. *Instead, the quality and surface termination of* *the HaP layer itself predefines the regime of energy level alignment at the interface.* The most important role for interfacial charge transfer lies in the saturation of interfacial defect states, which in many cases is either a minor issue due to the benign character of many HaP surfaces or could happen incidentally through the dopant molecules in the organic CTL. Hence, the success of alternative, specifically engineered organic hole transport materials,[283,284] could be due to secondary properties (e.g., being dopant-free, improved conductivity, morphology, moisture resistance) as long as the bounds to avoid the formation of a charge extraction barrier through interfacial energy level alignment are preserved, i.e. $IE_{HTL} < IE_{HaP}$.

**C.2.2. Inorganic semiconductor overlayers: Energy level alignment, reactivity, and band bending**

Following the discussion of the energy level alignment between HaP and organic CTLs, we turn to inorganic charge transport materials deposited on HaP surfaces. According to the materials described in section C.1.1. regarding chemistry at the HaP/substrate interface, more pronounced chemical reactions would be expected for the growth of transparent conductive oxide CTLs than for organic ETLs and HTLs on the HaP film.

Indeed, the increased reactivity makes it difficult to find examples of straightforward growth of a TCO CTL on a HaP semiconductor film that led to a functional device. Instead, a range of alternative inorganic hole transport materials was proposed and successfully tested for top layer CTLs in the conventional PSC device layout. These successes were primarily realized with the p-type inorganic salts copper iodide and copper thiocyanate (CuSCN).[285,286] Both compounds had been employed previously in DSSC research and hence the approach was easily applied to the PSC research field. However, the energy level diagram of the device structures was estimated from the energy levels of the isolated compounds (see Figure 36), and to date reports on measuring the actual energy level alignment remain sparse.[287,288] From the proposed energy diagrams, however, hole extraction should be as efficient as for the organic HTLs discussed in the previous section, since the IEs of CuI (5.2 eV) and CuSCN (5.3 eV) are smaller than that of the MAPbI$_3$ photoabsorber. However, this leaves open questions regarding the chemical reactivity of these interfaces. In the case of CuSCN, the claim was made that this p-type hole conductor could passivate trap states in the HaP. The claim was corroborated by optical (PL) and electrical (space charge limited current, SCLC) measurements for MAPbI$_{3-x}$Cl$_x$ films with CuSCN as an additive in the HaP precursor solution, in an inverted cell architecture on top of ITO.[289] Ye *et al.* further investigated the passivating capabilities of these Cu-salt HTMs and found that Cu(thiourea)I (Cu(Tu)I) could produce a PSC with high performance parameters, which they attribute to a trap passivating mechanism as depicted in Figure 36c.[290] While the Cu(Tu)I was used as an additive to the HaP precursors solution and is thus presumably incorporated at the grain boundaries in the HaP film, the mechanism could also be exploited to passivate trap states at HaP surfaces, and at interfaces with charge extraction layers and electrodes. Since neither of these inorganic HTMs would act as a strong electron acceptor on its own, a more detailed picture of the complexation chemistry at the interface would be required to better assess their role for surface passivation.

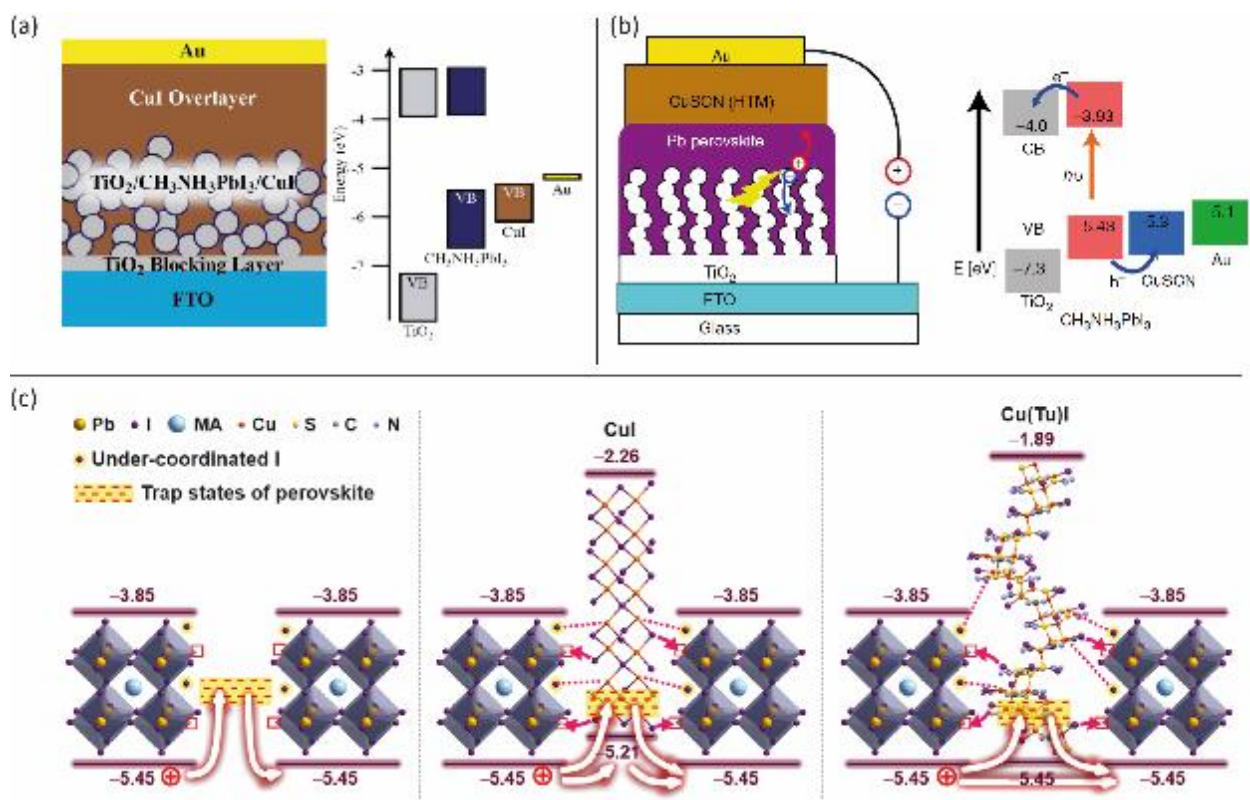

*Figure 36* – *Inorganic HTL on HaP absorber films in PSCs. Schematic of the conventional device layouts and energy level diagrams for devices with (a) CuI and (b) CuSCN as HTM. (a) Reprinted with the permission from Ref. 285. Copyright 2014 American Chemical Society. (b) Reprinted with permission from Ref. 286. Copyright 2014 by Springer Nature Publishing AG. Proposed model for the passivation of trap states in MAPbI$_3$ by CuI and Cu(thiourea)I (= Cu(Tu)I). Reprinted with the permission from Ref. 290. Copyright 2017 American Chemical Society.*

*The molybdenum oxide / HaP interface:*

Considering oxide-based HTMs, a natural first choice is molybdenum trioxide, which can be deposited on the HaP by evaporation. $MoO_3$ has found widespread application in organic electronics, where it is used to p-dope the organic semiconductor or is introduced as a thin layer between organic semiconductor and electrode to improve the extraction of holes.[291] These electronic mechanisms are rooted in the strong electron acceptor properties of $MoO_3$ (EA = 6.7 eV, WF = 6.8 eV, IE = 9.7 eV), which could play out similarly in HaP semiconductors. However, we recall the chemistry at the interface formed upon evaporation of $MAPbI_3$ on a $MoO_3$ substrate, presented in section C.1.1.: during layer formation, the HaP precursors decomposed and volatile compounds were desorbed. This led to a $PbI_2$-rich phase in the interface layers (<10 nm) close to the $MoO_3$.[187]

Wang *et al.* monitored the growth of evaporated $MoO_3$ on $MAPbIBr_2$ by PES measurements.[277] The UPS data are plotted in figure 37a. As the first $MoO_x$ layer (2 Å) is deposited on the HaP, gap states form, centered around ~0.8 eV and ~1.9 eV below $E_F$ ($MoO_3$, $E_g$ = 3 eV; while $MoO_2$ is a metallic compound). These states are commonly observed in $MoO_3$ layers and originate from an oxygen deficiency in sub-stoichiometric $MoO_x$ comprising $Mo^{6+}$ and $Mo^{5+}$ species as evidenced by XPS measurements. While Wang *et al.* were able to determine the formation of a strong interface dipole, no band bending occurred in the $MoO_3$ layer with increasing thickness. Furthermore, no distinct trend in core level shifts could be found in the MAPbIBr₂ film as a function of MoO₃ layer growth, and hence no clear assessment of band bending in the HaP could be made. However, in another instance, core level shifts in an MAPbI₃ film were observed after the evaporation of a MoO$_3$ layer, consistent with downward band bending in both layers, which is incompatible with a simple charge transfer process, but indicative of complex interface chemistry.[292]

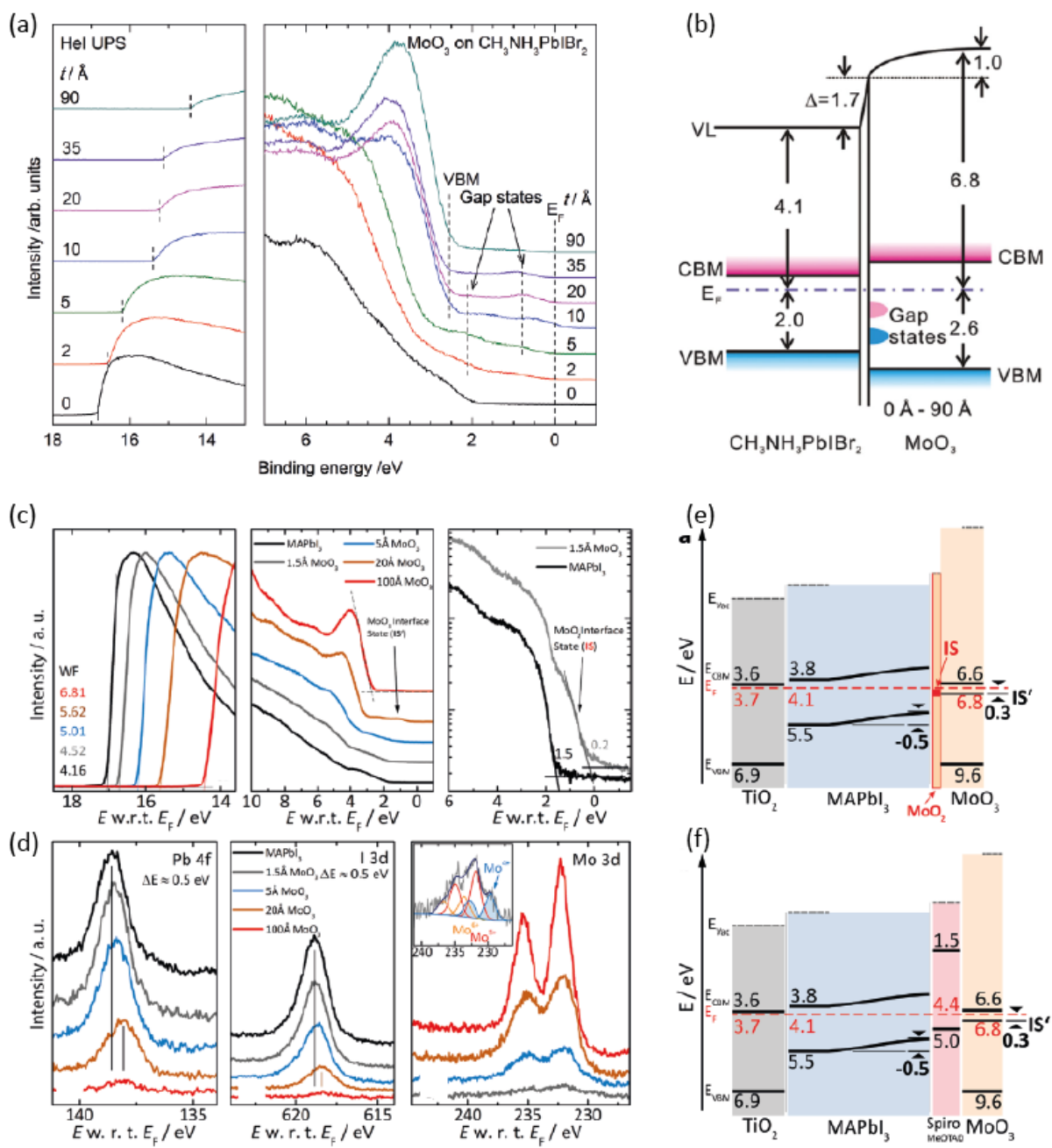

*Figure 37* – Incremental growth of MoO₃ on HaP films. (a) UPS data and (b) energy level diagram for MoO₃ on MAPbIBr₂. Reprinted wih the permission from Ref. 277. Copyright 2015 WILEY-VCH Verlag GmbH & Co. KGaA, Weinheim. (c) UPS (SECO region, VB region and close-up of VB on semi-logarithmic scale) and (d) XPS core level measurements (Pb 4f, I 3d and Mo 3d with inset showing a magnification for the 1.5 Å MoO₃ layer) for MoO₃ on MAPbI₃. Energy level diagrams for (e) MoO₃/MAPbI₃/TiO₂ and (f) MoO₃/spiro-OMeTAD(3nm)/MAPbI₃/TiO₂ samples. Reprinted with the permission from Ref. 188. Copyright 2016 American Chemical Society.

A closer examination of the HaP/MoO$_3$ interface chemistry and concomitant band bending in the perovskite was obtained by evaporating MoO$_3$ either directly on a pristine MAPbI$_{3-x}$Cl$_x$ surface or on the surface protected by an ultrathin (~3 nm) evaporated spiro-OMeTAD buffer layer.[188] The UPS measurements of the incrementally grown MoO$_3$ on top of the pristine HaP film are presented in Figure 37c and confirm the key findings from Wang *et al.*, i.e. an interface dipole is formed and interface states in the gap below the conduction band edge (onset at 0.2 eV below $E_F$) in the MoO$_3$ suggest a pronounced interfacial charge transfer and undercoordination of the molybdenum atoms. In contrast to previous findings, XPS measurements reveal a significant upward band bending in the HaP layer underneath the growing MoO$_3$ overlayer. Examination of the Mo 3d core level for the 1.5 Å thin MoO$_3$ film (Inset in the right panel of Figure 37d) yields more insight into the details of the chemical reaction at the HaP/MoO$_3$ interface. The emergence of Mo$^{4+}$ states are evidence for a reduction of the overlayer to MoO$_2$ and can chemically be explained by a Pb$^{2+}$ → Pb$^{4+}$ oxidation and/or 2I$^-$ → I$_2$ oxidation accompanying the Mo$^{6+}$ → Mo$^{4+}$ reduction. A summary over the energetics of the resulting interface is given in the energy diagram in Figure 37e. Note that the spatial extent of band bending in the HaP film, starting at the HaP/MoO$_3$ interface, cannot be further quantified in this measurement. While the magnitude of the Fermi level shift towards a mid-gap position at the immediate contact is well determined (ΔE = 0.5 eV), no further information is gathered with respect to the width and progression of the depletion zone into the bulk of the MAPbI$_{3-x}$Cl$_x$ film. Since the HaP film on a TiO$_2$ substrate is generally found to be n-type,[37,38] the scenario presented here could differ from the case of HAT-CN grown on the MAPbI$_3$/PEDOT:PSS samples discussed in section C.2.1..[190] In the latter case, the strong electron acceptor HAT-CN neutralized Pb$^0$ surface states of the HaP, returning the film to flat-band condition. In the MoO$_3$/MAPbI$_{3-x}$Cl$_x$/TiO$_2$ layer stack presented here, however, the HaP film would undergo band bending, bringing the Fermi level closer to mid-gap at the interface with $MoO_3$.

The energy diagram for the $MoO_3$/spiro-OMeTAD/$MAPbI_{3-x}Cl_x$/$TiO_2$ system featuring a ~3 nm thick evaporated spiro-OMeTAD buffer layer between $MoO_3$ and HaP is captured in Figure 37f. Interestingly, one observes the same band bending as in the previous case, but the reduction to $MoO_2$ at the interface is successfully inhibited, as evidenced by the XPS Mo 3d core level analysis.[188] Hence, the shift of $E_F$ at the HaP interface appears mainly driven by the high work function of the $MoO_3$ film. Note that the oxide is still sub-stoichiometric ($Mo^{5+}$ alongside $Mo^{6+}$ species) at the interface, as charge transfer to the adjacent organic/$MoO_x$ bi-layer takes place. However, when used in PSC devices, neither configuration improves the device parameters. Both interfaces lead to a significant deterioration of the performance, even though the ultrathin organic buffer helps mitigate interfacial charge carrier recombination. This situation underlines the point that, despite an energy level alignment that would nominally be favorable for carrier extraction at a PSC interface, the formation of interface defects plays a detrimental role in terms of device engineering. For comparison, in a more exotic device layout, the approach has been demonstrated to facilitate a narrow depletion region in a lateral perovskite heterostructure, i.e. via modulation doping on $TiO_2$/$MoO_3$ heterojunctions.[293] The results further explain why no prominent example for a high-performance PSC can be found that comprises an intimate HaP/$MoO_3$ contact. Nonetheless, $MoO_3$ can be employed successfully if well separated from the HaP layer in the device layout, but then rather serves a role to improve injection at the organic CTL/electrode interface or as a moisture barrier to improve the device stability.[77,294,295]

*Chemical reactivity in oxide overlayer growth processes:*

More generally, the application of oxide CTLs, which additionally protect against external stresses (mostly moisture) on the HaP layer, found more wide spread application, but usually requires a physical barrier (i.e. an inert organic layer) to the underlying HaP layers.[264,296] Aside from the reactivity of the oxide film, a major difficulty lies in finding deposition methods that are compatible with the HaP layer. Notable exceptions, including "soft" growth approaches such as low temperature evaporation, have been applied. Zardetto et al. give a comprehensive account on oxide thin film growth by atomic layer deposition (ALD) on HaP surfaces.[297] The ALD process relies on self-limiting reactions between the substrate surface and an organometallic precursor in the first half-cycle, and a co-reactant, often referred to as oxidizer in the second half-cycle. The oxidation in this second half-cycle can be done by $H_2O$ in a thermal ALD process or induced by an oxygen plasma in case of plasma-enhanced ALD (peALD).[298] The self-limitation of the growth and purging steps in each cycle enable a high level of control and accuracy for the thickness of the deposited layer. Key to this deposition method is the ability of the forming film to anchor to the substrate surface. While growth is usually well-behaved on an OH-terminated surface, such a scenario is seldomly found for most exposed HaP films (see section B.1.). Instead, the AX-terminated surface of the $ABX_3$ HaP could be subject to reactions with the oxidizer. In their study, Zardetto and coworkers attempted the deposition of various transition metal oxide layers, grown by peALD on $MAPbI_3$ and investigated the resulting chemistry by XPS (see Figure 38).[297] In a first pass, they compared the growth of 60 ALD cycles (about 6 nm) of $Al_2O_3$ from a trimethyl aluminum (TMA) precursor, either reacted with $H_2O$ or reacted at low-temperature in an $O_2$ plasma (i.e., in peALD). In both cases, an $AlO_x$ film was formed, but for the peALD-treated sample, no nitrogen signal of the underlying HaP was detected, indicating that the first layer of the $MAPbI_3$ film had decomposed. Such reaction was limited to the HaP surface layers, presumably during the growth of the first 2-3 cycles as the XRD patterns reveal only a small change in the intensity ratio of the observable crystalline fractions of PbI$_2$ to MAPbI$_3$. This reaction was expected given the known decomposition of MAPbI$_3$ by reactive oxygen species (see section B.1.2.), and could be further proven for peALD growth of other metal oxide (TiO$_2$, NiO and, MoO$_3$) films on MAPbI$_3$.[297] The XPS analysis of these oxide / HaP interfaces is shown in Figure 38. The peALD film growth from titanium isopropoxide (TTIP), Bis(ethylcyclopentadienyl)nickel(II), Ni(C$_5$H$_4$C$_2$H$_5$)$_2$, and bis(tert-butylimido)-bis(dimethylamido) molybdenum, (N$^t$Bu)$_2$(NMe$_2$)$_2$ precursors, led to a similar finding as in the deposition of Al$_2$O$_3$. However, in addition to the disappearance of the volatile nitrogen species and reaction products from the MA moiety, a signature peak related to oxy-iodo species emerged in the high binding energy range of the I 3d core level scans. Another attempt was made to grow ZnO from a diethyl zinc precursor and water as co-reactant. However, no persistent layer growth was achieved until after 100 cycles, attributed to the high reactivity between ZnO and MAPbI$_3$, which was already reported for MAPbI$_3$ growth on ZnO surfaces in terms of a thorough deprotonation of the MA$^+$ anion.[243]

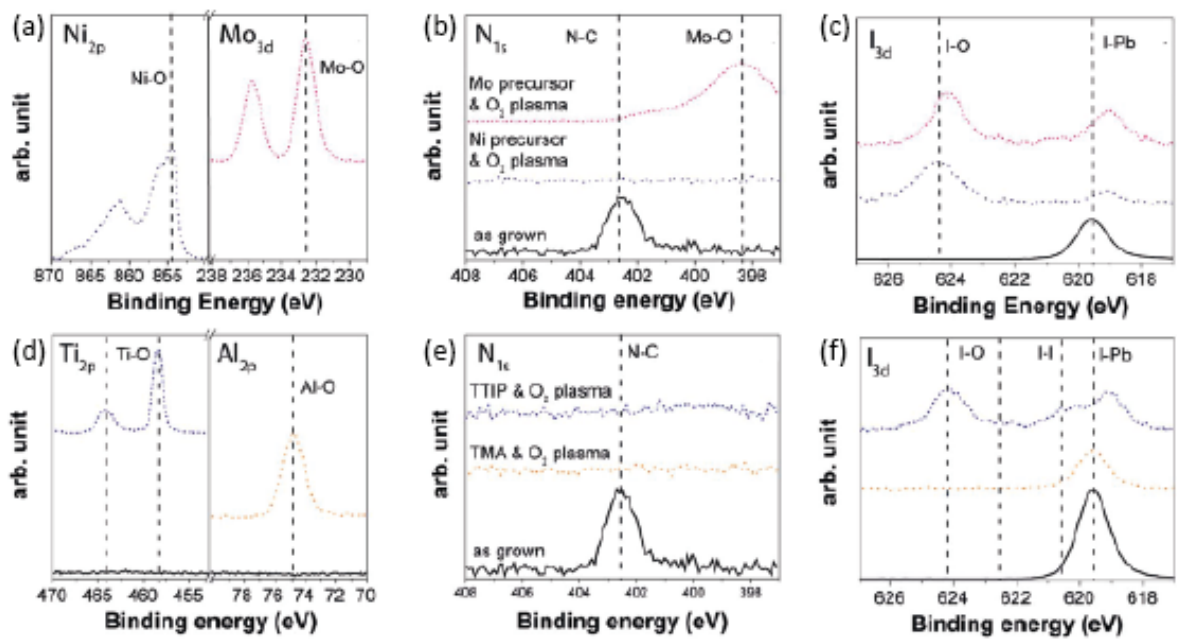

*Figure 38* – XPS analysis of transition metal oxide layers grown by peALD on MAPbI$_3$. (a,d) Ni 2p, Mo 3d, Ti 2p, and Al 2p core level regions. (b,e) N 1s core level region showing the loss of nitrogen, and hence MA. (c,f) I 3d core level region with oxy-iodo (I-O) species forming for all oxide-coated MAPbI$_3$ samples except for Al$_2$O$_3$. Color code for the curves remains the same throughout all figure panels: pink, MoO$_3$; purple, NiO; blue, TiO$_2$(TTIP), yellow, TiO$_2$(TMA), black, MAPbI$_3$ reference. Reproduced from Ref. 2017 with permission from The Royal Society of Chemistry.

Separating the damage to the HaP surface resulting from the growth process itself and from the general reactivity between the two compounds, the oxide and HaP, becomes very difficult. In fact, the two processes remain entangled for most oxide layer growth on HaPs – even for the gentle evaporation of MoO$_3$ – as the precursor phases of the respective oxides, and in particular the oxidizing agents, react with the HaP surface. In the case of MAPbI$_3$, this can be seen by the deprotonation of MA, formation of PbO$_x$ (which is not straightforward to distinguish from other Pb$^{2+}$ species within the resolution limit of most XPS experiments) and even to the extreme case of forming iodates and related metastable compounds under strongly oxidizing conditions.

### C.2.3. Conductive carbon transport layers

As an alternative to semiconducting organic or oxide charge transport materials, conductive carbon-based CTLs, such as carbon nanotubes (CNT) or graphene, carry the promise to combine mechanical and chemical robustness, with tailor-made optoelectronic properties (high conductivity and optical transmittivity). Their successful implementation in HaP devices, first and foremost in PSCs, has been demonstrated and documented in mini-reviews for carbon nanotubes CTLs,[265] as well as graphene layers.[266] Aside from their use in these nanomaterial formats, conductive carbon electrodes have been produced directly from carbon ink on HaP photoabsorbers to engineer HTM-free PSC layouts.[267] Yet, despite the successful use of a large swaths of carbon-based CTL materials and production routines for device fabrication, energy level alignment and chemical reactivity at HaP film interfaces was explored only for a few selected systems.

*CNT / HaP interfaces:*

The first instances of implementation of CNTs in HaP devices were realized in conventional PSC device layouts, where the CNT layer was deposited on the HaP and served to form an HTL. Li *et al.* laminated a free-standing CNT film from a floating catalyst chemical vapor deposition method (Figure 39a).[299] Habisreutinger *et al.*, on the other hand, used single walled carbon nanotubes (SWCNT) functionalized (i.e., individually wrapped) by the hole conducting polymer poly(3-hexylthiophene-2,5-diyl) (P3HT) in a spin-coating process and sealed the HTL by infiltrating PMMA into the SWCNT network after deposition (Figure 39b).[300] Interface formation and energy level alignment at the CNT/HaP interface were reported via PES measurements as part of an incremental growth study, in which semiconducting SWCNT films as thin as 2 nm (0.5 monolayer) were deposited by ultrasonic spray coating onto a $MAPbI_3$/$TiO_2$/FTO/glass sample.[301] Prior to deposition, the SWCNTs were wrapped by a bis(pyridine)-functionalized polyfluorene-based conjugated polymer (PFO-BPy) and sorted to be exclusively of (6,5) chirality, leading to a unique identification of valence band features in the UPS measurements shown in Figure 39c. Similar results were eventually obtained for laser-vaporized purely semiconducting SWCNT of varying chirality. The energy level alignment, projected from these measurements and summarized in Figure 39d, deviates substantially from the cases observed for organic semiconductor CTLs grown on HaP films. A small interface dipole seems to form and the Fermi level in the SWCNT layer, which is nominally intrinsic (or even slightly p-type) as seen for the reference sample on Au, is shifted closer to the CBM (C1 in Figure 39d) at the direct $MAPbI_3$/SWCNT interface. As the SWCNT layer thickness is increased, band bending is observed in the SWCNT film, which leads to a concomitant change in the vacuum level position. The effect can be explained by n-doping of the SWCNT at the immediate interface by (methyl)amine species from the HaP film. The XPS results in the study corroborate that no band bending occurs in the underlying $MAPbI_3$ layer, i.e. the Pb and I core levels remain fixed while the C and N signature of the SWCNT and co-polymer follow the band bending deduced from the UPS measurements. However, the mechanism of the interfacial doping of the SWCNT is not captured in the XPS data alone; a conjecture would be that excess methylamine on the surface could act as dopant, whereas the electronic structure of the MAPbI₃ film

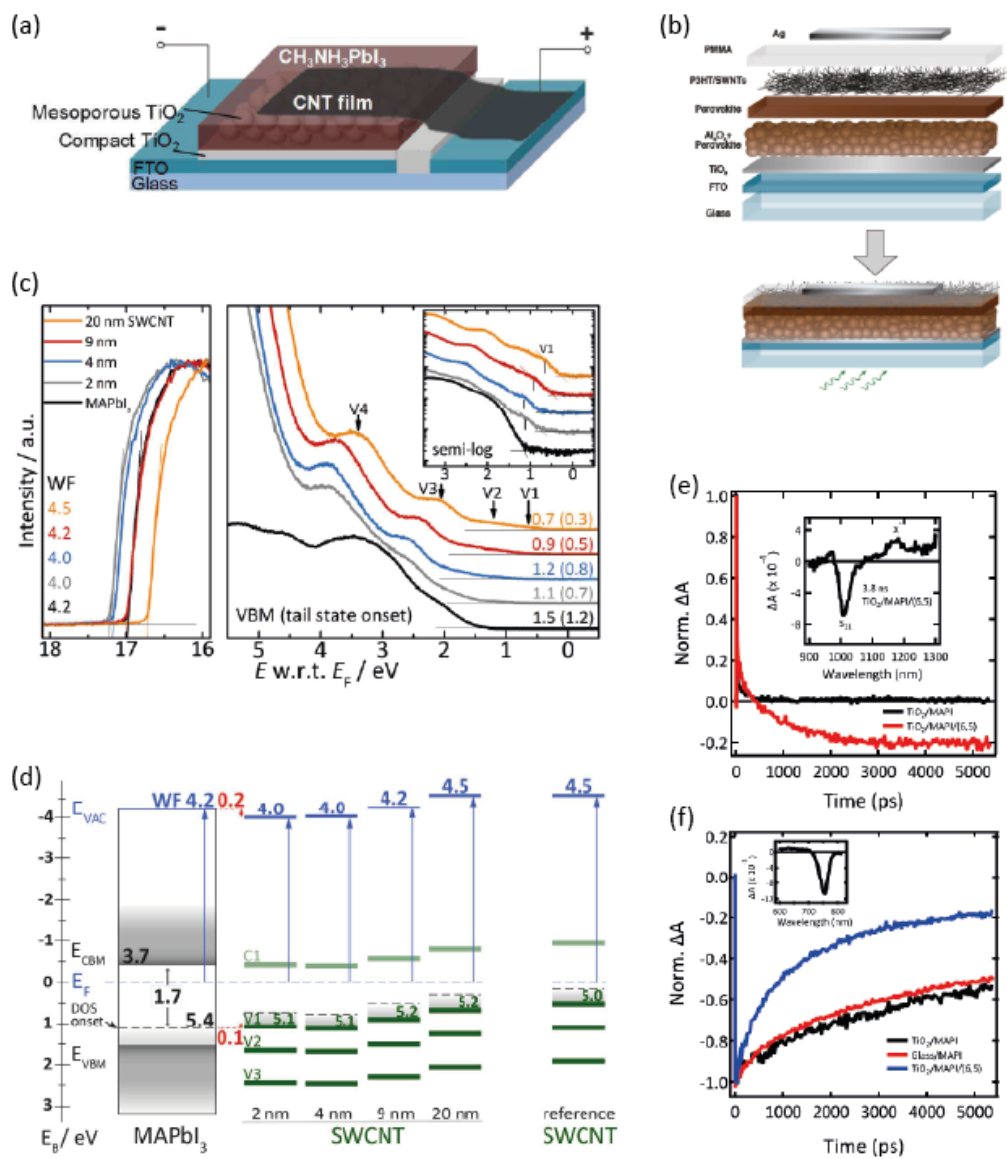

*Figure 39* – (a) Schematic of a PSC with free-standing CNT film as top electrode. Reprinted with the permission from Ref. 299. Copyright 2014 American Chemical Society. (b) Schematic of a PSC with a spin-coated HTL of P3HT-coated SWCNT and subsequent PMMA sealing. Reprinted with the permission from Ref. 300. Copyright 2014 American Chemical Society. (c) UPS data showing secondary electron cut-off and valence band region (inset in semi-log scale) of SWCNT layers deposited on MAPbI₃ by ultrasonic spray coating. V1-V4 denote valence band features (Van-Hove singularities) of the SWCNT with uniquely (6,5) chirality. (d) Energy level diagram of SWCNT on MAPbI₃ and reference SWCNT layer on Au (right). (e,f) Transients of absorption spectra of MAPbI₃ films with and without SWCNT layer of (e) the $S_{11}$ transition in the SWCNT film and (f) ground state bleach in the MAPbI₃ film. Reprinted with the permission from Ref. 301. Copyright 2016 American Chemical Society.

remains unperturbed.[301]

The interfacial energy level alignment and band bending in the SWCNT layer have direct implications on the charge carrier transfer across the SWCNT/MAPbI$_3$ interface. Transient absorption spectroscopy data are presented in Figures 39e and 39f, for the singlet ($S_{11}$) excitation on the SWCNT, which is accompanied by a trion state ($X^+$, in Figure 39e) and the ground state bleach of the band gap absorption on the MAPbI$_3$. The ground state bleach recovery of MAPbI$_3$ occurs much faster with the SWCNT transport layer deposited on top than without, indicating that photoexcited carriers are efficiently extracted. On the other side of the interface, populating the $S_{11}$ state occurs on the same timescale, which lets us assign the accelerated ground state bleach recovery to hole transfer onto the SWCNT film. The configuration was further tested in PSC devices, for which a thin SWCNT interlayer sandwiched between the MAPbI$_3$ absorber layer and a conventional spiro-OMeTAD HTL led to enhanced device performances.[302] The overall improvement is based on a combination of faster charge extraction through the SWCNT HTL, suppressed back-recombination due to the band bending in the HTL, and concerted enhanced electron extraction through the TiO$_2$ interface at the backside, which was evidenced by time-resolved microwave conductivity measurements.

With respect to a generalization of the conclusions regarding the energy level alignment at the CNT/HaP interface, many open topics remain to be addressed: (i) the interaction with other HaP surfaces, in particular with purely inorganic HaPs featuring a higher density of surface states, (ii) the energy level alignment for metallic and multi-walled CNTs, (iii) the impact of the deposition process (solvent, temperature), even though reference samples indicate only minor changes to the MAPbI$_3$ surface,[301] and (iv) the role of the co-polymer wrapping. In the case presented here, the energy levels of the PFO-BPy are thought to be well outside the gap of the SWCNT transport levels. Yet, we cannot rule out, that the polymer plays a role in the energy level alignment and interfacial chemistry between SWCNT layer and HaP film.

*Graphene / HaP interfaces:*

To date the use of graphene as a top CTL on HaPs had been explored in only a few studies. Most notable examples are the realization of PSCs with graphene oxide (GO),[303] or thiolated nanographene as top HTLs.[304] Similar to the use of CNT HTLs, devices with graphene-based HTLs worked reasonably well, i.e., performance parameters were comparable to those of state-of-the-art devices using organic HTLs at the time. However, there are no detailed experimental studies that elucidate the electronic level alignment at HaP/graphene interface. Energy level diagrams, usually assuming vacuum level alignment, suggest the absence of any hole extraction barrier, as the IE of the graphene-based compounds, used in the studies, is expected to be on the order of 5.2 for reduced graphene oxide,[303] and 5.3 eV for thiolated nanographene (perthiolated tri-sulfurannulated hexa-peri-hexabenzocoronene with twelve 4-t-butylphenyl groups).[304] This is compatible with efficient hole extraction when compared to organic semiconductor HTMs (see section C.2.1.), but these numbers disregard the energy level alignment processes as evidenced, for instance, for interfaces between HaPs and CNT-based HTLs.[301] Thus, the following example indicates that the chemistry at the HaP/graphene interface is complex and can easily dominate the electronic energy level alignment: Acik *et al.* grew $MAPbX_3$ (X = I,Br,Cl) layers, by spin-coating, on a 3-5 layer thick GO film on quartz or on $SiO_2$ on Si, and compared these samples to $MAPbX_3$ films, directly grown on $SiO_2$/Si substrates.[305] They adjusted the $MAPbX_3$ layer thickness to be a few 100 nm for a structure and morphology analysis (an example of a $MAPbBr_3$/GO sample is shown in Figure 40a,b), and modified the spin-coating procedure to fabricate 10-20 nm thick $MAPbX_3$ films to enable probing the interface region via FTIR and XPS. The most striking finding from the structural analysis – in particular XRD measurements as shown in Figure 40 – is that the $MAPbX_3$ growth is strongly dependent on the choice of the X anion in the HaP: while nucleation of $MAPbBr_3$ occurs readily, the yield for $MAPbI_3$ is very poor and only happens in a very narrow temperature window. In the study, these differences are linked to the reactivity at the GO/MAPbX$_3$ interface, with H$_2$O (here: adventitious) playing a dominant role as a mediator. FTIR and XPS data for MAPbI$_3$ growth on GO (Figure 40e-k) were used to examine the interfacial reaction and compare the resulting films to pristine MAPbI$_3$ films formed on SiO$_2$/Si (Figure 40f,g). The IR spectra reveal the presence of (HCO$_3^-$):(H$_2$O)$_n$ clusters and Pb(CO$_3$)$_2$ up to temperatures as high as 150 °C, and are consistent with the XPS data of films annealed to 130 °C. Acik *et al.* suggest that the occurrence of Pb(CO$_3$)$_2$ species can be explained by the reduction of the GO substrate driven by the MAI and PbI$_2$ precursors in an highly acidic environment, caused by HI by-products. Iodide anions are readily hydrated – e.g. through traces of water and oxygen in the precursors – and can subsequently react with the epoxides and hydroxyl groups on the GO surface in an exothermic reaction occurring at room temperature, according to their Gibbs free energy calculations.[305] This mechanism prevents the growth of MAPbI$_3$ on GO, while the lead carbonate by-products are formed. The proposition is summarized in the following reaction schemes:

$$2\ Pb_{(s)} + O_{2(aq)} + 2\ CO_3^{2-}{}_{(aq)} + 2\ H_2O \leftrightarrow 2\ PbCO_{3(s)} + 4\ OH^-{}_{(aq)}$$

$$Pb^{2+}{}_{(aq)} + I_3^-{}_{(aq)} + 2\ H_2O \leftrightarrow PbO_{2(s)} + 3\ I^-{}_{(aq)} + 4\ H^+{}_{(aq)}.$$

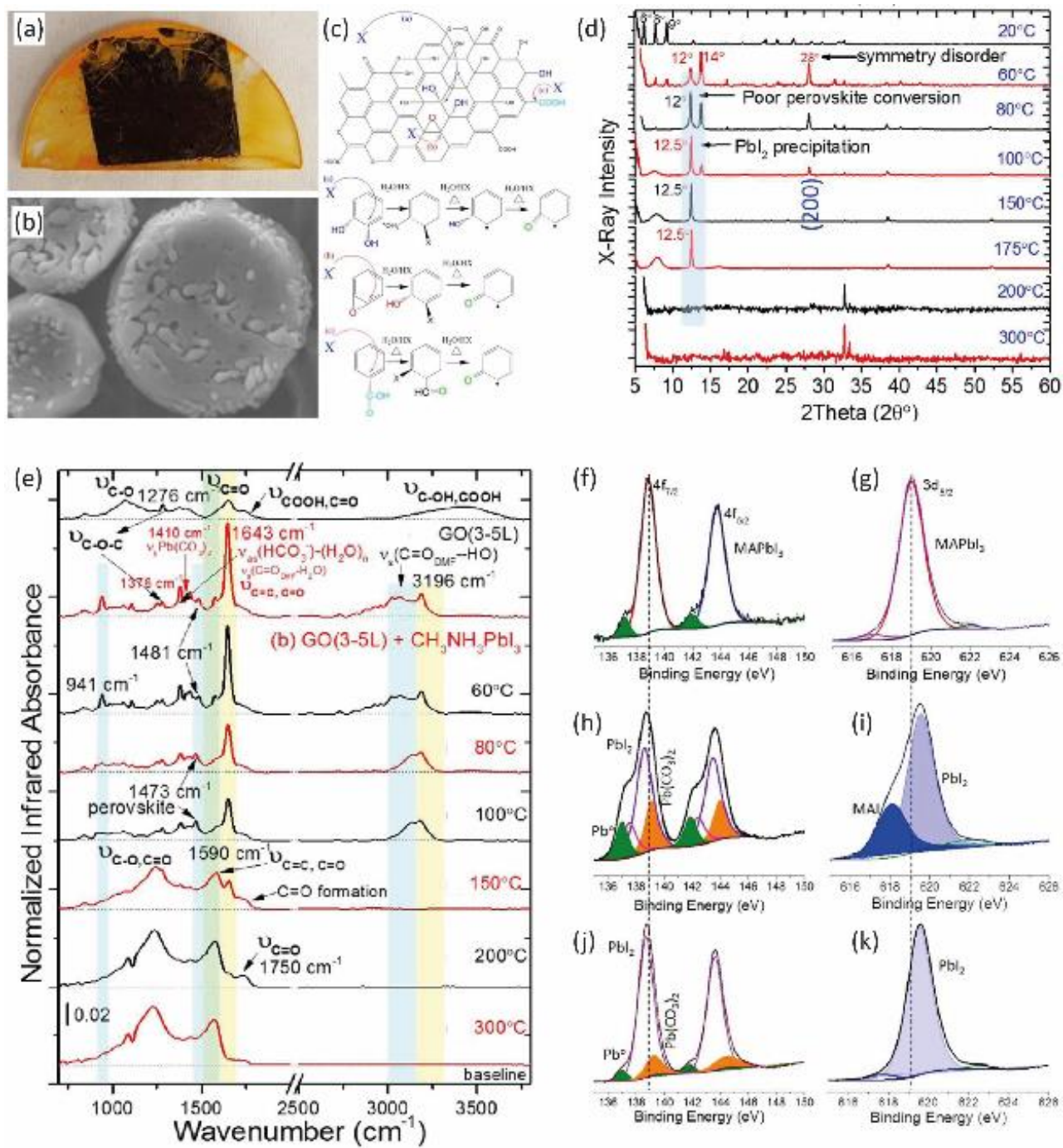

*Figure 40* – *(a) Photograph and (b) SEM image (magnification) of MAPbBr$_3$ grown on GO. (c) Potential reaction schemes of the MAPbX$_3$ precursors with GO. (d) XRD pattern showing inhibited MAPbI$_3$ formation. (e) FTIR spectra. (f-k) XPS Pb 4f and I 3d core level spectra for MAPbI$_3$ grown on (f,g) SiO/Si, (h,i) GO/SiO/Si at room temperature, and (j,k) GO/SiO/Si annealed to 130 °C. Reproduced from Ref. 305 with permission from The Royal Society of Chemistry.*

In these equations, solid compounds and precipitates are denoted with an index (s), whereas species in aqueous solution are indexed by (aq).

Acik *et al.* conclude that the combination of a GO surface with a hygroscopic environment would lead to poor HaP film formation, but found it to be less dominant, the faster perovskite film nucleation occurred (e.g. in the case of MAPbBr$_3$). If we now translate these findings to graphene oxide layers deposited on top of HaP surfaces, we can project that chemical driving forces for HaP dissociation are not only detrimental for long term stability, but will also dominate any energy level alignment process. From an application perspective, a mitigation of this effect could be achieved if the reactivity was decreased, for instance by employing reduced graphene oxide (rGO) films as electrodes.[266,305]

**C.3. The metal/HaP interface**

Aside from selected rare examples, such as the conductive carbon electrodes in the previous section, metal contacts are ubiquitously employed in HaP-based devices at the cell terminals. The perovskite film is often sandwiched between semiconducting CTLs and is not in direct contact with the metal layer. However, various scenarios exist that make the metal/HaP interface an important subject for dedicated investigations:

- Extrinsic species migration (e.g. from the metal electrode) can occur through the CTL towards the HaP film. This also includes migration and accumulation of dopants from the CTLs such as $Li^+$ in spiro-OMeTAD to the CTL/HaP interface.[113]
- Pinholes can form in the semiconducting CTL, potentially leading to direct physical contact between adjacent metal and HaP films.
- By design, device layouts might omit the CTL and directly implement the metal electrode on top of the HaP film, e.g. to reduce fabrication costs. One of the earlier examples for this geometry has been the "HTM-free PSC", for which a gold contact was directly evaporated on top of a $MAPbI_3$ film (see Figure 41a).[306,307]

In the following, we will mostly focus on the latter case, the deliberately designed HaP/metal contact, which allows for the most straightforward attribution of the energy level alignment and potential chemical reactions. Not mentioned in the three cases presented above is a fourth category: migration of the individual components of the HaP film and its decomposition products to the metal electrode. This scenario does not well represent a true HaP/metal interface, since it is unlikely that intact HaP fragments would migrate through the CTL to the metal layer. However, gaining insight into the reactivity of the individual HaP constituents with the metal is helpful to understand the chemical process(es) at well-defined HaP/metal interfaces. From a technological perspective, however, this case of migrating halide anions could be even more relevant as it pertains to one of the most prevalent degradation mechanisms that could lead to device failure. Especially iodine species can diffuse towards the electrode and corrode the metal film by oxidizing it, which is well documented for Ag and Al electrodes.[189,295,308,309]

The complexity of the HaP/metal contact can be estimated from capacitance voltage (C-V) measurements of the "HTM-free" device as demonstrated by Huang *et al.* for a TiO$_2$/MAPbI$_3$/Au layer stack (Figure 41b-d).[310] They simulated a physical model to fit the C-V data and found good agreement between experimental data and a fit for which the HaP film in the model was assumed to be n-type and a *strongly n-doped region* was introduced in the MAPbI$_3$ layer at the Au interface. The physical nature of the interfacial layer was not specified, but might be related to metallic-like surface states and charged species as discussed for the organic and oxide overlayers in sections C.2.1. and C.2.2. In equilibrium, the model suggests type inversion in the MAPbI$_3$ film induced by charge transfer from the Au overlayer, as in an idealized Schottky-contact between a metal and a semiconductor. Furthermore, Huang *et al.* calculate that the highly doped interfacial layer would reduce the built-in field and hence $V_{oc}$ of a device. This leads them to project that the solar cell characteristics could be improved by minimizing the impact of the interfacial layer, while enhancing the inversion in the cell, i.e. by raising the work function of the metal layer.[310]

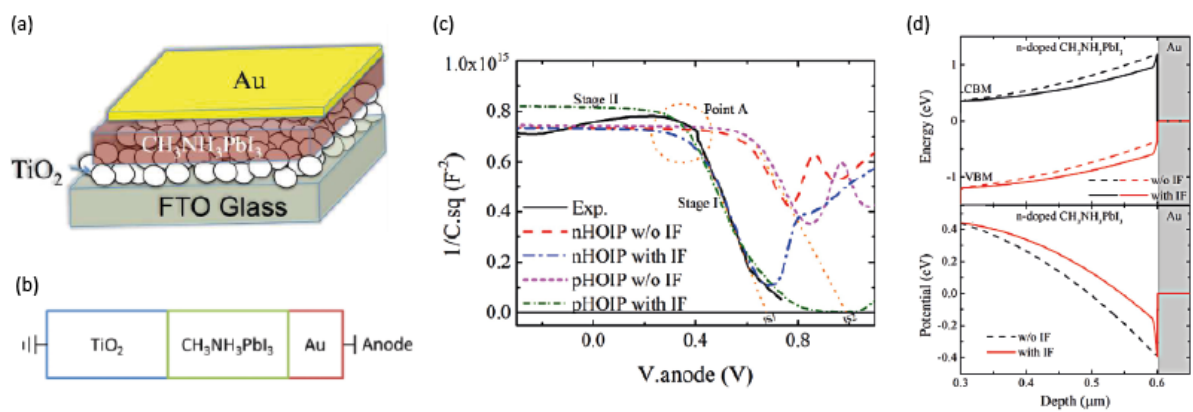

*Figure 41* – (a) Schematic of an "HTM-free" PSC configuration with Au top contact. Reprint from ref 307. (b) TiO$_2$/MAPbI$_3$/Au layer stack used in capacitance-voltage (C-V) measurements. (c) Experimental C-V data (black line) and model fits (dashed lines) for assumed n- or p-type hybrid organic-inorganic perovskite (HOIP) films, with and without the assumption of an interfacial depletion layer (IF). "Stage I" and "Stage II" indicate regimes of carrier depletion at high voltages above the transition "Point A". (d) Corresponding static band alignment (top panel) and potential profile (bottom panel). Reprinted figure with permission from Ref. 310. Copyright 2017 by EDP Science.

Interestingly, a related scenario was demonstrated in a different instance without a direct HaP/metal interface, but with a thin (< 20 nm) layer of the organic HTM 4,4'-Bis(N-carbazolyl)-1,1'-biphenyl (CPB) between a MAPbBr$_3$ film and metal top contact in an FTO/TiO$_2$/MAPbBr$_3$/CBP/metal cell.[311] In this study, Kedem *et al.* used metals with different work functions (Pb: 4.2 eV, Au: 5.1 eV, Pt: 5.6 eV) and found via SPV and EBIC measurements the work function to be strongly correlated to the device parameters, explaining this by charge transfer from the metal through the organic HTL to the HaP film to bring the layer stack in thermodynamic (electronic) equilibrium. As a result, the energy bands in the MAPbBr$_3$ film remain flat for the Pb top contact, while the high work function metals Au and Pt invert the HaP layer, which becomes p-type except next to the TiO$_2$ ETL, i.e., a *p-n* homojunction is formed.[311]

### C.3.1. Chemical reactions between metal film and HaP

Energy level alignment and chemical reactions at HaP/metal interfaces have been investigated via PES measurements for various combinations of HaP films and top metal layers. We begin with the Au/MAPbI$_3$ interface, as gold contacts are among the most relevant study systems due to their ubiquitous use in HaP-based devices. No strong chemical reaction is expected, in view of the chemical inertness of Au, corroborated by earlier experiments, such as growth of several monolayer thin, intact MAPbI$_3$ films on Au (see section B.2.),[100] or through the formation of ohmic contacts in Au/MAPbBr$_3$/Au devices.[312] We will find later that this hypothesis is not valid under all circumstances.

Liu *et al.* incrementally deposited Au layers on MAPbI$_3$ films and monitored energy level positions in the HaP film and the Au overlayer with a combination of UPS/XPS experiments.[313] The results are illustrated best with the Pb 4f and Au 4f binding energy positions, respectively, as shown in Figure 42a,b. As the first

Au sub-monolayer (0.5 Å) is deposited, the Pb core levels shift to higher binding energies, then gradually shift back towards lower binding energies with increasing Au film thickness. The Au core levels shift slightly to lower binding energies by about 0.4 eV over the first 3 Å. Liu *et al.* suggest that this shift can be interpreted as charging of the Au particles until an interconnected layer is formed, as the Fermi edge in the UPS data follows the same trend.[313] Interestingly, this trend does not correspond well to Huang's conjecture that for the n-type MAPbI$_3$ layer with the strongly n-doped interface layer, the Fermi level is shifted towards a mid-gap position, while a strongly p-type region forms at the Au/MAPbI$_3$ interface once the contact to the Au is established.[310] Instead, the scenario that Liu *et al.* encountered could point to the initial formation of charged surface states, similar to those proposed by Zu *et al.* for HaPs with a high concentration of Pb$^0$ at the surface,[164] with subsequent formation of a Schottky contact, i.e. upwards band bending in the HaP film towards the interface with the Au layer, if the initially formed surface states were donor-like. No further evidence of an interfacial chemical reaction was apparent from the XPS data, except that the HaP film closer to the interface, i.e. at higher Au coverage, appears more lead-deficient than deeper in the volume, i.e. compared to the measurement without the top Au layer and hence less attenuation of the PES signal.[313] We note, that a more direct experimental assessment and conditioning of the interface appears to be required to establish a conclusive picture.

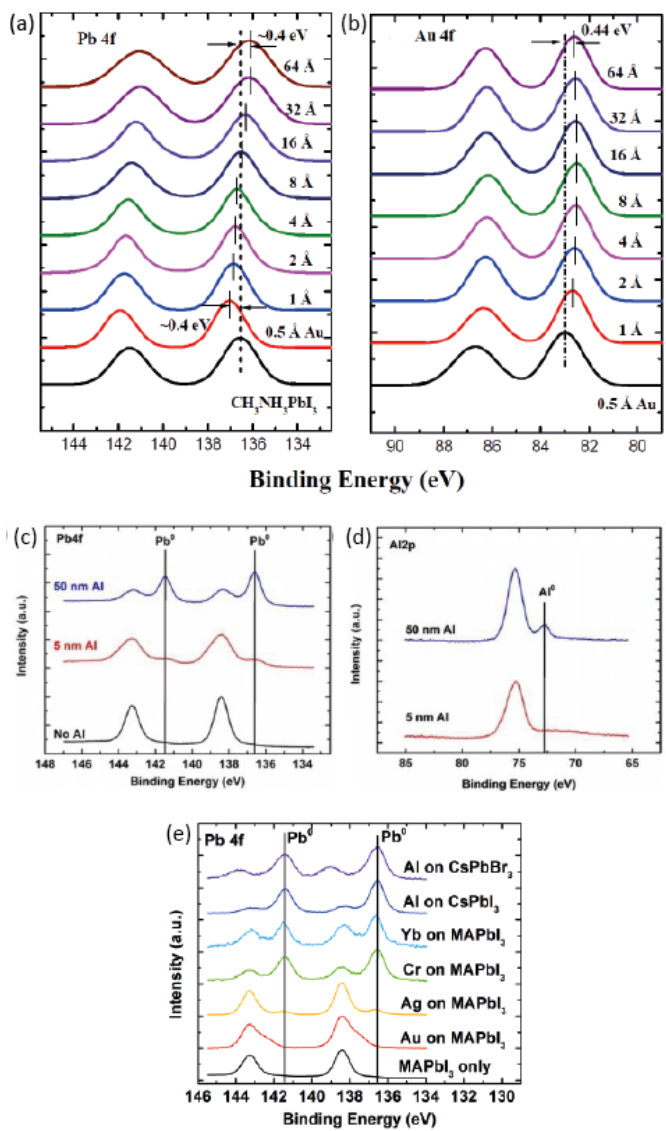

*Figure 42* – XPS data for metal layers deposited on HaP films. (a) Pb 4f and (b) Au 4f core levels for incrementally grown Au layers on MAPbI$_3$. Note that the color coding between the two plots is not matched for each gold overlayer thickness. Reproduced from Ref. 313 with permission from the PCCP Owner Societies. (c) Pb 4f and (d) Al 2p core levels for 5, and 50 nm thick Al layers on MAPbI$_3$. (e) Pb 4f core level for a set of various metal films deposited on either, CsPbBr$_3$, CsPbI$_3$, MAPbI$_3$. Reprinted with the permission from Ref. 189. Copyright 2016 American Chemical Society.

A different picture is envisioned for the Al/HaP interface, for instance for metallic Al evaporated on a TiO$_2$/MAPbI$_3$ sample. XPS data in Figure 42c,d reveal a redox reaction in which Pb$^{2+}$ is reduced to Pb$^0$, while Al is oxidized.[189] Notably, the reaction occurred spontaneously in vacuum without further external stresses such as light, moisture or oxygen. Zhao and coworkers further tested their hypothesis on redox chemistry between the metallic cation, Pb$^{2+}$ and a neutral metal film for other HaP/metal interfaces and found reduced Pb for Al deposited on CsPbI$_3$ and CsPbBr$_3$, as well as for Yb and Cr layers deposited on MAPbI$_3$ (Figure 42e). In case of Ag on MAPbI$_3$, the Pb$^0$ concentration was significantly smaller than in the case of the (Y,Cr,Al)/HaP interfaces and essentially disappeared for the Au/MAPbI$_3$ interface. However, Zhao's spectra of the Au/MAPbI$_3$ interface differ significantly from those by Liu *et al.*, who did not find evidence for any redox reaction products.[313] In contrast to that, Zhao *et al.* find shoulders at slightly lower binding energies in the Pb core levels and suggest that this finding could be attributed to partial charge transfer at the HaP/Au interface.[189] Importantly, the reaction partners in the HaP for the redox chemistry with the deposited metal species are the B-site metallic cation (here $Pb^{2+}$) and the halide anion ($X^-$) and not the A-site cations.

The origin of the discrepancy between various studies remains unclear but could be linked to experimental details. First, the HaP layers could exhibit different stoichiometries and surface terminations, which would be more or less likely to undergo or promote a redox reaction, i.e. organic surface terminations (MA, FA, …) could act as a buffer layer. Second, as in the earlier section C.2.2. on the HaP/semiconductor interfaces, measurement conditions (i.e. light bias, X-ray intensity) might be substantially different and induce changes in the HaP film through desorption or drive the redox reaction via photovoltage.

### C.3.2. Metal ion migration and diffusion barriers

The case of a (partial) redox reaction observed at the $MAPbI_3$/Au interface presented above hints to the fact that an HTM-free contact in this configuration would be metastable and at least pose a challenge for the long-term stability of the HaP device. Hence, even with an organic HTL as buffer between Au electrode and HaP absorber, the scenario of Au migrating to the HaP film could lead to a degradation of the device functionality and thus chemically inert Au electrodes could prove problematic for devices. Domanski and coworkers investigated the migration of Au into an $FA_{1-x-y}MA_xCs_yPbI_{3-z}Br_z$ film through a spiro-OMeTAD HTL in a conventional PSC device configuration, by performing ToF-SIMS measurements of cells before and after operation.[110] In their case, aging at elevated temperatures (75 °C) led to more pronounced performance degradation than aging at moderate temperature (20 °C). While no macroscopic structural or morphological changes were seen for the HaP film itself, the ToF-SIMS measurements done on the sample aged at high temperatures (here: 70 °C) revealed that gold had migrated to the bottom of the HaP layer down to the $TiO_2$/HaP interface, which was not seen in a control sample or in another sample aged at 30 °C.[110]

In order to limit Au diffusion into the HaP film, Domanski *et al.* applied a thin chromium layer between the HTL and the Au electrode. They further suggested that the addition of a metal oxide diffusion barrier (e.g. $Al_2O_3$) could further inhibit metal migration and hence improve the cell stability. Similar strategies have been proposed to overcome ion migration in the other direction, i.e. mobile halide species that could first migrate into the HTL and then into the metal electrode. As laid out in the previous sub-section, this would lead to strong redox chemistry and finally to complete corrosion of the metal film. This can be prevented by introducing a thin protective layer between HTL and metal electrode. While highly reactive with an adjacent HaP film (see section C.2.2.), even $MoO_3$ can fulfill this role, as demonstrated for thin $MoO_3$ layers sandwiched between spiro-OMeTAD HTL and metal top electrodes.[294] Sanehira *et al.* demonstrated that the $MoO_3$ layer between the HTL and a Ag or Al electrode formed dense and stable $AgO_x$ or $AlO_x$ layers that inhibit the diffusion of reactive halide species into the electrode.[295] Preventing metal migration through the HTL into the HaP, e.g. by reduced graphene oxide between HTL and Au electrode,[76] proved to be a successful strategy to mitigate the interface formation.

### C.4. Passivation strategies at the HaP top interface

The previous sections, which describe the growth of functional thin films on HaP layers, demonstrate that the energy band alignment depends on many subtle aspects of the interfacial chemistry and can lead to pronounced band bending in the perovskite layer, e.g., in case of HaP/oxide interfaces. Hence, similar to the earlier discussion on the substrate/HaP interface (section C.1.4), we need to consider strategies to passivate trap states and inhibit chemical reactions at the top HaP/CTL or HaP/electrode interface. As mentioned in the introduction, interfacial electronic states and recombination processes are major considerations when quantifying operational parameters of HaP-based devices,[69] although this view does not fit well with assessments that HaP surfaces are relatively inert, compared to those of inorganic semiconductors such as III-V compounds.

One likely reason for the low carrier recombination at exposed HaP surfaces is the preferential termination with an organic compound (e.g. MA or FA), and hence a closed shell electronic system without dangling bonds, thus exhibiting low trap state density; a feature that is also prevalent in organic semiconductor technologies such as in organic photovoltaics (OPV). This characteristic should be absent for purely inorganic HaPs, as suggested by the non-negligible trap DOS tailing into the band gap.[101] Interestingly, even with HaP compositions featuring mixed Cs and organic cations on the A-site, the tendency for an organic surface termination is maintained. This is also apparent for $CsPbI_3$ quantum dot systems, which despite their nominal inorganic composition are usually capped with organic anionic ligands that can be exchanged in tailored ligand exchange reactions to achieve an FA surface termination, for instance.[78–80,314] At the same time, tailoring the organic surface termination can serve to make an organic-inorganic HaP film (surface) more robust and suppress surface recombination: by introducing a larger organic anion, such as butylamine or aminovaleric acid ($HOOC(CH_2)_4NH_2$, on the A-site, similar to the example of sheet terminations in mixed 2D/3D compounds. Even for a system in which a long-chained or bulky A-site molecule is added in the precursor stage, phase segregation (e.g. of a 2D phase nucleating at the interface and a 3D bulk phase with the small-sized cation) can occur for the resultant film and lead to substantial enhancement of the passivation and long-term stability, presumably by suppressing chemical reactions.[75] However, the structural analysis and microscopic identification of the molecules in the film is not straightforward. Hence, it remains unclear whether passivation effects can be attributed to passivation of the bottom interface with the substrate, the top surface, or grain boundaries in the film. It is particularly difficult to assess whether the additives result in better HaP crystallization and growth, or if residual additives in the film reduce non-radiative recombination. In the following, we choose a few selected examples, for which the passivation is thought to leave the film morphology and structure unchanged, and which is thought to be localized mostly at the HaP surface or the top interface between HaP and CTL/electrode.

We note that the nature of defect states in the HaP film and the mechanisms of defect passivation are complex and are treated in a separate review on the topic,[31] which provides a comprehensive account of passivation and means to assess properties such as defect types and densities or recombination rates, determined experimentally with the support of theory. In the following we give only a few selected examples of passivated HaP interfaces.

**C.4.1. Self-assembled monolayer deposition on top of HaP**

Passivation of HaP surfaces can be achieved by coating the HaP layer with monolayers of organic molecules. A straightforward way to realize such a system is to first grow the pristine HaP film and subsequently apply the passivating layer on top. While procedures for self-assembling monolayer (SAM) growth on oxide surfaces (see section C.1.4.) have been well-established, grafting mechanisms for molecular layers on HaP surfaces had to be developed separately.

One of the most promising ideas has been to form supramolecular compounds at the surface through a Lewis acid-base reaction as previously explored in many semiconductor fields.[255,256] Abate *et al.* demonstrated this method for growing iodopentafluorobenzene (IPFB) on $MAPbI_{3-x}Cl_x$ films as depicted in Figure 43a by immersing the HaP layers in a solution of IPFB.[131] IPFB acts as a Lewis acid and binds to under-coordinated halide anions on the HaP surface. In this example, the IPFB treatment is applied immediately before the deposition of a spiro-OMeTAD HTL and completion of a PSC device. The effect is that hole recombination at that interface is reduced and the diode characteristic improved.[131]

This type of approach has been explored in several other instances – and also for the passivation of other defect states. When describing the deposition of organic acceptor molecules (e.g. HAT-CN) on HaP in section C.2.1., we discussed the mechanism of passivating Pb surface defects on halide-deficient HaP surfaces, which act as donor states.[190] Conceptually, the same goal can be achieved by applying an organic molecular layer, for example by spin-coating, on an HaP surface. Noel *et al.* used thiophene and pyridine molecules to passivate $MAPbI_{3-x}Cl_x$ surfaces and performed TR-PL measurements on those samples (figure 43b).[315] From the increase in PL decay time and quantum efficiency, they conclude that non-radiative recombination is decreased for the treated samples and propose that the molecules act as a Lewis base and passivate under-coordinated Pb atoms. We note, that these optical measurements serve as a proxy to estimate trap state densities. In this case, the model fit, which considered dynamic interplay between free charges, excitons, and electronic sub-gap states for HaPs,[316] led to the conclusion that trap state densities were reduced after the SAM treatment; thus, presumably these states are mostly at the surface of the HaP or at the interface with the subsequently deposited organic HTL.[315] A more detailed and accurate identification of type, density, and spatial location of recombination centers in the sample can be refined by an adjustment of measurement conditions and model parameters.[317]

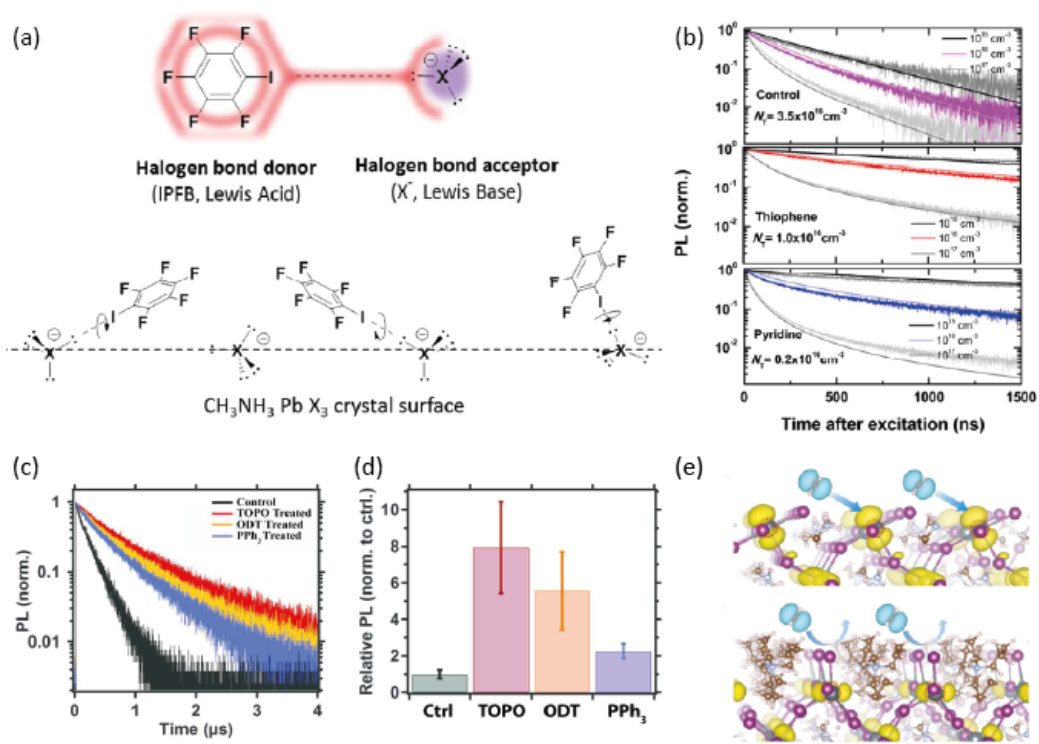

*Figure 43* – *(a) Schematic of iodopentafluorobenzene (IPFB) anchored as a Lewis acid to a $MAPbX_3$ surface with the undercoordinated X anion as Lewis base. Reprinted with the permission from Ref 131. Copyright 2014 American Chemical Society. (b) PL transients for $MAPbI_{3-x}Cl_x$ films without surface treatment (top), thiophene surface treatment (middle) and pyridine surface treatment (bottom). The curves were fitted to a model described in Ref. 316 to derive the trap state densities $N_T$. Reprinted with the permission from Ref. 315. Copyright 2014 American Chemical Society. (c) PL decay transients and (d) integrated PL intensity of $MAPbI_3$ films treated with various surface modifications. Reprinted with the permission from Ref. 133. Copyright 2016 American Chemical Society. (e) Schematic of water adsorption on an MA-terminated (top) and a TEA-terminated (bottom) $MAPbI_3$ surface. Reprinted with permission from Ref. 319. Copyright 2016 by Springer Nature Publishing AG.*

Further targeted experiments focused on the optical characterization of HaP films with and without surface treatment. 1,2-dithiol (EDT) was used by Stewart *et al.* to substantially reduce recombination in MAPbI$_3$ nanocrystals, presumably by passivating lead-rich surfaces. Given that the occurrence of such lead-rich species at the perovskite surface is not necessarily common, it is suggested in the study that EDT might etch iodide-rich surfaces, thereby exposing lead-rich surfaces, onto which the molecules eventually bind.[132] deQuilettes *et al.* expanded on similar experiments using trioctylphosphine oxide (TOPO), octadecanethiol (ODT), and triphenylphosphine (PPh$_3$) as post-treatment modifiers for MAPbI$_3$ films.[133] The PL data are summarized in Figure 43c,d and follow the trend observed for the previously described surface treatments: PL decay time and integrated intensity increase, which points to a reduction in non-radiative recombination rates. For these samples, the character of a surface treatment without substantial percolation of the surfactant into the HaP film due to steric hinderance is evidenced by glow discharge optical emission spectroscopy, which probes the chemical depth profile. For instance, in case of the ODT surface treatment, sulfur was exclusively found at the sample surface. Additionally, chemical shifts in the solid-state NMR data (e.g. demonstrated for the phosphorous moiety for the samples treated with PPh$_3$) revealed that the molecules donate an electron to the adjacent perovskite layer and hence would likely act as a ligand on the film surface.[133]

In the next iteration of this approach, molecules used as Lewis bases and serving as surface termination would not only improve the optoelectronic performance by passivating defect sites (in this case under-coordinated Pb atoms) in the HaP film, but also shield it from environmental stresses and mitigate chemical reactions that would otherwise cause decomposition and finally degradation. For this purpose, hydrophobic molecules can be employed, which reduce the uptake in moisture in the HaP film and inhibit the oxygen-induced degradation mechanisms. Generally, the atomistic molecular adsorption mechanism can be modeled in first-principle simulations of the HaP surface and the respective adsorbates. For instance, DFT calculations suggest that thiophene molecules preferentially anchor to the MAI-terminated MAPbI$_3$ surface, while MD simulations with additional H$_2$O molecules reveal the water-repellent properties of the so-modified HaP surface.[318]

The most obvious choice for this next iteration falls onto amine molecules with extended alkyl branches. Such decoration of MAPbI$_3$ surfaces has been pursued with a range of large ammonium compounds, such as tetra-ethyl ammonium (TEA) and longer-chained tetra-alkyl ammonia,[319] which preserve the valence with respect to the MA in the bulk HaP on the TEA adsorption site, while the long side chains prevent diffusion of extrinsic species. Aging studies then show that HaP films capped with TEA molecules withstand environmental impact, such as moisture ingress (see Figure 43e), while the unprotected films degrade into PbI$_2$ due to a loss of MAI. This enhanced stability was predicted in DFT calculations and used to achieve higher and long-term stability in PSC devices.[319]

Adopted from the silicon solar cell playbook, where a thin insulating oxide or amorphous silicon layer is inserted between the intrinsic Si film and the highly doped Si or ITO contact, e.g. in the heterostructure with intrinsic layer (HIT) cell geometry, a thin buffer layer can also be incorporated between HaP film and CTL. Compared to earlier examples in which substitution of the surface MA moiety was explored, the absence of anchoring of this layer on the perovskite film promises to make it simpler than SAM growth, as lattice matching requirements between buffer layer and HaP are more relaxed. This approach has been demonstrated in inverted devices with a thin insulating layer, e.g. polystyrene or fluoro-silane, between the perovskite and a $C_{60}$/PCBM ETL.[320] That study reports that insertion of this buffer layer between HaP and ETL increased the performance of PSCs, supposedly mainly by suppressing recombination, while at the same time serving as encapsulation to prevent moisture-related HaP film degradation.

**C.4.2. Inert oxide lamination**

We conclude this review by discussing other passivating materials than the ultra-thin organic surface modifiers presented in the previous section. More generally speaking, various encapsulation methods have been proposed to shield the completed PSC device from external factors, from basic early approaches using organic materials such as Surlyn,[321] thermoplastic polymers with integrated adhesive,[322,323] or ethylene vinyl acetate, which were shown to maintain higher PSC mechanical stability after thermal cycling than Surlyn encapsulation.[324]

As a promising alternative to these organic encapsulants, amorphous $TiO_2$ or $SnO_x$ layers made by atomic layer deposition (ALD) have been used as barriers against moisture infiltration in PSCs.[325,326] A similar use could be attributed to ALD-$Al_2O_3$ coatings, which have shown high versatility, being grown over both hydrophilic and hydrophobic substrates, and tuned to present either hydrophilic or hydrophobic surfaces themselves.[327] ALD-$Al_2O_3$ films provide good optical transmittivity, which is a requirement for many PV device configurations and exhibit water vapor transmission rates on the order of $1 \times 10^{-6}$ g m$^{-2}$ day$^{-1}$. However, those films are typically obtained from trimethylaluminum (TMA) and water as precursors, at high process temperatures (80 - 300 °C),[328] i.e., close to or well above "safe" PSCs processing temperatures.

Consequently, several of these protocols need to be modified and adapted to PSC encapsulation. One of the first examples was demonstrated by Kim and Martinson, who compared different precursors and procedures for ALD- $Al_2O_3$ coatings directly on $MAPbI_{3-x}Cl_x$ for deposition temperatures around 100 °C.[329]

Their results (Figure 44a) show that at these elevated temperatures any combination of TMA and ozone or water as oxidizer would result in a discoloration of the HaP layer and hence degradation. This situation was eventually corrected by using a non-hydrolytic low temperature process with aluminum triisopropoxide (AIP) and acetic acid (AA) as the oxygen source. A similar preservation of the HaP film was obtained by an analogous process to grow ALD-TiO$_2$ from titanium tetraisopropoxide (TTIP).[329] Around the same time, Dong *et al.* explored the introduction of a thin ALD-Al$_2$O$_3$ layer between MAPbI$_3$ and HTL using O$_3$, but found it to act as a strong oxidizer that degraded the perovskite and converted it to PbI$_2$.[330] This set of results is in line with our earlier remarks on the deposition of oxide CTLs on HaP films (section C.2.2.) by peALD.[297] Clearly, the growth of an inert passivating oxide layer on HaP requires milder conditions than those used normally, to preserve the HaP surface and suppress the generation of defects as a result of interfacial chemistry.

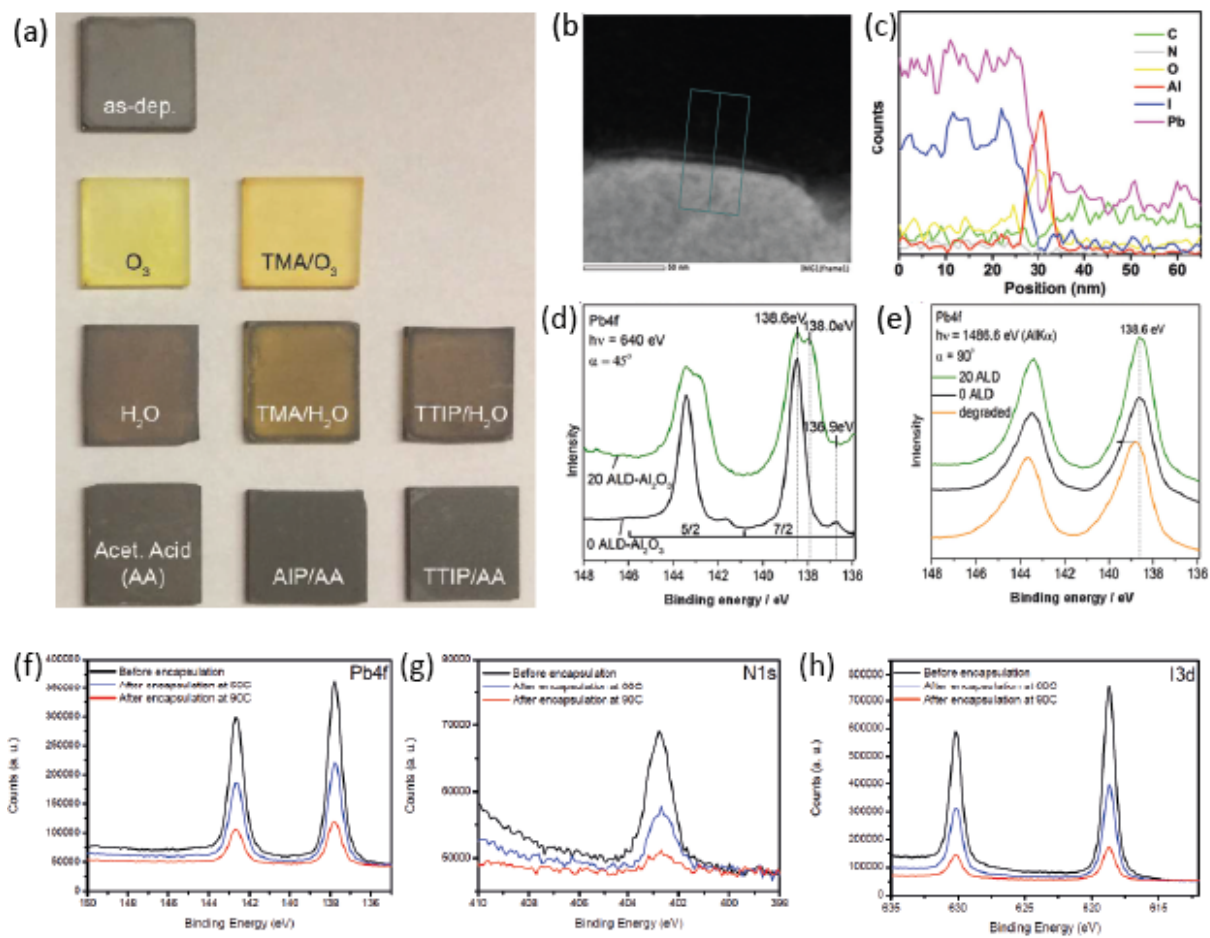

*Figure 44* – Oxide lamination of HaP films by ALD. (a) Photograph of $MAPbI_{3-x}Cl_x$ films with and without ALD overlayers: first row – plain, second row – with $O_3$ treatment and $AL_2O_3$ from ($O_3$/TMA); third row – $H_2O$ treatment, $Al_2O_3$ from ($H_2O$/TMA), and $TiO_2$ from ($H_2O$/TMA); fourth row – Acetic acid (AA) treatment, $AL_2O_3$ from (AA/AIP), and $TiO_2$ from (AA/TTIP). HaP film degradation is avoided for non-hydrolytic processes using isopropoxides and AA precursors. Reproduced from Ref. 329 with permission from The Royal Society of Chemistry. (b) Cross-section TEM image of a $MAPbI_3$/ALD-$Al_2O_3$ (0.8 nm)/spiro-OMeTAD layer stack and (c) elemental profiles from the EDX analysis of the area, marked in (b). Reproduced from Ref. 331 with permission from The Royal Society of Chemistry. (d) Synchrotron-based surface sensitive and (e) lab-source-based XPS measurements of the Pb core level region for $MAPbI_3$/ALD-$Al_2O_3$(~2.8nm) interfaces. Reprinted wih the permission from Ref. 332. Copyright 2016 WILEY-VCH Verlag GmbH & Co. KGaA, Weinheim. (f-h) XPS core level analysis of $MAPbI_3$/ALD-$Al_2O_3$(~2nm) interfaces. Reproduced from Ref. 333 with permission from The Royal Society of Chemistry.

In a later study, Koushik *et al.* used an ultrathin (0.8 nm) Al$_2$O$_3$ layer deposited from TMA/H$_2$O at 100 °C directly on an HaP film to make a hydrophobic tunneling insulating layer and enhance the air stability of a PSC in planar architecture.[331] They confirmed the formation of a dense Al$_2$O$_3$ film by high angle annular dark field (HAADF) scanning TEM of a sample cross-section and verified by XRD that the ALD-Al$_2$O$_3$ layer prevents decomposition of the underlying bulk MAPbI$_{3-x}$Cl$_x$ into MAI and PbI$_2$. To elucidate the interface chemistry for ALD-Al$_2$O$_3$ growth on HaP films, Kot *et al.* performed XPS measurements on an ALD-Al$_2$O$_3$ layer (20 cycles, ~2.8 nm), grown at room temperature from TMA and H$_2$O precursors on MAPbI$_3$.[332] Measurements were taken at low excitation energy ($\hbar\omega$= 640 eV) at a synchrotron light source under a 45° take-off angle, and with a higher excitation energy Al K$_\alpha$ X-ray source ($\hbar\omega$= 1486.6 eV) at normal emission to increase the probing depth. The comparison of the more surface-sensitive Pb 4f core level spectra (via synchrotron source) with the larger probing depth signal (laboratory X-ray source) led to several marked differences (figure 44d,e). The former revealed a metallic lead component for the bare MAPbI$_3$ surface, and the signature of a Pb-O bond following the application of the ALD-Al$_2$O$_3$ layer. Since this signature is absent in the data generated in the more sub-surface sensitive experiment, the authors assigned this contribution to a thin layer of PbO$_x$ at the HaP/Al$_2$O$_3$ interface.[332] We note at this point that a clear discrimination between this effect originating from the ALD processing or from transient changes to the film during the measurement remains difficult.[160] The high X-ray flux of the synchrotron light source can lead to the formation of metallic Pb, similar to the white light illumination experiment performed by Zu *et al.*,[164] and could thus induce chemical reactions between these newly formed species and any oxide overlayer, in particular if partially converted ALD precursors are still present at the interface.

In a different study, Ramos *et al.* employed low-temperature (60 °C) ALD deposition to encapsulate completed PSC devices in a glass/FTO/$TiO_2$/$MAPbI_3$/spiro-OMeTAD/Au geometry, increasing the long-term stability of the device.[333] They also investigated the direct deposition of ultra-thin (~2 nm) ALD-$Al_2O_3$ layers on the glass/FTO/$TiO_2$/$MAPbI_3$ layer stack from the same fabrication procedure to investigate the influence of the ALD process on the HaP film, as it could occur either due to thermal stress during the ALD layer fabrication or through intimate contact of the precursors with the $MAPbI_3$, e.g. at the sample edges or through pin-holes in the organic HTL/electrode overlayer. The Pb, I and N core level spectra, presented in figure 44g-h, give no indication of any new interface chemical species. However, the elemental Pb/I and Pb/N ratios revealed that, at higher processing temperatures (90 °C) in the ALD layer production step, the film became more iodine deficient, whereas the Pb/I ratio remained unchanged in the low temperature (60 °C) process, suggesting that $MAPbI_3$ decomposes and volatile MAI species evaporate at higher temperatures.

In summary, the encapsulation of HaP films with inert oxide layers with the objective of passivating defect states on the surface and protecting the perovskite from extrinsic chemical species seems to be possible, if the processing conditions to grow the oxide overlayer are sufficiently mild: no excessive temperature, no reactive oxygen species (e.g. ozone, $O_2$-plasma), and no exposure to energetic radiation such as UV light. The requirement to run a low-temperature process is directly linked to the evaporation of volatile components in the HaP composition (especially MA and MAI) and could potentially be relaxed for purely inorganic perovskites such as Cs-based HaPs. However, the growth of oxide films on these inorganic HaPs could present a different set of challenges as the A-site cation presumably does not present a closed shell electronic system and is thus prone to undergo oxygenation itself.

**D. Conclusion and Outlook**

In this review we introduced the role of interfaces for HaP-based semiconductors, elucidated the surface chemistry, and finally investigated the energy level alignment and interface chemistry between HaP compounds and the most prevalent materials used for charge transport, cell termination or interface passivation. The field of HaP-based optoelectronics experienced a significant surge in the past six years, during which the community was able to reach a considerable understanding of fundamental materials properties of the HaPs. However, one must concede at this point that the best candidates for adjacent transport layers did not follow clear-cut guidelines and – quite frankly – their choice is currently based on rudimentary scientific insight. One of the reasons is that we still lack a good description of the exact function of each component in the HaP material itself, and of their interplay regarding macroscopic physical properties such as electronic transport characteristics. This applies in particular to the bulk and surface defects, which are central to the main questions and topics of this review. From a theoretical view point, many credible yet contradictory findings on defect types and energies were reported for these materials. And from the experimental side, the unambiguous identification of defect types and properties has been very difficult, as already shown over the past decades for simpler and more stable semiconductor materials. Thus, we are presently in a situation where we cannot be sure that old and trusted models and approaches for defects remain applicable.

With respect to interfacing the HaP absorber to CTLs, the recent research activities focused on assessing the impact of the electronic energy level alignment on the device functionality. Yet, this could only be pursued for the most common layer combinations due to the extensive and time-consuming nature of these studies. The derivation of a set of universal guidelines on electronic level alignment for any combination of HaP film and CTL material faces a major obstacle directly from the onset: the chemical composition of the HaP film and surface, which remains mostly unknown and undisclosed, but could vary substantially between studies. Finding a clear set of standard starting conditions remains a major challenge, since the HaP layer can comprise highly volatile species, such as methylammonium, which could desorb easily, and mobile reactive ions, such as iodine, which can migrate and reconstruct the surface. The preferential surface termination of an $ABX_3$ compound presented in the majority of literature reports, i.e. an AX-terminated surface, could hence be easily disrupted and lead to the exposure of under-coordinated B-site and X-site atoms as reaction partners for any adjacent functional material. On top, the complexity and instability of the surface conditions could change during most experiments that involve high fluences of radiation or electric fields, which makes it particularly difficult to rely on theoretical calculations for the interpretation of each experiment.

Looking at the basic question considered in this review, i.e. **if key optoelectronic functionalities of the HaP semiconductor device are determined or limited by the HaP interface**, we must admit at this point that the response can only be ambiguous at best: yes, controlling the interface is central to obtaining the maximum optoelectronic performance and stability of HaP thin films, which translates into the requirement of tailored interfaces in HaP-based devices. However, to gain control over this set of parameters further research to understand and condition the interfacial chemistry is required, a goal that remains elusive due to the high degree of chemical complexity of the HaP film itself, which is at present beyond the complexity of technologically established semiconductor surfaces.

**D.1 Interface as a critical control parameter for HaP film properties**

In spite of these difficulties, several findings can be used to inform us about the qualitative importance of interfaces and guide us to the next targeted interface studies. One of the most striking results from the various interface studies presented here is the apparent motion of the Fermi level in the HaP film as a function of the electronic structure of adjacent layers. Beginning with a change of the substrate underneath, $E_F$ in the HaP film can be found either close to the CBM or at mid-gap. This property could be related to the interfacial chemistry and the subsequent creation or passivation of defect states in the HaP film. The general rule, however, seems to be that growth of HaP films on oxide substrate leads to the unintended formation of a defective interlayer, which dominates charge extraction and recombination.

That interface is thus also the principal locus for charge accumulation, ion migration due to the compositional gradient and chemical pseudo-capacitances due to chemistry between the substrate and the constituents of the HaP film that migrate to this interface. The key to solving hysteresis effects in PSCs and other HaP devices probably rests in understanding the charge and species exchange at the interface to the active HaP film.

A similar scenario is observed, when other materials such as semiconducting CTLs or metal electrodes are grown on top of the HaP film. One can certainly state that the surfaces of the most regularly employed HaP layers are relatively inert compared to those of inorganic semiconductors (Si or III-V compounds). This property originates from the fact that in an organic-inorganic HaP, the surface termination likely corresponds to organic A-site cations without unsaturated bonds. However, because of the volatility of some constituents described above, surface defects and associated electronic states can form readily, and the surface can exhibit undercoordinated species with unsaturated bonds. This transition could also be reached during the deposition process of the overlayer, if the overlayer material is too reactive or the process conditions are too harsh.

Looking at organic CTLs, either deposited on top of HaP films or used as a substrate, one of the issues discussed earlier in this review is the correlation, or lack thereof, between the expected electronic level alignment at the interface and the performance parameters of a PSC device. Theoretically, the attainable photovoltage should be limited by the band offsets between the photoabsorber and the adjacent CTLs. We have seen, however, that the photovoltage is by and large independent of the mismatch between the energy band positions as long as the offsets are not too extreme to cause a carrier extraction barrier, in which case the diode characteristics are skewed and the fill factor collapses. This holds only true in part, since the presence of defect states at the surface of the HaP can change the band alignment significantly.

It is also clear at this point that no unified picture has been established for the band alignment mechanisms between HaP layers and more exotic CTLs and electrode configurations, interfacing the HaP film to conductive carbon CTLs, metal oxide CTLs and metal electrodes. Processes at these interfaces are dominated by interface chemistry, which can come in many combinations of Lewis acid-base and redox reactions and dramatically change the interface composition as the interface is formed, over time or under stress through extrinsic catalysts, electric fields or light.

Overall, the full implications of the interface energetics on recombination and barrier formation are convoluted with changes in the interface chemistry. At the same time, high mobility of some ions in the HaP film could even lead to a change in bulk properties, and therefore attempts to assess the related device physics only as a function of interface formation – or even more narrowly in terms of the electronic level alignment at the interface – remain somewhat futile or at least, premature. Thus, the challenge to formulate the desired energy band alignment between these electronically dissimilar compounds limits us in deriving useful guidelines for prospective HaP/CTL combinations.

**D.2 Interfacial Design meets HaP technology: Future research roadmap**

Ultimately, the main path to further technological advances in the field of HaP-based optoelectronics is robust and durable passivation of interfaces and, hence, suppression of interface chemistry. To overcome the limitations in assessing the interface materials properties, we propose to expand the current body of work by focusing first on the fundamental electronic processes at the HaP surface and across interfaces to adjacent transport materials while constantly monitoring the chemical composition. Conceptionally, this can be achieved in an experiment by establishing *operando* measurement protocols, that track chemical and electronic information at the same time.

We need therefore to develop reliable experimental methods that capture the most relevant physical properties (work function, defect density, etc.) and chemical properties (stoichiometry, oxidation states, etc.) pertaining to charge transport across the interface and at time scales compatible with the chemical processes that are associated with ion migration or desorption. This can most likely be achieved by combining advanced optical and electron spectroscopies/microscopies, to correlate charge carrier dynamics with the interface energetics and species formation. The focus of this endeavor should be on a precise measurement of interface energetics, energy barriers, and associated charge carrier dynamics (transfer and recombination rates). While doing so, we need to keep in mind that, due to the metastability of HaP-related interfaces, the measurement probes and parameters (light, bias, etc.) can cause chemical and structural reorganization of the interface, which needs to be taken into consideration when interpreting the results: what are potential precursors at the interface that would lead to the observed products and could such precursors be tracked qualitatively under less invasive measurement conditions (lowered flux)?

Once these tools are developed, we could begin to establish more comprehensive models for electronic processes at these interfaces and correlate them to the device physics. This step would involve a broad screening of potential CTL materials beyond the classical organic semiconductor and transparent conductive oxides, while evaluating their interaction with the HaP film. The connection between the atomistic picture of interface chemistry and the device functionality would require concomitant electrical measurements of the full device or sub-devices to verify the estimated charge carrier dynamics (i.e. transfer barriers and recombination velocities). Ideally, this device fabrication and characterization effort would be coupled *in situ* to the spectroscopic tools to achieve a hand-shake routine to countercheck the results.

The next challenge would then be to use this set of methods to differentiate between effects driven by the interface chemistry, from those that originate from the bulk properties of (hybrid) HaP compounds. Only then could the HaP film stoichiometry and CTL interlayer choices be fine-tuned to meet the needs of electronic heterostructures in novel devices. This effort will require further characterization and analysis routines with variable depth resolution (such as synchrotron-based XPS and XAS methods) as well as synthesis routes for dedicated layer-by-layer grown model systems, e.g. through CVD or PVD processes with improved precision and reproducibility. We project that this avenue would lead the research community to a more comprehensive picture, that encompasses the atomistic origin of the electronic interaction and chemistry between the material systems as well as the energy band alignment effects on the electrical properties of the device. From a technological point of view, this gain in insight would deliver a better assessment for the choice of materials and guide synthesis routines and film processing, with the immediate goal of achieving more efficient and stable devices. In the long run, a better understanding of the interface chemistry in HaPs can be leveraged for novel electronic applications, e.g. in the form of HaP heterostructures. We stress that the insight gained from better understanding defect chemistry and carrier dynamics in such HaP heterostructures could enable novel technologies, such as hot-carrier solar cells, tunable quantum-well lasers, and room-temperature qubits. At the moment, however, our main focus is on determining a reliable assessment route of the interface chemistry, which dominates any electronic process across the interface.

**ACKNOWLEDGEMENTS:**

P. Schulz thanks the French Agence Nationale de la Recherche for funding under the contract number ANR-17-MPGA-0012. D. Cahen and A. Kahn acknowledge support by US-Israel Binational Science Foundation (Grant No. 2014357). D. Cahen also thanks the Institute Photovoltaïque d'Île-de-France for a visiting professorship and the Ullmann family foundation (via the Weizmann Inst) for support.

**GLOSSARY:**

AFM: Atomic Force Microscopy

ALD: Atomic Layer Deposition

ARPES: Angle Resolved PhotoEmission Spectroscopy

BG: Band Gap

BZ: Brillouin Zone

CB: Conduction Band

CBM: Conduction Band Minium

CNT: Carbon NanoTubes

CPD: Contact Potential Difference

CTL: Charge Transport Layer

CTM: Charge Transport Material

CVD: Chemical Vapor Deposition

DLTS: Deep Level Transient Spectroscopy

DSSC: Dye Sensitized Solar Cell

DOS: Density Of States

DFT: Density Functional Theory

EA: Electron Affinity

$E_F$: Fermi Level

EQE: External Qunatum Efficiency

ETL: Electron Transport Layer

ETM: Electron Transport Material

FA: Formamidinium

FTO: Fluorine doped Tin Oxide

FY-XAS: Fluorescence Yield X-ray Absorption Spectroscopy

$\Phi$: Work Function

GGA: Greatest Gradient Approach

GO: Graphene Oxide

HAXPES: HArd X-ray PhotoEmission Spectroscopy

IE: Ionization Energy

IQE: Internal Qunatum Efficiency

IPCE: Incident Photon-to-electron Conversion Efficiency

IPES: Inverse PhotoEmission Spectroscopy

ITO: Indium doped Tin Oxide

HOMO: Highest Occupied Molecular Orbital

HTL: Hole Transport Layer

HTM: Hole Transport Materials

LEED: Low Energy Electron Diffraction

LUMO: Lowest Unoccupied Molecular Orbital

MA: Methylammonium

MD: Molecular Dynamics

PES: PhotoEmission Spectroscopy

PL : Photoluminescence

PSC: Perovskite Solar Cell

rGO: reduced Graphene Oxide

HaP: Halide perovskite

QFLS: Quasi-Fermi Level Splitting

QSGW: Quasiparticle Self-consistent GW

SAM: Self-Assembled Monolayer

SOC: Spin-Orbit Coupling

STM: Scanning Tunneling Microscopy

SWCNT: Single-Walled Carbon NanoTube

TFSI: bis-(TriFluoromethaneSulfonyl) Imide

ToF-SIMS: Time-of-Flight Secondary Ion Mass Spectrometry

UPS: Ultraviolet Photoemission Spectroscopy

VB: Valence Band

VBM: Valence Band Maximum

TCO: Transparent Conductive Oxide

XPS: X-ray Photoemission Spectroscopy

XRD: X-Ray Diffraction

XRR: X-Ray Reflectrometry